\documentclass[aps,prd,10pt,groupedaddress,nofootinbib,reprint]{revtex4-2}
\usepackage[utf8]{inputenc}
\usepackage{textcomp}
\usepackage[dvipsnames]{xcolor}
\usepackage{amssymb,amsfonts,amsmath}
\usepackage[sort&compress]{natbib}
\usepackage[english]{babel}
\usepackage{braket}
\usepackage{graphicx}
\usepackage{bm}
\usepackage{bbm}
\usepackage{braket}
\usepackage{afterpage}
\usepackage{gensymb}
\usepackage{dcolumn}
\usepackage{mathtools}
\usepackage{IEEEtrantools}
\usepackage{pstricks,pst-coil,pst-node,pst-circ,pst-plot,pst-3dplot,
pst-solides3d,pst-sigsys,pstricks-add,pst-eucl,pst-grad}
\usepackage[normalem]{ulem}
\allowdisplaybreaks[1]

\usepackage{dcolumn}
\usepackage{lmodern}
\usepackage{microtype}
\usepackage{hyperref}
\hypersetup{
	breaklinks=true,
    colorlinks=true,
    linkcolor=Blue,      
    urlcolor=Blue,
    citecolor=Blue
}

\newcommand{\beq}{\begin{eqnarray}}
\newcommand{\eeq}{\end{eqnarray}}
\newcommand{\nn}{\nonumber\\}
\newcommand{\del}{\partial}

\newcommand{\rme}{{\rm e}}
\newcommand{\rmd}{{\rm d}}
\def\bfalpha{{\boldsymbol \alpha}}
\def\bftau{{\boldsymbol \tau}}

\def\x{{\boldsymbol x}}
\def\p{{\boldsymbol p}}

\usepackage{uri}

\begin{document}

\title{Thermodynamics of the parity-doublet model:\\Symmetric nuclear matter and the chiral transition}

\author{J\"urgen Eser}
\email[]{juergen.eser@ipht.fr}

\author{Jean-Paul Blaizot}
\email[]{jean-paul.blaizot@ipht.fr}

\affiliation{Universit\'e Paris-Saclay, CNRS, CEA, Institut de physique th\'eorique,
91191, Gif-sur-Yvette, France}

\begin{abstract}
We present a detailed discussion of the thermodynamics of the parity-doublet nucleon-meson model
within a mean-field theory, at finite temperature and baryon-chemical potential, with special
emphasis on the chiral transition at large baryon densities and vanishing temperature. We consider isospin-symmetric 
matter. We systematically compare the parity-doublet model to a related singlet model obtained by 
disregarding the chiral partner of the nucleon. After studying the ground state properties of nuclear matter, 
the nuclear liquid-gas transition, and the density modifications of the nucleon sigma term which govern the
low-density regime, we give new insight into the underlying mechanisms of 
the zero-temperature chiral transition occurring at several times the nuclear 
saturation density. We show that the chiral transition is driven by a kind of  symmetry energy that tends to equilibrate the populations of opposite parity baryons. This symmetry energy dictates 
the composition of matter at large baryon densities, once the phase space for the appearance of the 
negative-parity partner is opened. We furthermore highlight the characteristic role, within the thermodynamics, of the chiral-invariant mass 
of the parity-doublet model. We include the chiral limit into all of our discussions 
in order to provide a complete picture of the chiral transition.
\end{abstract}

\date{\today}

\maketitle

\section{Introduction}

Recent observations of gravitational waves from neutron stars and neutron star mergers have triggered a renewal interest in the study of the equation of state of dense matter at finite density and moderate temperature (for a review see Ref.~\cite{Baym:2017whm} and references therein). There is indeed hope that such observations can provide useful constraints on the equation of state of dense matter, complementing the empirical information that can be obtained from relativistic heavy-ion collisions at various facilities. 

One important question is whether nuclear matter turns into quark matter under the conditions that prevail in the interior of neutron stars or in a neutron star merger. To answer this question, one needs a good knowledge of the equation of state over a wide range of baryonic densities. From a theoretical point of view, a determination of the equation of state at finite baryon density is difficult. Standard lattice techniques cannot be applied and most theoretical studies therefore rely on models whose range of validity is difficult to control. Often, the equation of state is built via an interpolation procedure between the low-density region, where low-energy nuclear physics provides information, to the very high density region, where QCD perturbation theory becomes applicable. This paper will be concerned with the general question of up to how large density  one can reliably extrapolate models that reproduce well the properties of nuclear matter near its ground state. 

Dense matter is expected to undergo a number of phase transitions (possibly reduced to smooth crossovers) as the temperature or  the baryon density increases. In this context chiral symmetry  plays a special role. Its explicit breaking is manifest in many  low-energy nuclear physics phenomena, while  lattice calculations indicate that it is restored at high temperature,  and it is likely that the same feature shows up at high baryon density. Chiral symmetry restoration is accompanied  by the vanishing of an order parameter, the quark condensate, or equivalently the expectation value of a scalar field. The chirally symmetric phase may be also characterized by the presence of degenerate parity doublets. It is the identification of such massive parity doublets in early lattice calculations that motivated the development of the so-called parity-doublet model \cite{detar_linear_1989}. Since then, the evidence for parity doubling in the baryon spectrum has received further support from lattice calculations \cite{datta_nucleons_2013,aarts_light_2017}.  It should be noted though that most of these evidences concern finite temperature and zero baryon density.

In its simplest version, the parity-doublet model is a generalisation of the linear sigma model \cite{detar_linear_1989} (see also Refs.~\cite{Jido_2000,jido_chiral_2001} for a detailed formulation). The chiral field, composed of a scalar field and a pion field, is coupled to a baryon parity doublet. A natural identification of the partner of the nucleon, the dominant degree of freedom in nuclear matter, is the $N^* (1535)$. This is what we shall use in this paper, being aware of the fact that this particular choice may not quite fit with the present understanding of the couplings of the $N^*$ to $\pi$ and $\eta$ (see e.g.\ Refs.~\cite{Jido_2000,Zschiesche:2006zj}). A remarkable feature of the model is to accommodate a mass term for the baryon, that is compatible with chiral symmetry. Thus, once chiral symmetry is restored, the members of the doublet become degenerate, but remain massive. This is a distinctive feature of the model, as compared for instance to extensions of the original Walecka model \cite{WALECKA1974491} where the baryon mass is entirely given by the scalar field, and therefore vanishes in the chirally symmetric phase.  Thus the parity model offers us a novel perspective on how chiral symmetry is realized in various environments. In spite of shortcomings, it is indeed a nice model, offering a playground for many detailed calculations. We have used it recently in an analysis of $\pi$-$\pi$ scattering, where it was found to yield remarkably accurate results \cite{Eser:2021ivo}. 

The parity-doublet model, in its original version or in various extensions,  has been used in numerous dense-matter studies, including neutron-star matter, see  e.g.\ Refs.~\cite{hatsuda_parity_1989,Zschiesche:2006zj,dexheimer_nuclear_2008,Weyrich_2015,marczenko_chiral_2018,yamazaki_constraint_2019,minamikawa_quark-hadron_2021,marczenko_chiral_2022,minamikawa_chiral_2023,kong2023study,Fraga:2023wtd}. In this paper, we shall restrict ourselves to the simplest version of the model, keeping only the nucleon parity-doublet degree of freedom as described above, and ignoring other possible degrees of freedom such as hyperons \cite{Steinheimer_2011,fraga2022strange,minamikawa2023parity}
or $\Delta$ excitations of the nucleon \cite{Marczenko_2022}. We shall also leave aside interesting aspects of chiral symmetry, in particular those associated with the $U(1)_A$ anomaly \cite{minamikawa_chiral_2023}. Our main concern in this paper is to understand in detail the dynamics of the chiral transition in the model, in particular at finite baryon density, how this is related via the parameter determination to low-energy nuclear matter properties, and learning from such an analysis how reliable can be an extrapolation to the high-density regime where the chiral transition is predicted to occur in the model. 

In our analysis, we shall find it instructive to compare the results obtained in the parity-doublet model with those of simpler models that can be viewed as extensions of the Walecka model that account for chiral symmetry. A generic example is  the so-called chiral nucleon-meson model 
 \cite{brandes_fluctuations_2021} (see also Refs.~\cite{berges_quark_2003,Floerchinger_2012}).   Many features  of this model are indeed shared by  the parity-doublet model. Our study will be limited to a mean-field approximation, i.e.\ a classical field approximation for the mesons and a one-loop calculation of the fermion determinant. The fermionic fluctuations included in the fermion determinant play an essential role in chiral symmetry restoration, and cannot be ignored, while at high density the meson fluctuations are presumably corrections that can  be accounted for by a modification of the effective potential for the mesonic fields, without introducing additional qualitative changes.  Note that the effect of fluctuations in both the chiral nucleon-meson model and the parity-doublet model have been studied within functional renormalisation group approaches \cite{brandes_fluctuations_2021,Tripolt_2021,Weyrich_2015}. 
 One important conclusion of Ref.~\cite{brandes_fluctuations_2021} is that chiral symmetry restoration appears to take place at very high density. The same prediction holds in the parity-doublet model \cite{Zschiesche:2006zj}, which exhibits in fact a stronger stability, with symmetry restoration taking place only for density at least ten times that of nuclear matter. Understanding the origin of this important feature is part of the motivation for the present study.

Although the present setup allows us to study the finite-temperature chiral transition, and we shall indeed present results for this, the approximations that we use prevent us to get a fully quantitative or even qualitative picture. This is because, at finite temperature and low baryon density,  meson fluctuations, in particular those of the pions,  are expected to play a major role. Treating correctly these fluctuations would be essential to make contact for instance with chiral perturbation theory \cite{gerber_hadrons_1989,brandt_chiral_2014}, as well as with  the resonance gas model (for a recent review  see Ref.~\cite{Andronic_2018}). 

Another important aspect of the present work is the systematic comparison with the chiral limit, where the explicit symmetry breaking term is made to vanish and the pion becomes effectively massless. The chiral limit enters in particular the discussion of the nucleon sigma term, whose magnitude provides a hint about the magnitude of the chiral condensate in a baryon, and more generally in moderately dense matter. It also happens that the nature of the chiral transition differs in the chiral limit from what it is for the physical pion mass. In fact, much can be understood about the detailed dynamics of the chiral transition at finite baryon density and vanishing temperature by contrasting the results obtained for the physical pion mass with those of the chiral limit.

The outline of the paper is as follows. In the next section, we recall the basics of the parity-doublet model, its symmetry properties, and the chiral-invariant mass term. We also discuss the phenomenological bosonic potential that complements the fermionic part. In the following section, we review the parameter determination, trying to clarify the correlations between the different parameters of the model, and the constraints coming from empirical data on nuclear-matter ground state properties, as well as its liquid-gas transition. Section \ref{sec:sigma_term} is devoted to a discussion of the nucleon sigma term, focusing in particular to uncertainties in the extrapolation between the chiral limit and the physical point. We also study corrections to the sigma term coming from the presence of baryonic matter, and show that an expansion in powers of the density is not well converging. The section~\ref{sec:chiraltrans} is devoted to a detailed study of the chiral transition. We provide a detailed analysis of the transition in the parity-doublet model (and the related singlet model), either at vanishing chemical potential and finite temperature, or at vanishing temperature and finite baryon-chemical potential. We emphasize the very different natures of the transition in the two cases. The present study is restricted to isospin-symmetric matter. A continuation of the present work to asymmetric, and in particular, neutron matter will be presented in a forthcoming publication \cite{nextpaper}.

%\clearpage

\section{Parity-doublet model}

In this section, we review the main features of the parity-doublet model, fixing the notation and providing a first qualitative discussion of its thermodynamics. We follow here the general presentation given in Ref.~\cite{Eser:2021ivo} (with slightly different notation). 

The model that we consider consists of a baryon parity doublet that will be eventually identified with the nucleon $N(939)$  and the $N^* (1535)$ (with mass $1510\ \mathrm{MeV}$ \cite{ParticleDataGroup:2022pth}), which are coupled to a set of meson fields: an isoscalar meson field $\sigma$ and its chiral partner, an isovector  pion field $\pi$,  and in addition an isoscalar vector field $\omega_\mu$. At moderate densities, the scalar field $\sigma$ provides attraction between the baryons, while the vector field  provides repulsion, the competition between both leading eventually to the ground state of nuclear matter as a self-bound system. Since we are concerned in this paper solely with isospin-symmetric  matter, we ignore a potential coupling of the baryons to  an isovector vector field (which would play a role in isospin-asymmetric matter, see e.g.\ Refs.~\cite{dexheimer_nuclear_2008,brandes_fluctuations_2021}). 

The overall Lagrangian takes the form of a generalized sigma model. A characteristic feature of the parity-doublet model is to allow the baryon to acquire a mass while respecting chiral symmetry. Such a mass survives  chiral symmetry restoration at high temperature or high baryon density, both members of the doublet becoming then degenerate with the same mass $m_0$. We shall also compare the results obtained with the parity-doublet model to those obtained with a similar model involving only the positive-parity nucleons. Such a model, which can be seen as a chirally symmetric extension of Walecka-type models \cite{WALECKA1974491}, is sometimes referred to as a chiral nucleon-meson model \cite{Drews:2013hha,Floerchinger_2012}. Here we shall refer to it simply as to the ``singlet'' model as opposed to the ``doublet'' model from which it derives trivially by leaving out the negative-parity partner. As will be seen, we shall learn much about the dynamics of these models from the comparison between their singlet and doublet versions. Note that in the singlet model, the entire mass of the baryon is generated by the coupling of the nucleon to the scalar field, while in the doublet model only the deviation of the mass from $m_0$   is generated by such a coupling.

\subsection{The model}

In order to construct the parity-doublet model, we start with two massless Dirac spinors $\psi_a$ and $\psi_b$  of  opposite parities, with $\psi_a$ having positive parity. Each of these spinors can be decomposed into left and right components, e.g.\ $\psi_a^L$ and $\psi_a^R$, such that $\gamma^5\psi_a^R=\psi_a^R$ and $\gamma^5\psi_a^L=-\psi_a^L$. Both $\psi_a^L$ and $\psi_a^R$ are isospin doublets which transform independently under the flavor transformations of  $SU(2)_R\times SU(2)_L$, viz. 
\beq
\psi_a^R\rightarrow \rme^{i\frac{\bfalpha_{_R}\cdot \bftau}{2}} \psi_a^R, \qquad \psi_a^L\rightarrow \rme^{i\frac{\bfalpha_L\cdot \bftau}{2} } \psi_a^L,
\eeq
where $\bftau=(\tau_1,\tau_2,\tau_3)$ denotes the usual Pauli matrices, and $\bfalpha=(\alpha_1,\alpha_2,\alpha_3)$ are the (real) parameters of the transformation. 
The isospin transformations correspond to transformations where $\bfalpha_{_R}=\bfalpha_{_L}$, while in a chiral transformation, the left and right components transform in opposite ways, that is
\beq
\psi_a^R\rightarrow \rme^{i\frac{\bfalpha\cdot \tau}{2}} \psi_a^R, \qquad \psi_a^L\rightarrow \rme^{-i\frac{\bfalpha\cdot \tau}{2} } \psi_a^L,
\eeq
which we may write more compactly as 
\beq\label{chiralpsia}
\psi_a\rightarrow \rme^{i\frac{\bfalpha\cdot \tau}{2} \gamma^5 }\psi_a\, .
\eeq
The same properties hold  for $\psi_b$, with the essential new feature that  the chiral transformations of $\psi_a$ and $\psi_b$ are correlated in a special fashion. In the so-called 
``mirror assignement'', the left and right components of $\psi_b$ transform, respectively, in the same ways as the right and left components of $\psi_a$. That is, if in a chiral transformation $\psi_a$ transforms as indicated in Eq.~(\ref{chiralpsia}), then $\psi_b$ transforms as 
\beq
\psi_b\rightarrow \rme^{-i\frac{\bfalpha\cdot \tau}{2} \gamma^5 }\psi_b.
\eeq
This construct allows us to include  in the Lagrangian a  mass term  of  the following form \cite{detar_linear_1989}
\beq\label{eq:masschiral}
-m_0(\bar\psi_a\gamma^5\psi_b-\bar\psi_b\gamma^5\psi_a),
 \eeq
 while preserving its chiral symmetry. 
 It is indeed easily verified, using the specific rules for the chiral transformation of the parity partners discussed above, that this expression (\ref{eq:masschiral}) is invariant under a chiral transformation. 
 
 We write the Lagrangian of the model as the sum of two contributions, 
${\cal L}_{\rm F}$ and ${\cal L}_{\rm B}$ which denote respectively the fermionic and bosonic parts of the total Lagrangian.
%\begin{widetext}
%\begin{IEEEeqnarray}{rCl}
%	S\!\left[\bar{\psi}_{a},\psi_{a}, \bar{\psi}_{b},\psi_{b},\varphi\right] =  
%	\int_{0}^{\beta} \mathrm{d}\tau \int\mathrm{d}^{3}x \; & \bigg\lbrace &
%	\bar{\psi}_{a} \left[\gamma_{\mu}\partial_{\mu} 
%	+ y_{a} \left(\sigma + i \gamma^5 \vec{\pi} \cdot \vec{\tau} \right)\right] \psi_{a}
%	+ \bar{\psi}_{b} \left[\gamma_{\mu}\partial_{\mu} 
%	+ y_{b} \left(\sigma - i \gamma^5 \vec{\pi} \cdot \vec{\tau} \right)\right] \psi_{b}
%	\nonumber\\[-0.05cm] & & 
%	-\, i g_{v} \left[  \bar{\psi}_{a} \gamma_{\mu} v_{\mu}  \psi_{a} + \bar{\psi}_{b} \gamma_{\mu} v_{\mu}  \psi_{b}\right]
%		+\, m_{0} \left[\bar{\psi}_{a}\gamma^5\psi_{b} - \bar{\psi}_{b}\gamma^5\psi_{a}\right]
%	\nonumber\\ & & 
%	+\, \frac{1}{2} \left(\partial_{\mu}\varphi\right) \cdot \partial_{\mu}\varphi
%	+ V\!\left(\varphi^{2}\right) - h_{\mathrm{ESB}}\left(\sigma - f_{\pi}\right) 
%	+ \frac{1}{2} \left(m_{v}^{2}v_{\mu}^{2} \right)
%	\bigg\rbrace ,
%	\label{eq:action}
%\end{IEEEeqnarray}
%\end{widetext}
The fermion Lagrangian is of the form
\begin{widetext}
\beq\label{eq:actionF}
{\cal L}_{\rm F}=\left(\bar{\psi}_{a}\ \bar{\psi}_{b}\right)
\left(\begin{array}{cc}
  \gamma^{\mu}(i\partial_{\mu} -g_v \omega_{\mu})
	- y_{a} \left(\sigma + i \gamma^5 \vec{\pi} \cdot \vec{\tau} \right) & -m_0 \gamma^5 \\[0.1cm]
  m_0 \gamma^5& \gamma^{\mu}(i\partial_{\mu}-g_v \omega_{\mu}) 
	- y_{b} \left(\sigma - i \gamma^5 \vec{\pi} \cdot \vec{\tau} \right)
\end{array}\right)
\left( \begin{array}{c}
 {\psi}_{a}\\[0.1cm]
{\psi}_{b}
\end{array}
  \right) .
\eeq
\end{widetext}
Aside from the kinetic term, and the mass term just discussed, this Lagrangian exhibits the coupling of the fermions to the mesonic fields
$\sigma$,  $\vec{\pi}$, and $\omega_\mu$. In this paper, these mesonic fields will be treated in the classical approximation (i.e.\ as classical background fields for the fermions), and their  Lagrangian will be specified below. Let us just note at this point that in the states to be considered, which are assumed to be both rotationally and parity invariant,  only the sigma field $\sigma$ and the zeroth component of the vector field, denoted $\omega$, acquire a classical value. From now on we shall therefore set $\vec{\pi}=0$.  The choice of a unique coupling strength $g_{v}$  of the vector meson to both 
$\psi_{a}$ and $\psi_{b}$ is convenient and in line with previous works on the subject (see e.g.\ Ref.~\cite{Motohiro:2015taa}).  The  Yukawa couplings $y_a$ and $y_b$  between the baryons and the chiral fields are distinct and their values will be fixed by physical constraints. 

\begin{figure}[t]
	\centering
	\includegraphics[scale=1.0]{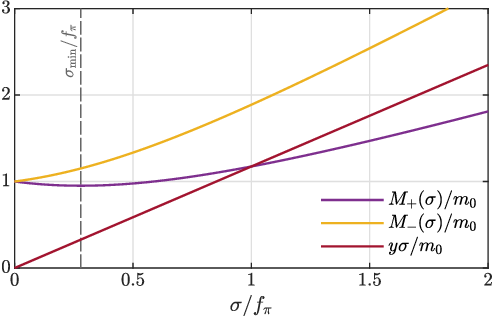}
	\caption{Fermion mass $M_{\pm}$ and singlet mass $y\sigma$ as function 
	of the $\sigma$ field. The mass $M_{+}$ features a minimum at $\sigma_{\mathrm{min}}$ (cf.\ also Refs.~\cite{Gallas:2009qp, Gallas:2013ipa}).}
 	\label{fig:Mplusminus}
\end{figure}

The physical fermion states are obtained as linear combinations of states with the same parity, e.g.\ $\psi_a$ and $\gamma^5 \psi_b$. The coefficients of these linear superpositions are determined by diagonalizing the mass matrix with the help of the following orthogonal transformation 
\begin{equation}\label{eq:rot_phys}
	\begin{pmatrix}
	\psi_{+} \\ \psi_{-}
	\end{pmatrix}
	= \begin{pmatrix}
	\cos\theta &\gamma^5 \sin\theta \\[0.1cm]
	 -\gamma^5 \sin\theta & \cos\theta
	\end{pmatrix} \begin{pmatrix}
	\psi_{a} \\ \psi_{b}
	\end{pmatrix} .
\end{equation}
A simple calculation yields
\begin{equation}
	2\theta = \arctan\left[\frac{2 m_{0}}{(y_{a} + y_{b})\sigma}\right],
\end{equation}
and the 
%  $\psi_{\pm}$ 
%carry positive ($+$) and negative ($-$) parity, respectively. The rotation to the physical 
%basis, $(\psi_{a}, \psi_{b}) \longrightarrow (\psi_{+}, \psi_{-})$, reveals 
 masses of  $\psi_{+}$ and   
$\psi_{-}$, respectively $M_+$ and $M_-$,  are given by
\begin{equation}	M_{\pm} = \frac{1}{2} \left[\pm \sigma \left(y_{a} - y_{b}\right)
	+ \sqrt{\sigma^{2} (y_{a} + y_{b})^{2} + 4 m_{0}^{2}}\right].
	\label{eq:massphys}
\end{equation}
The fields $\psi_+$ and $\psi_-$ are the physical states that we shall associate respectively to the nucleon $N(939)$ and its parity partner $N^*(1535)$, at this level of approximation. Note that $M_+<M_-$ implies that $y_a<y_b$, irrespective of the value of $m_0$,  that is   $\psi_b$ is more strongly coupled  to the chiral field than $\psi_a$.

%The physical states $\psi_{+}$ and $\psi_{-}$ are isospin doublets
%to which we assign the proton $\mathfrak{p}$, the neutron $\mathfrak{n}$, and the 
%chiral partners $\mathfrak{p}^{\ast}$ and $\mathfrak{n}^{\ast}$,
%\begin{equation}
%	\psi_{+} = \begin{pmatrix} \mathfrak{p} \\ 
%	\mathfrak{n} \end{pmatrix}, \qquad
%	\psi_{-} = \begin{pmatrix} \mathfrak{p}^{\ast} \\ 
%	\mathfrak{n}^{\ast} \end{pmatrix}.
%\end{equation}
%The associated rotation operator $O$ is parametrized by the rotation angle
%that is,
%\begin{widetext}
%\beq
% {\cal L}_F=\left(\bar{\psi}_{+}\bar{\psi}_{-}\right)
%\left(\begin{array}{cc}
 % \gamma_{\mu}(\partial_{\mu}-ig_v \omega_{\mu})+M_+ +i \gamma^5 \vec{\pi}\cdot \vec{\tau} \,A_+(\theta)
%	  & -i\vec{\pi}\cdot \vec{\tau} \,B(\theta) \\
% i\vec{\pi}\cdot \vec{\tau} \,B(\theta) & \gamma_{\mu}(\partial_{\mu} -ig_v \omega_{\mu})+M_- -i \gamma^5 \vec{\pi}\cdot \vec{\tau}\,A_-(\theta)
%\end{array}\right)
%\left( \begin{array}{c}
% {\psi}_{+}\\
%{\psi}_{-}
%\end{array}
%  \right)
%  \eeq
%  \\ \nonumber\\ \nonumber\\
%{\cal L}^{1}_F=\left(\bar{\psi}_{+}\bar{\psi}_{-}\right)
%\Big( \y_\sigma \sigma + i\y_\pi \gamma_{5} \vec{\pi} \cdot \vec{\tau}\Big)
%\left( \begin{array}{c}
% {\psi}_{+}\\
%{\psi}_{-}
%\end{array}
%  \right),\qquad 
%\y_{\sigma,\pi}=\left(\begin{array}{cc}
%	y_{\sigma,\pi}^{++}&-y_{\sigma,\pi}^{+-}\\
%	y_{\sigma,\pi}^{-+}&y_{\sigma,\pi}^{--}
%\end{array}\right)

%\end{widetext}
% where $A_\pm (\theta)$ and $B(\theta)$ are given by 
%\beq
%A_\pm(\theta)&=&\frac{1}{2} \left[\pm(y_a+y_b) +(y_a-y_b) \cos 2\theta\right]\nn
%B(\theta)&=&\frac{1}{2} (y_a-y_b) \sin 2\theta.
%\eeq
In this physical basis the  two fields $\psi_+$ and $\psi_-$ decouple (formally, as both baryons remain coupled to the same mesonic fields). The fermionic Lagrangian can then be written as $ {\cal L}_{\rm F}= {\cal L}_{\rm F}^+ + {\cal L}_{\rm F}^-$, with 
 \beq
  {\cal L}_{\rm F}^\pm=\bar{\psi}_{\pm}  \left(i\gamma^{\mu}\partial_{\mu}-g_v \gamma^{0}\omega -M_\pm \right){\psi}_{\pm} ,
  \eeq
  whose spectrum is given by 
  \beq
 ( E^\pm_\p-g_v\omega)^2= \p^2+M_\pm^2 .
  \eeq
The sigma field modifies the masses $M_\pm$ while the vector field produces a constant (independent of the three-momentum $\p$) shift of the single particle energies (opposite for particles and antiparticles). To make things clearer, we set $\varepsilon^{\pm}_\p=\sqrt{\p^2+M_\pm^2}$. The energies $E^\pm_\p$ of the particles,  and $\bar E^\pm_\p$ of the antiparticles, are then given respectively by 
\beq\label{eq:spenergy}
E^\pm_\p=\varepsilon^\pm_\p+g_v\omega,\qquad  \bar E^\pm_\p=\varepsilon^\pm_\p-g_v\omega.
\eeq

We readily recover the singlet model, such as for instance the one used in Ref.~\cite{brandes_fluctuations_2021}, by dropping the 
parity-odd fermion $\psi_{b}$ in Eq.~(\ref{eq:actionF}) and setting $m_{0} =0$. 
The rotation in  Eq.~(\ref{eq:rot_phys}) becomes obsolete and we may identify 
$\psi_{+} \equiv \psi_{a}$. The nucleon mass then reduces to $M_{+} = y_{a}\sigma \equiv y\sigma = M$. 
Hence, in contrast to the parity-doublet model, the nucleon mass in the singlet model is entirely generated by the condensation of the $\sigma$ field, and it vanishes when chiral symmetry is restored.  

The dependence of the baryon masses in both the singlet and the doublet models is shown in Fig.~\ref{fig:Mplusminus}. The parameters used for this plot are those determined  in the next section (see Tables~\ref{tab:parameter_choice_PD} and \ref{tab:parameter_choice_WT} below). For a vanishing value of $\sigma$, the masses of the doublet members are  degenerate at the value $m_0$. The splitting observed between $M_+$ and $M_-$  as $\sigma$ increases is a robust feature of the parity-doublet model. It is a direct consequence of the  diagonalisation of the $2\times 2$ mass matrix in Eq.~(\ref{eq:actionF}), which is involved in the definition of the physical fields. Since, when $m_0<M_N$, both $M_+$ and $M_-$ eventually increase  with $\sigma$ at large values of $\sigma$ (in order to reach their physical values at $\sigma=f_\pi$), this initial splitting implies that $M_+(\sigma)$ exhibits a shallow minimum at a value $\sigma_{\rm min}$ given by 
\beq\label{eq:sigmamin}
\sigma^2_{\rm min}=\frac{m_0^2 (y_a-y_b)^2}{y_a y_b (y_a+y_b)^2}.
\eeq
With our choice of parameters, the value of $\sigma_{\rm min}$ is  $\sigma_{\rm min}/f_\pi\approx 0.28$. 
The role of this minimum of $M_+(\sigma)$ in the chiral transition will be discussed later. It is also worth noticing that, with the present choice of parameters,  $M_+$ never deviates too much from $m_0$, in contrast to $M_-$.

We now complete the discussion of the Lagrangian of the model, by specifying its 
 mesonic part ${\cal L}_{\rm B}$. Since we are treating the meson fields in the classical approximation, only the potential terms of ${\cal L}_{\rm B}$ are relevant. We set
\beq\label{mesonL}
{\cal L}_{\rm B}= -V\!\left(\varphi^{2}\right) +h\left(\sigma - f_{\pi}\right)+ \frac{1}{2} m_v^2 \omega^2,
\eeq
where $\varphi^2\equiv\sigma^2+\pi^2$ and $f_{\pi}$ the pion decay constant. 
Following previous works \cite{Floerchinger_2012,brandes_fluctuations_2021}, we express the potential $V(\varphi^{2})$ as a fourth-order polynomial in 
$\varphi^2-f_{\pi}^{2}$,
\begin{equation}
	V\!\left(\varphi^{2}\right) = \sum_{n\, =\, 1}^{4} 
	\frac{\alpha_{n}}{2^{n} n!} \left(\varphi^{2} - f_{\pi}^{2}\right)^{n},
	\label{eq:potential}
\end{equation}
where $\alpha_{n}$ will be referred to as a  Taylor coefficient.  As we shall see in the next section, the higher-order coefficients $\alpha_3$ and $\alpha_4$ are needed in order to be able to reproduce nuclear matter properties. 
The  term $h\sigma$ accounts for the explicit symmetry breaking in the direction of the $\sigma$ field. It confers the pion a finite mass. It also prevents chiral symmetry to be restored at high temperature and density. We shall often refer in this work to the so-called ``chiral limit'': this is obtained by letting $h\to 0$, while keeping all other parameters fixed. 

\begin{figure}[t]
	\centering
	\includegraphics[scale=1.0]{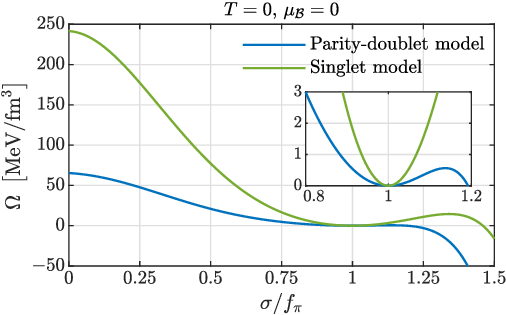}
	\caption{Thermodynamic potential in vacuum ($T = 0$, $\mu_{\mathcal{B}} = 0$), i.e.\ 
	$\Omega \equiv U$.}
 	\label{fig:OmegaVacuum}
\end{figure}

The mean-field approximation that is used in this paper consists in treating the mesonic fields as classical fields, while keeping the fermion fluctuations to order one-loop. These fermion fluctuations are functions of the $\sigma$ field, and contribute therefore to the mesonic effective potential. We call $U(\sigma,\omega)=U(\sigma)-\frac{1}{2} m_v^2 \omega^2$ the full resulting effective potential in vacuum, noticing that  $\omega$ is non vanishing only in the presence of matter (see below).  Note that in contrast to previous works (see e.g.\ Ref.~\cite{Zschiesche:2006zj}) we do not include in the meson Lagrangian self-interactions of the vector field. The renormalisation of the one-loop fermion contribution is detailed in  Appendix~\ref{sec:vacuum} (see Eq.~(\ref{Ubosons})). The renormalisation conditions are chosen so that the first and second derivatives with respect to $\sigma^2$ of the fermionic one-loop contribution vanish at $\sigma=f_\pi$. It follows that the corresponding  derivatives of $U(\sigma)$ coincide with those deduced from ${\cal L}_{\rm B}$ in Eq.~(\ref{mesonL}), i.e.\ they are given by $\alpha_1$ and $\alpha_2$. Note however that the renormalisation conditions entail a modification of the potential in the vicinity of $\sigma=0$  which receives in fact a large contribution from the fermion loop. There are indeed  large cancellations between the fermion loop and the potential $V(\varphi)$, these cancellations being more important in the parity-doublet model than in the singlet model (see the discussion in Appendix~\ref{sec:vacuum}). Note finally  that $\alpha_4$ being negative,  the potential is not bounded from below \cite{minamikawa_chiral_2023}. However, this occurs at values of $\sigma$ that are larger than the values that are relevant to the physics that we want to study.\footnote{In Ref.~\cite{minamikawa_chiral_2023}, a fifth-order term is added to make the potential bounded from below, but we do not find it necessary to do so here.} 

The potentials used in our calculations are displayed in Fig.~\ref{fig:OmegaVacuum}. One sees that the two potentials corresponding respectively to the singlet and doublet models overlap in the region $\sigma\simeq f_\pi$, as they should in order to fit the same nuclear matter properties. The  large difference between the two potentials can be traced back in part  to the large difference in the values of $m_\sigma$ in the two models (640 MeV for the singlet model, and 340 MeV for the doublet model). This entails in particular large differences in the vicinity of $\sigma=0$ with a strong impact on the chiral transition, as we shall see in Sec.~\ref{sec:chiraltrans}.

\subsection{Thermodynamics}

In studying the thermodynamics of the parity-doublet model, we limit ourselves in this paper to uniform systems that are  isospin symmetric (all isospin members of the doublets are equally occupied). We want however to explore the properties of equilibrium states as a function of the baryon density. To do so, we  introduce a chemical potential $\mu_{\mathcal B}$ coupled to the baryon density  $n_{\mathcal{B}}$:
\beq
 n_{\mathcal{B}} =    
	\left\langle \bar{\psi}_{+}\gamma^{0}\psi_{+} 
	+ \bar{\psi}_{-}\gamma^{0}\psi_{-}\right\rangle .
	\eeq
	Note that the chemical potential enters the Lagrangian density in the same way as the component $\omega$ of the vector potential, whose role as we have seen is to shift the single particle energies by a constant amount. It is then convenient at some places to absorb this shift into a modified chemical potential ${\tilde \mu}_{\mathcal{B}}$,
	\beq
	{\tilde \mu}_{\mathcal{B}} = \mu_{\mathcal{B}} - g_v \omega.
	\eeq
	With this convention, the combination $E_\p^\pm-\mu_{\mathcal{B}}$ that enters for instance the expression of the statistical factors can be written as $\varepsilon_\p^\pm-\tilde \mu_{\mathcal{B}}$. Similarly, for the antiparticles, whose chemical potential is opposite to that of the particles, $\bar E_\p^\pm +\mu_{\mathcal{B}}\mapsto \varepsilon_\p^\pm+\tilde \mu_{\mathcal{B}}$.

The  grand canonical potential density $\Omega$ contains, in addition to the vacuum contribution $U$ discussed in the previous section, a matter contribution. The latter is the contribution of independent fermion quasiparticles whose energies depend on the mesonic fields. We have
\begin{equation}\label{eq:OmegaUBMat}
	\Omega = U -4T\sum_{\substack{i\,=\,\pm 1,\\ r\, =\, \pm 1}}\int \frac{\rmd^3 p}{(2\pi)^3}\ln\left[ 1+\rme^{-\beta \left(\varepsilon^i_\p-r\tilde \mu_{\mathcal B}\right)}  \right],
\end{equation}
where 
the index $i$ runs over the two  parity states and $r$ refers to particles and antiparticles. 
The overall factor 4 accounts for the sum over spin and isospin. In the above equation,  the term proportional to $T$ has a finite limit as $T\to 0$, equal to  $\tilde{{\cal E}}_{\rm qp}-\mu_{\mathcal B} n_{\mathcal B}$, where $\tilde{{\cal E}}_{\rm qp}$ denotes the quasiparticle contribution to the energy density, the total energy density being ${\cal E}=U+\tilde{{\cal E}}_{\rm qp}$.

The  grand canonical potential density $\Omega$ is a function of the chemical potential $\mu_{\mathcal{B}}$ and the temperature $T$. In addition, it depends on the values of the fields $\sigma$ and $\omega$. These constant fields $\sigma$ and $\omega$ are to be considered as internal variables that need to be determined, for given $T$ and $\mu_{\mathcal{B}}$ by the requirement that $\Omega$ be stationary with respect to their variations. This leads to the equations 
\begin{equation}\label{Omegamin}
	\left. \frac{\partial\Omega}{\partial\omega}\right|_{\mu_{\mathcal{B}},T;\sigma}=0, \qquad \left.\frac{\partial\Omega}{\partial\sigma}\right|_{\mu_{\mathcal{B}},T;\omega} = 0.
\end{equation}
The first equation (\ref{Omegamin}) is  essentially the equation of motion for the field $\omega$ in the classical approximation where all derivatives of the field vanish:
\beq\label{eq:Gv}
g_v\omega = G_v n_{\mathcal B},\qquad G_v\equiv \frac{g_v^2}{m_v^2}.
\eeq
It relates the $\omega$ field to its source,  the baryon density $n_{\mathcal{B}}$.  The elimination of the $\omega$ field in favor of the baryon density allows us to express the shift in the single particle energies in Eq.~(\ref{eq:spenergy}) as $\pm G_v n_{\mathcal{B}}$. It also yields a repulsive interaction between the baryons, i.e.\ a contribution $(G_v/2) n_{\mathcal{B}}^2$ to the energy density (see Eq.~(\ref{eq:Gvnb2}) below).

The baryon density, $n_{\mathcal B}=-\partial\Omega/\partial\mu_{\mathcal B}|_{\sigma,\omega}$ can be decomposed as a sum of densities $n_{\mathcal B}=n_{\mathcal B}^++n_{\mathcal B}^-$ of positive ($n_{\mathcal B}^+$) and negative ($n_{\mathcal B}^-$) parity baryons: 
\beq \label{nBpm}
n_{\mathcal B}^\pm  
	=4  \sum_{\substack{
	r\, =\, \pm 1}}r \int \frac{\rmd^3 p}{(2\pi)^3}\,  n_{\rm F}( \varepsilon^{\pm}_\p - r\tilde{\mu}_{\mathcal{B}}) ,
\eeq
where  $n_{\rm F}$ is  the Fermi-Dirac distribution
\beq
	n_{\rm F}( \varepsilon_\p ) = \left(\rme^{\beta  \varepsilon_\p} + 1 \right)^{-1}.
	\eeq
Note that only the total baryon density $n_{\mathcal B}$ is controlled by the chemical potential $\mu_{\mathcal B}$.  However, in the present approximation, the baryon density naturally splits into separate contributions coming from each member of the parity doublet, with the relative sizes of each contribution being determined by the different energies of the positive ($\varepsilon_\p^+$) versus negative ($\varepsilon_\p^-)$ parity baryons.

The second equation (\ref{Omegamin}) is akin to a gap equation. It  can be written as 
\beq\label{eq:gap}
	&&\left.\frac{\partial U}{\partial \sigma} \right|_{\omega}
=- y_{+} \,n_{\rm s}^{+} 
	- y_{-}\,n_{\rm s}^{-},\label{eq:MF_eq_sigma}
	\eeq
where 	the scalar densities are given by 
\beq  
	n_{\rm s}^{\pm} = \left.\frac{\partial\Omega}{\partial M_{\pm}} \right|_{\mu_{\mathcal{B}},T;\omega}
	=4 \sum_{\substack{
	r\, =\, \pm 1}} \int \frac{\rmd^3 p}{(2\pi)^3}\,  \frac{M_{\pm}}{\varepsilon^{\pm}_\p}
	n_{\rm F}(\varepsilon^{\pm}_\p - r\tilde{\mu}_{\mathcal{B}}) . \nn
\eeq 
In writing Eq.~(\ref{eq:MF_eq_sigma})  we have set 
\begin{equation}\label{eq:derivsgimaterm}
	\frac{\mathrm{d}M_{\pm}(\sigma)}{\mathrm{d}\sigma} \equiv y_{\pm}(\sigma),\end{equation}
with
\begin{equation}
	y_{\pm} = \frac{1}{2} \left[\pm \left(y_{a} - y_{b}\right)
	+ \frac{\sigma (y_{a} + y_{b})^{2}}
	{\rule[0ex]{0ex}{2ex}\sqrt{\rule[0ex]{0ex}{2ex}\sigma^{2} (y_{a} + y_{b})^{2} 
	+ 4 m_{0}^{2}}}\right].
	\label{eq:yukawa_phys}
\end{equation}
Note the relation
\beq\label{eq:jurgenformula}
y_{+} M_++y_{-} M_-=(y_a^2+y_b^2)\sigma,
\eeq
which will be used later.

The  scalar densities  play an essential role in the restoration of chiral symmetry, in balancing, within the gap equation,  the source of spontaneous symmetry breaking that is included in the  effective potential of the scalar field ($U''(\sigma=0)<0$). 
The effect of the presence of matter on the $\sigma$ field can be understood qualitatively from the gap equation (\ref{eq:gap}): the potential in vacuum has a minimum at $\sigma=f_\pi$. The right-hand side of Eq.~(\ref{eq:gap}) is generically negative, so that the solution of this equation  is to be found in the region where  $\rmd U/\rmd\sigma <0$, that is for values of sigma smaller than $f_\pi$.   In other words the presence of matter generically tends to decrease the $\sigma $ field. 

Finally, let us recall that the value of  $\Omega$ calculated with the fields $\omega$ and $\sigma$ that solve Eqs.~(\ref{Omegamin}) is equal to $-P(\mu_{\mathcal{B}},T)$ where $P$ is the thermodynamic pressure. 

%%%%%%%%%%%%%%%%%%%%%%%%%%%%%%%%%%%%%%%%%%%%%%%%%%%%%%%%%%%%%%%%%%

\section{Determination of parameters}
\label{sec:results}

%%%%%%%%%%%%%%%%%%%%%%%%%%%%%%%%%%%%%%%%%%%%%%%%%%%%%%%%%%%%%%%%%%

In this section we determine the parameters of the parity-doublet model, as well as those of the singlet model,  by relating them to some well established properties of the vacuum and of symmetric nuclear matter in its ground state or in its liquid-gas phase transition.  The values of the physical quantities that we aim to reproduce are indicated in Tables~\ref{tab:parameters} and \ref{tab:experiment} below. The doublet model contains the following nine parameters: the four Taylor coefficients $\alpha_1, \alpha_2,\alpha_3, \alpha_4$ of the potential $V(\varphi)$, Eq.~(\ref{eq:potential}),  the parameter $h$ of the symmetry breaking term, the parameter $m_0$ in Eq.~(\ref{eq:masschiral}), the vector coupling $G_v$ in Eq.~(\ref{eq:Gv}), and the Yukawa coupling constants $y_a$ and $y_b$. The singlet model contains only seven parameters (obtained by eliminating $m_0$ and $y_b$ from the list of the doublet model parameters).

\subsection{Vacuum state}
\label{sec:parameters}

Let us first consider the vacuum state. In this case, $\Omega=U(\varphi,\omega)$ where  the internal variables $\varphi$ and $\omega$  need to be adjusted so as to satisfy Eqs.~(\ref{Omegamin}).  The vector field $\omega$ is  related to the baryon density via Eq.~(\ref{eq:Gv}) and vanishes in the vacuum. Regarding $\varphi$, we recall that  $f_\pi$ denotes the minimum of the potential in the presence of the explicit symmetry breaking term $-h(\sigma-f_\pi)$. Since the fermion loop contribution is chosen so as to give vanishing contributions to the first and second derivatives of $U$ with respect to $\sigma^{2}$, one may study the vicinity of the physical vacuum by keeping only the first two terms in the potential $V(\varphi)$, namely
\beq\label{Vvarphi}
V(\varphi)\simeq \frac{\alpha_1}{2} (\varphi^2-f_\pi^2)+\frac{\alpha_2}{8}(\varphi^2-f_\pi^2)^2  .
\eeq
The extrema of $U(\sigma) $ are then given by the solution of the following gap equation
\beq\label{eq:minVphih}
\frac{\rmd U}{\rmd\sigma}= \alpha_1\sigma +\frac{\alpha_2}{2}\sigma(\sigma^2-f_\pi^2)-h=0,
\eeq
where $h$ is to be chosen so that the solution is $\sigma=f_\pi$. Setting $\sigma=f_\pi$ in the equation above, one finds  $\alpha_1=h /f_\pi$. 
A further differentiation of $V(\varphi)$  yields the meson masses
\beq\label{alpha12}
m_\sigma^2= \alpha_1 +\alpha_2 f_\pi^2 ,\qquad m_\pi^2= \alpha_1, 
\eeq 
from which we get $\alpha_2= (m_\sigma^2-m_\pi^2)/f_\pi^2$.
It follows in particular that $h=m_\pi^2 f_\pi$. 
For the physical pion mass $m_{\pi} = 138\ \mathrm{MeV}$,
we thus have
\begin{equation}\label{eq:hvalue}
	h= m_{\pi}^{2} f_{\pi} \simeq 1.77 \times 10^{6}\ \mathrm{MeV}^{3},
\end{equation}
with the pion decay constant $f_{\pi} = 93\ \mathrm{MeV}$. Note that here and throughout this paper $m_\pi$ and $f_\pi$ refer to the physical  pion mass and decay constant in vacuum. Note also that while $\alpha_1$ and $h$ are directly fixed by these physical quantities, this is not the case of $\alpha_2$ which depends on $m_\sigma^2$, for which exists a range of acceptable values. The Yukawa coupling constants $y_a$ and $y_b$ are functions of $\sigma$ and $m_0$. However, the dependence on $\sigma$ can be eliminated by using the relations $M_+(f_\pi)=M_N$ and $M_-(f_\pi)=M_{N^\ast}$. We obtain then
\begin{multline}\label{eq:YukawaMN}
	y_{a,b} = \frac{1}{2f_{\pi}}\bigg[\pm (M_{N} - M_{N^{\ast}}) \\
	+ \sqrt{(M_{N} + M_{N^{\ast}})^{2} - 4m_{0}^{2}}\bigg],
\end{multline}
so that $y_a$ and $y_b$  depend effectively only on $m_0$. In the singlet model, the Yukawa coupling is simply given by $y=M_N/f_\pi$. 

\begin{table}
	\caption{\label{tab:parameters}Input parameters for the initialization of
	the model \cite{ParticleDataGroup:2022pth, Drischler:2015eba,
	brandes_fluctuations_2021}.}
	\begin{ruledtabular}
		\begin{tabular}{lr}
		Parameter & Numerical value\\
		\colrule\\[-0.25cm]
        Pion decay constant $f_{\pi}\ [\mathrm{MeV}]$ & $93$ \\[0.05cm]
        Pion mass $m_{\pi}\ [\mathrm{MeV}]$ & $138$ \\[0.05cm]
        Nucleon mass in vacuum $M_{N}\ [\mathrm{MeV}]$ & $939$ \\[0.05cm]
        Mass of the chiral partner $M_{N^{\ast}}\ [\mathrm{MeV}]$ & $1510$ \\[0.05cm]
		Nuclear saturation density $n_{0}\ [\mathrm{fm}^{-3}]$ & $0.16$ \\[0.05cm]
		Binding energy $E_{\mathrm{bind}}\ [\mathrm{MeV}]$ & $-16$
		\end{tabular}
	\end{ruledtabular}
\end{table}

At this point, we have determined two parameters, $\alpha_1$ and $h$ and traded $\alpha_2$ in favor of $m_\sigma^2$. We have also related $y_a$ and $y_b$ to $m_0$. We are therefore left with six parameters to be determined (four for the singlet model), for which we now  turn to symmetric nuclear matter in its ground state. 

\subsection{Symmetric nuclear matter}
Nuclear matter in its ground state at $T = 0$ exists as a self-bound system, with vanishing pressure, at a given density $n_0$, commonly  referred to as the ``saturation'' density. At this density or below, only nucleons are present (the density of the negative-parity partners is completely negligible). So the singlet and doublet models differ solely by the dependence of the nucleon mass on $\sigma$. In this section we shall alleviate the notation and denote the baryon density simply as $n$ instead of $n_{\mathcal{B}}$, and similarly for the chemical potential $\mu_{\mathcal{B}}\mapsto \mu$. 

\subsubsection{Digression on the gap equations}

The grand potential at zero temperature is given by 
\beq\label{eq:OmegacalEt0}
\Omega(\mu;\sigma,\omega)=U(\sigma,\omega)+\tilde {\cal E}_{\rm qp}(\mu;\sigma,\omega) -\mu n(\mu;\sigma,\omega), \qquad
\eeq
where 
\beq
U(\sigma,\omega)=U(\sigma)-\frac{1}{2} m_v^2 \omega^2.
\eeq
The baryon density $n$ and the quasiparticle energy density $ \tilde {\cal E}_{\rm qp}$ are obtained by filling all the quasiparticle levels with energy smaller than the chemical potential, which leads to the expressions
\beq
n=4\int_\p \theta(\tilde \mu-\varepsilon_\p^+),\qquad \tilde{\cal E}_{\rm qp}=4\int_\p \theta(\tilde \mu-\varepsilon_\p^+) \,E_\p^+, \qquad
\eeq
with the shorthand notation
\begin{equation}
	\int_\p = \int \frac{\rmd^3 p}{(2\pi)^3}, 
\end{equation}
which we shall use occasionally throughout the rest of the paper.
In order to verify that  $n=-\del \Omega/\del \mu|_{\sigma,\omega}$, we note that the constraint $E_\p\lesssim \mu$ translates into a constraint on the Fermi momentum $p_F$ such that $E_{p_F}=\sqrt{p_F^2+M^2}+g_v \omega=\mu$. This relates the Fermi momentum, and hence the density $n=2p_F^3/(3\pi^2)$, to the chemical potential.

In fact, at $T=0$, the quasiparticle energy density $ \tilde {\cal E}_{\rm qp}$ is more naturally expressed in terms of the density than in terms of the chemical potential. We have 
\beq
\tilde {\cal E}_{\rm qp}(n;\sigma,\omega)=4 \int_{|p|\, <\, p_{F}} 
	\frac{\rmd^3 p}{(2\pi)^3}\, \left[\sqrt{p^2 + M(\sigma)^2}+g_v \omega\right].\nn
	\eeq
The last term in this equation gives a contribution equal to $g_v\omega n$. 
The chemical potential is now obtained as
\beq\label{eq:musigmaomegapF}
\mu=\left.\frac{\del \tilde {\cal E}_{\rm qp}}{\del n}\right|_{\sigma,\omega}=\sqrt{p_F^2+M^2(\sigma)}+g_v \omega,
\eeq
which  coincides with the expression given just above.

In addition to this relation, we have the two equations (\ref{Omegamin})  that express the stationarity of the grand potential, or equivalently  the energy density ${\cal E}=\tilde {\cal E}_{\rm qp}+U(\sigma,\omega)$,  with respect to the fields $\omega$ and $\sigma$. The first equation relates $\omega$ to the baryon density, $\omega=({g_v}/{m_v^2}) n$,  and, when combined to the contribution $g_v\omega n$ contained  in $\tilde {\cal E}_{\rm qp}$,  leads to the following contribution to the total energy density
\beq\label{eq:Gvnb2}
-\frac{1}{2}m_v^2 \omega^2+g_v \omega n=\frac{1}{2}G_v n^2.
\eeq
In the following we shall denote by ${\cal E}_{\rm qp}$ the contribution of the quasiparticle energies without the repulsive vector contribution. That is we shall set
\beq\label{eq:calEnsign}
{\cal E}(n;\sigma)= {\cal E}_{\rm qp}(n;\sigma)+\frac{1}{2}G_v n^2+U(\sigma),
\eeq
where 
\beq
 {\cal E}_{\rm qp}(n;\sigma)=4 \int_{|p|\, <\, p_{F}} 
	\frac{\rmd^3 p}{(2\pi)^3}\,\sqrt{p^2 + M(\sigma)^2}.
	\eeq

The  gap equation  reads
\beq\label{eq:gapsigma}
\frac{\rmd U(\sigma)}{\rmd \sigma}+yn_{\rm s}=0,
\eeq
where $n_{\rm s}=\del {\cal E}_{\rm qp}{/\del M}|_{n;\sigma}$ is a function of $\sigma$ and $p_F$ (an analytic expression is provided in Eq.~(\ref{eq:nsanal}) below).  In fact for a density smaller than normal nuclear matter density, the dependence on $\sigma$ is weak and, to a very good approximation, $n_{\rm s}\simeq n$. 

\begin{figure}[h]
	\centering
	\includegraphics[scale=1.0]{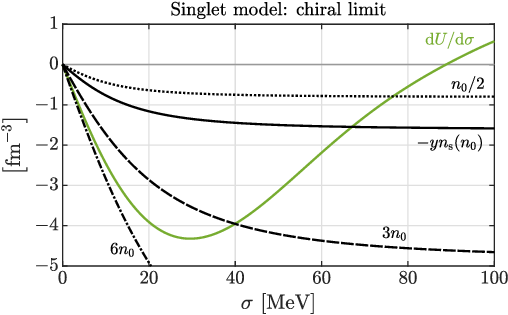}
	\caption{Graphical solution of the gap equation. The green curve represents $\rmd U/\rmd \sigma$ as function of $\sigma$. This curve is independent of $n$. The black curves give the contribution from the scalar density $-yn_{\rm s}$ as function of $\sigma$ for given density $n$ (solid: for $n=n_0$; dotted $n = n_0/2$; dashed: $n=3n_{0}$; dash-dotted: $n=6n_{0}$). Note that in the range of densities and values of $\sigma$ relevant to the present discussion, $n_{\rm s}\simeq n$. The intersection point between the green curve and a black one gives the solution $\sigma_n$ of the gap equation for the corresponding values of $n$.  }
	\label{fig:gap_equation_singlet_1}
\end{figure}

An illustration of the graphical solution of the gap equation for the singlet model (in the chiral limit) is provided in Fig.~\ref{fig:gap_equation_singlet_1}. In the region of interest, as we just said, $n_{\rm s}\simeq n$  and the solution of the gap equation provides a smooth relation between $n$ and $\sigma$. Calling $\sigma_n$ the solution of the gap equation for a given density $n$, one can calculate the total energy density ${\cal E}(n;\sigma_n)=U(\sigma_n)+{\cal E}_{\rm qp}(n;\sigma_n)+\frac{1}{2} G_v n^2$. This is the strategy that we shall use to calculate the  nuclear matter energy per particle as a function of density. 

%\begin{figure}[h]
%	\centering
%	\includegraphics[scale=0.8]{EoverA.pdf}
%	\caption{The energy per particle in nuclear matter at small densities. }
 %	\label{fig:EoverA}
%\end{figure}

\subsubsection{The saturation mechanism}

At this point it may be useful to recall the basic mechanism that leads to the so-called saturation of nuclear matter, that is how the equilibrium zero-pressure state is achieved in the present models.  In non relativistic calculations saturation is understood as the equilibrium state obtained when attractive forces balance the kinetic energy and the repulsive forces. In the present relativistic models, the  repulsion is due to vector-meson exchange.  The baryon density constitutes a source for the vector meson field, according to Eq.~(\ref{eq:Gv}) which may be used to eliminate the vector field in favor of the baryon density. This yields, as we have seen,  a repulsive contribution to the energy density (see Eq.~(\ref{eq:Gvnb2})), characteristic of a two-body short-range repulsive interaction.

\begin{figure}[t]
	\centering
	\includegraphics[scale=1.0]{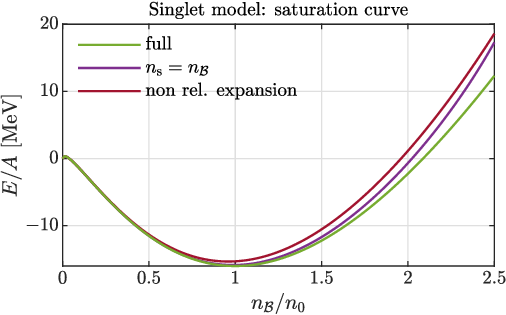}
	\caption{Saturation curve, $E/A={\cal E}(n)/n-M_N$. This is obtained by solving the gap equation with the full potential, with in the right-hand side either $-yn_{\rm s}$ or $-yn=-yn_{\mathcal{B}}$. The third curve shows the non relativistic approximation to the kinetic energy. }
	\label{fig:saturation_curve_nonrel}
\end{figure}

The mechanism of attraction is somewhat different and it involves the variation of the nucleon mass with the strength of the sigma field. 
Let us consider a system with low baryon density. Then the sigma field is close to $f_\pi$, and the nucleon mass does not differ much from its value in the vacuum.  The  non relativistic approximation for the kinetic energy is  valid, and yields the following expression for the energy density
\beq
{\cal E}(n;\sigma_n)=n\left[M(\sigma_n)+\frac{3}{5}\frac{p_F^2}{2 M(\sigma_n)}\right] +\frac{G_v}{2} n^2 +U(\sigma_n).\nn
\eeq
To obtain this equation, we have used the equation of motion (\ref{eq:Gv}) relating the field $\omega$ to the baryon density. As for the field $\sigma_n$, this is the solution of  the gap equation (\ref{eq:gap}). 
In the vicinity of $\sigma\simeq f_\pi$,   $U(\sigma)\simeq ({m_\sigma^2}/{2} )\tilde \sigma^2$, 
where $ \tilde \sigma=\sigma-f_\pi$. The gap equation reads then 
\beq
m_\sigma^2\tilde\sigma_n\simeq -yn
\eeq
where we have used $n_{\rm s}\simeq n$.
At this point, we consider for simplicity the singlet model where the mass is given by $M=y\sigma$. Then 
\beq\label{eq:msigmaGs}
M(\sigma_n)=M_N-G_s n,
\eeq 
where we have set $G_s\equiv {y^2}/{m_\sigma^2}$ and we have identified $M(f_\pi)$ to the nucleon mass $M_N$. The energy per particle then becomes
\beq\label{eq:calEovern}
\frac{{\cal E}(n;\sigma_n)}{n}\simeq M_N+\frac{3}{10}\frac{p_F^2}{M_N} +\frac{n}{2}\left( G_v-G_s   \right)  .
\eeq
This expression, valid only for small density, allows us to understand the mechanism of attraction. In fact,  at this level of approximation, the attraction between nucleons can be seen as the result of a simple scalar exchange between the nucleons, leading to a contribution to the energy density analogous to that of the vector exchange, Eq.~(\ref{eq:Gvnb2}), but with an opposite sign. The behavior of the energy per particle is plotted in Fig.~\ref{fig:saturation_curve_nonrel}. There is a small increase (hardly  visible on the figure) at very small density which comes from the kinetic energy contribution $\sim n^{2/3}$. Then the linear behavior of Eq.~(\ref{eq:calEovern}) sets in. Since $G_v<G_s$ (for the singlet model, $G_s\simeq 9.7$ fm$^2$, while $G_v\simeq 5.4$ fm$^2$) the resulting slope is negative and eventually leads to a negative energy per particle. 

The linear behavior of the scalar contribution is eventually suppressed by higher-order terms, and a minimum of the energy per particle is reached. In the original Walecka model \cite{WALECKA1974491}, the saturation comes from the taming of the attraction due to a modification of the relation between $n_{\rm s}$ and $n$ as $n$ increases. However, in the present case, an additional factor, in fact the dominant one, comes from the non linear meson interactions coded in the potential $V(\sigma)$, via the Taylor coefficients $\alpha_3$ and $\alpha_4$. These high-order terms are necessary in order to optimize nuclear matter properties. Their presence requires a fine tuning, involving a cancellation of large contributions in the vicinity of $\sigma=f_\pi$.  This impacts not only the potential in the vicinity of $\sigma=f_\pi$, but also in the vicinity of  $\sigma=0$. It affects therefore  the chiral transition (see also Fig.~\ref{fig:OmegaVacuum} and the associated discussion). In this context it is important to keep in mind that some of the properties of the chiral transition that we shall discuss in Sec.~\ref{sec:chiraltrans} depend on an extrapolation on which we have little physical control.  

To emphasize the role of $\alpha_3$ and $\alpha_4$, we have plotted in Fig.~\ref{fig:saturation_curve_nonrel} the saturation curve (i.e.\ the binding energy per nucleon $E/A$ vs.\ the density $n$) obtained in the singlet model. The full result is compared to those obtained using either the non relativistic approximation, as in Eq.~(\ref{eq:calEovern}), or the approximation $n_{\rm s}\simeq n$ when solving the gap equation. We see that the deviations are nearly negligible for density $n\lesssim n_0$, indicating that both approximations are quite accurate in this density range.  However, if we were to ignore the contribution of $\alpha_3$ and $\alpha_4$, that is  use for the potential the expansion $U(\sigma)\simeq ({m_\sigma^2}/{2} )\tilde \sigma^2$, as we did above to get the small density behavior, we would not be able to reproduce the saturation curve (as already observed in previous studies, see e.g.\ Ref.~\cite{berges_quark_2003}).

The minimum of the energy per particle ${\cal E}/n$ corresponds to a state of zero pressure. Recall indeed that the pressure is related to the energy density ${\cal E}$ by 
\beq
P=n\frac{\rmd {\cal E}(n)}{\rmd n}-{\cal E}(n)=n^2\frac{\rmd {\cal E}/n}{\rmd n}.
\eeq
At  saturation, the chemical potential $\mu=\rmd {\cal E}/\rmd n$ is equal to the energy per particle,  $\mu_0={\cal E}(n_0)/n_0$, so that $P(n_0)=0$. 
A plot of the pressure as a function of the nucleon density can be seen in Fig.~\ref{fig:pressure} below. The curve corresponding to zero temperature indeed reveals the existence of a point of vanishing pressure, where furthermore the 
compressibility, 
\beq
\chi=\frac{1}{n}\left( \frac{\rmd P}{\rmd n}\right)^{-1},
\eeq
is positive. Note that $P(n)=P(n;\sigma_n)$ where 
$\sigma_n$ is the solution of the gap equation corresponding to the density $n$. This solution can be followed by continuity in the regions where the pressure is negative or even in regions where the compressibility is negative and the system is unstable. The knowledge of the energy density for all values of $n$ between zero and $n_0$ is useful  for   estimating  the surface energy, as we shall see shortly.

\subsubsection{The ground state properties}

We return now to the determination of the parameters.  Since the pressure vanishes in the ground state of nuclear matter, the chemical potential $\mu_0=\rmd {\cal E}/\rmd n|_{n_{0}}$ is equal to the energy per baryon ${\cal E}(n_0)/n_0$, which differs from $M_N$ by the binding energy per nucleon,  $E_{\mathrm{bind}}=-16$ MeV. It follows that 
\begin{equation}
	\mu_{0} = M_{N} + E_{\mathrm{bind}} = 923\ \mathrm{MeV}.
\end{equation}

Inside nuclear matter, the nucleons behave as quasiparticles with energies given as a function of momentum by (see Eq.~(\ref{eq:spenergy}))
\beq\label{eq:Epplus}
E_\p^+=\sqrt{\p^2+M_+^2}+G_v n=\varepsilon_\p^++G_v n,
\eeq
where  $M_+$ is the mass of the nucleon in matter, given  by Eq.~(\ref{eq:massphys}). 
 It is convenient here to introduce  the Landau effective mass $M_*$, defined as 
\beq
\left.\frac{\partial E_\p}{\partial p}\right|_{p_F}\equiv \frac{p_F}{M_*},
\eeq
where $p_F/M_*$ is the Fermi velocity. A simple calculation yields
\beq\label{MLGv}
M_*=\mu_0-G_v n_0= \sqrt{p_{F}^{2} + M_{+}^{2}(\sigma_0,m_0)}.
\eeq
 The first equation (\ref{MLGv}) provides a direct relation between  the effective mass $M_*$ and the vector interaction strength $G_{v}$.\footnote{ Note that the formula (\ref{MLGv}) implies that the effective mass is the same for the positive and negative baryons. However, for negative baryons, the effective mass exists only when there is a Fermi sea of negative baryons, which is not the case at the normal nuclear matter density that we consider in this section. }
 This formula may be also interpreted in the context of Fermi liquid theory (see e.g.\ Refs.~\cite{Baym:1975va,Friman:2019ncm}) where it can be written in the form 
  \beq
 M_{\ast}=\mu_0\left( 1+\frac{F_1}{3}\right),
 \eeq
 where the Landau parameter $F_1$ is directly related to the vector coupling strength $G_v$:  $F_1=-3G_v n_0/\mu_0$. 
 
 The second equation (\ref{MLGv}) provides a constraint on the model parameters. It allows us in particular to determine $\sigma_0$ as a function of $m_0$,   given a value of $M_*$.

 As shown in Eq.~(\ref{alpha12}) the Taylor coefficients $\alpha_1$ and $\alpha_2$ are related to $m_\pi^2$ and $m_\sigma^2$. The other  coefficients, $\alpha_{3}$ and $\alpha_{4}$, 
are determined  from the two conditions
\begin{equation}
	{\cal E}(n_0;\sigma_0) = \mu_0 n_0, \qquad
	\left.\frac{\del {\cal E}(n_0;\sigma)}
	{\del\sigma}\right|_{\sigma_0} = 0,
	\label{eq:determine_taylor}
\end{equation}
where $\sigma_0$ is the value of the sigma field in nuclear matter, i.e.\ $\sigma_0=\sigma_{n_0}$. Note that these conditions are analogous to those which were used to fix the values of $\alpha_1$ and $\alpha_2$ from the local properties of the potential in the vicinity of $\sigma=f_\pi$.  The first condition is the condition of vanishing pressure. The second condition is the gap equation. 

\begin{table}[t]
	\caption{\label{tab:parameter_choice_PD}Set of chosen parameter values 
	(if not stated otherwise) and approximate values of derived parameters for the parity-doublet model.}
	\begin{ruledtabular}
		\begin{tabular}{lr}
		Parameter & Numerical value\\
		\colrule\\[-0.25cm]
        Chiral-invariant mass $m_{0}\ [\mathrm{MeV}]$ & $800$ \\[0.05cm]
        Isoscalar mass $m_{\sigma}\ [\mathrm{MeV}]$ & $340$ \\[0.05cm]
        Landau effective mass $M_{\ast}$ & $0.93 \times M_{N}$ \\[0.05cm]
        \colrule\\[-0.25cm]
		Yukawa coupling $y_{a}$ & $6.9$\\[0.05cm]
		Yukawa coupling $y_{b}$ & $13.0$\\[0.05cm]
		Yukawa coupling $y_{+}(f_{\pi})$ & $4.5$\\[0.05cm]
		Yukawa coupling $y_{-}(f_{\pi})$ & $10.6$\\[0.05cm]
        In-medium condensate $\sigma_{0}\ [\mathrm{MeV}]$ & $65.9$\\[0.05cm]
		Taylor coefficient $\alpha_{3}\ [\mathrm{MeV}^{-2}]$ & $4.4 \times 10^{-1}$\\[0.05cm]
		Taylor coefficient $\alpha_{4}\ [\mathrm{MeV}^{-4}]$ & $-7.8 \times 10^{-5}$\\[0.05cm]
		Vector coupling $G_{v}\ [\mathrm{fm}^{2}]$ & $1.58$\\[0.05cm]
	   \colrule\\[-0.25cm]
	   Compression modulus $K\ [\mathrm{MeV}]$ & 242.8\\[0.05cm]
	   Surface tension $\Sigma\ [\mathrm{MeV\,fm}^{-2}]$ & 1.28
		\end{tabular}
	\end{ruledtabular}
\end{table}
\begin{table}
	\caption{\label{tab:parameter_choice_WT}Set of chosen parameter values
	for the singlet model (if not stated otherwise) and approximate values 
	of derived parameters.}
	\begin{ruledtabular}
		\begin{tabular}{lr}
		Parameter & Numerical value\\
		\colrule\\[-0.25cm]
        Isoscalar mass $m_{\sigma}\ [\mathrm{MeV}]$ & $640$ \\[0.05cm]
        Landau effective mass $M_{\ast}$ & $0.8 \times M_{N}$ \\[0.05cm]
        \colrule\\[-0.25cm]
		Yukawa coupling $y_{a} \equiv y$ & $10.1$\\[0.05cm]
		In-medium condensate $\sigma_{0}\ [\mathrm{MeV}]$ & $69.7$\\[0.05cm]
		Taylor coefficient $\alpha_{3}\ [\mathrm{MeV}^{-2}]$ & $2.2 \times 10^{-1}$\\[0.05cm]
		Taylor coefficient $\alpha_{4}\ [\mathrm{MeV}^{-4}]$ & $-4.3 \times 10^{-5}$\\[0.05cm]
		Vector coupling $G_{v}\ [\mathrm{fm}^{2}]$ & $5.44$\\[0.05cm]
			   \colrule\\[-0.25cm]
	   Compression modulus $K\ [\mathrm{MeV}]$ & 299.2\\[0.05cm]
	   Surface tension $\Sigma\ [\mathrm{MeV\,fm}^{-2}]$ & 1.43
		\end{tabular}
	\end{ruledtabular}
\end{table}

At this point, all parameters have been either fully determined, or related to  $m_0$, $m_\sigma$, and $G_v$ (or $m_\sigma$ and $G_v$ in the singlet model). We shall then explore the range of acceptable values of these parameters by looking at other physical properties of nuclear matter. These concern additional  ground state properties beyond the ground state density and binding energy, and the characteristics (pressure, density, and temperature) of the critical point of the liquid-gas transition.  

Among the ground state properties that are commonly referred to in this context are the surface tension $\Sigma$ and the compression modulus $K$. A further quantity is the nucleon sigma term, which will be the object of the next section. We find it  useful  here to review briefly the derivations of  $\Sigma$ and  $K$ as this provides insight into their dependence on the parameters and the uncertainties in their determination.

The surface tension is typically computed  by considering a semi-infinite slab of nuclear matter, whose density is uniform in the $x$ and $y$ direction, but varies from $n_0$ to 0 in the $z$ direction, i.e., the density is a function $n(z)$.  Surface properties have been studied in relativistic models similar to the present ones for a long time, solving field equations, or using semi-classical approximations (see e.g.\ Refs.~\cite{boguta_relativistic_1977,centelles_semiclassical_1993,Hua:2000gd} for some representative calculations). Here we shall follow a phenomenological approach, based on semi-classical and non relativistic approximations. We write the total energy per unit surface area as the following functional of the density $n(z)$ (see e.g.\ Ref.~\cite{blaizot1981breathing})
\beq\label{eq:Sigma}
\Sigma[n]=\int_{-\infty}^{+\infty} \rmd z \left[ {\cal E}(n)+\frac{C}{2n} \left(\frac{\rmd n}{\rmd z}  \right)^2 -\mu_0 n  \right].
\eeq
In this expression, ${\cal E}(n)={\cal E}(n;\sigma_n)$ is the energy density of uniform nuclear matter, calculated as indicated earlier in this section (see Eq.~(\ref{eq:calEnsign})), with $\sigma_n$ the solution of the gap equation for the density $n$, and $\mu_0$ is the saturation chemical potential, equal to ${\cal E}(n_0)/n_0$. The gradient term may be seen as a phenomenological contribution, which takes value only in the surface region.  The parameter $C$ controls the shape of the density profile and the particular functional form $(1/n) \rmd n/\rmd z$ finds its origin in the so-called extended Thomas Fermi approximation (see Ref.~\cite{blaizot1981breathing} and references therein). Finally, the term $\mu_0 n$,  once integrated over $z$, is the energy the system would have if all the nucleons were carrying the same energy per particle as in the bulk.

\begin{figure}[t]
	\centering
	\includegraphics[scale=1.0]{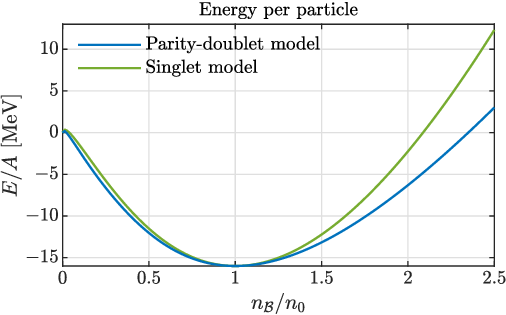}
	\caption{Saturation curves for both the singlet and doublet models. Only the part between $n=0$ and $n=n_0$ contributes to the surface energy. The small difference between the blue and green curves in the region $0\le n\le n_0$ entails a significant difference in the estimate of the surface tension, that correlates to the different values of the compression modulus in the two models. }
	\label{fig:saturation12}
\end{figure}

The surface energy $\Sigma[n]$ in Eq.~(\ref{eq:Sigma}) is  a functional of the density $n(z)$. By requiring this functional to be stationary with respect to variations of $n$, one gets 
\beq
\mu_0=\frac{\rmd {\cal E}}{\rmd n}+\frac{C}{2 n^2}\left( \frac{\rmd n}{\rmd z} \right)^2-\frac{C}{n}\frac{\rmd^2 n}{\rmd z^2},
\eeq
which is easily seen to be equivalent to the equation 
\beq\label{eq:Sigma2}
\frac{C}{2n} \left(  \frac{\rmd n}{\rmd z}  \right)^2 +\mu_0 n-{\cal E}(n)= {\rm cst.}
\eeq
To determine the constant in the right-hand side of this equation, we note that as $z\to -\infty$, the density goes to the normal nuclear matter density $n_0$, which is constant. The derivative drops, and we are left with $\mu_0-{\cal E}(n_0)/n_0$ which vanishes. Thus the constant is zero.  It follows therefore that, when calculated with the solution of the above differential equation, the surface tension can be written in the form  
\beq
\Sigma=\int_{-\infty}^{+\infty} \rmd z\; 2[{\cal E}(n)-\mu_0 n].
\eeq
On the other hand, from Eq.~(\ref{eq:Sigma2}), we get
\beq\label{eq:dndz}
\frac{\rmd n}{\rmd z} =- \sqrt{(2n/C)[{\cal E}(n)-\mu_0 n]},
\eeq
so that, finally\footnote{In the recent literature (see e.g.\ Refs.~\cite{berges_quark_2003,Floerchinger_2012,Drews:2013hha}), the surface tension has been estimated  as
$\Sigma = \int_{\sigma_0}^{f_{\pi}}
	\mathrm{d}\sigma\,  \sqrt{2 \Omega(\mu_0;\sigma)}$, where $\Omega$ is the grand potential evaluated for the saturation chemical potential. 
The connection to the  calculation presented here is unclear to us.}
\beq\label{eq:Sigman}
\Sigma=\int_{0}^{n_0} \rmd n\; \sqrt{2C\left[\frac{{\cal E}(n)}{n}-\mu_0 \right]}.
\eeq
With the particular choice that we have made for the gradient contribution in Eq.~(\ref{eq:Sigma}), the formula above shows that the surface energy is obtained by integrating the deviation of the energy per particle with respect to its saturation value. The final result is then sensitive to the details of the saturation curve  in the region $0\le n\le n_0$  (see Fig.~\ref{fig:saturation12}). Note that this involves regions where the pressure is negative and also regions where the system is mechanically unstable with a negative compressibility. In the present case, the stabilizing agent is the gradient term in Eq.~(\ref{eq:Sigma}). 

To estimate the surface tension, we need not only the saturation curve, but also the parameter $C$. This is estimated as follows. It turns out that the solution of the differential equation 
(\ref{eq:Sigma2}) overlaps very precisely with the function $n(z)=n_0/(1+\rme^{z/a})$. One may then chose $C$ so that $a$, which measures the surface thickness, takes a given value. We have chosen $a=0.5$ fm, which is within  the range of values extracted from nuclear densities (see e.g.\ Refs.~\cite{bohr1998nuclear,friar1975theoretical}). With this choice, we get $\Sigma=1.43$ MeV/fm$^2$ for the singlet model, and $
\Sigma=1.28$ MeV/fm$^2$ for the doublet model. These values are larger than the typical values extracted from the analysis of the masses of nuclei which would favor a value $\Sigma\lesssim 1.2 $ MeV/fm$^{2}$ \cite{bohr1998nuclear}.  Note that the  values of $\Sigma$ are approximately  proportional to the compression modulus (see Tables \ref{tab:parameter_choice_PD} and \ref{tab:parameter_choice_WT}), which appears to be a generic feature of this kind of models. For instance in Ref.~\cite{Hua:2000gd} it is argued that both $\Sigma$ and $a$ are inversely proportional to $m_\sigma$, the ratio being roughly proportional to the compression modulus. Here, since we keep the surface thickness constant,  $\Sigma $ itself becomes proportional to $K$.

We turn now to the compression modulus $K$, whose calculation can be done in two complementary ways. Some elements of these calculations will be useful later when we discuss the chiral transition in Sec.~\ref {sec:chiraltrans}. Let us recall that $K$ is defined as 
\beq	K & = & 9 n_0 \left.\frac{\rmd\mu}{\rmd n}
	\right|_{n_{0}},
	\eeq
where the derivative is to be evaluated at the saturation density $n_0$. The calculation of the derivative proceeds as follows
\beq
\frac{\rmd \mu}{\rmd n}=\left.\frac{\del \mu}{\del n}\right|_\sigma +\left.\frac{\del \mu}{\del \sigma}\right|_n \,\frac{\rmd \sigma}{\rmd n},
\eeq
where $\mu$ is given by
\beq
\mu= \sqrt{p_F^2+M_+^2} + G_v n.
\eeq
We have then
\beq
\left.\frac{\del \mu}{\del \sigma}\right|_n =y_{+} \frac{M_+}{M_{\ast}},\qquad \left.\frac{\del \mu}{\del n}\right|_\sigma=G_v+\frac{\pi^2}{2p_F M_{\ast}}.
\eeq
In order to calculate ${\rmd \sigma}/{\rmd n}$ we differentiate the gap equation $\rmd U/\rmd \sigma+y_{+} n_{\rm s}^+=0$ and obtain
\beq\label{eq:disgmadn0}
\frac{\rmd \sigma}{\rmd n}=-\frac{y_{+}}{m_{\sigma_0}^2} \left.\frac{\partial n_{\rm s}^+}{\partial n}\right|_\sigma ,
\eeq
where
\beq
m_{\sigma_0}^2= \frac{\rmd^2 U}{\rmd \sigma^2}+n_{\rm s}^+ \frac{\rmd^2 M_+}{\rmd \sigma^2}+y_{+}\left.\frac{\del n^+_s}{\del \sigma}\right|_n ,
\eeq
all derivatives being evaluated for $\sigma=\sigma_0$. Collecting all these results, we obtain
\beq\label{eq:dmuBdnB1}
\left.\frac{\rmd \mu}{\rmd n}\right|_{n_{0}}=G_v+\frac{\pi^2}{2p_F M_{\ast}}-\frac{y_+^2}{m_{\sigma_0}^2} \frac{M_+}{M_{\ast}} \left. \frac{\del n_{\rm s}^+}{\del n}\right|_{\sigma_0}.
\eeq

The second way to proceed amounts to calculate $\rmd n/\rmd \mu$, with $n$ given by 
\beq
n=4\int_\p \theta(\mu-E_\p^+).
\eeq
Taking the derivative with respect to the chemical potential, one gets
\beq\label{eq:K1}
\frac{\rmd n}{\rmd\mu}&=& 4\int_\p \delta(\mu-E_\p^+) \left(1-\frac{\rmd E_\p^+}{\rmd n}\frac{\rmd n}{\rmd\mu}  \right)\nn[0.1cm]
&=& N_0\left(1-f_0^+ \frac{\rmd n}{\rmd\mu}\right),
\eeq
where
\beq\label{eq:K2}
N_0=\frac{2p_F M_{\ast}}{\pi^2}
\eeq
is the density of state at the Fermi surface, and   $F_0=N_0 f_0^+$ is a Fermi liquid parameter, defined as   
\beq\label{eq:K3}
f_0^+\equiv \frac{\rmd E_\p^+}{\rmd n}=G_v+y_{+} \frac{M_+}{M_{\ast}}\frac{\rmd \sigma}{\rmd n},
\eeq
where the expression (\ref{eq:Epplus}) of $E^+_\p$ has been used. 
At this point we have obtained
\beq\label{eq:compressibility0}
\frac{\rmd n}{\rmd \mu}=\frac{N_0}{1+F_0}.
\eeq
By substituting the expression (\ref{eq:disgmadn0}) of $\rmd\sigma/\rmd n$ in Eq.~(\ref{eq:K3}), one easily shows that this expression agrees with that given in Eq.~(\ref{eq:dmuBdnB1}) for ${\rmd \mu}/{\rmd n} $. The 
  familiar form  of the compression modulus follows \cite{Baym:1975va}:
\beq	K =3\frac{p_F^2}{\mu_0}\frac{1+F_0}{1+F_1/3} .
\eeq
The value of the compression modulus  of the singlet model is larger than the value extracted from the analysis of giant monopoles excitations of large nuclei (see e.g.\ Ref.~\cite{Blaizot:1995zz}\footnote{The value adopted in the present paper $K=230\pm20$ MeV is based on a rough update of the analysis of Ref.~\cite{Blaizot:1995zz}, taking into account the most recent data \cite{Youngblood:1999zza,Patel:2014cxa}.}), even when compared to the largest values suggested in Ref.~\cite{shlomo_deducing_2006}. For the parity-doublet model, the value obtained is compatible with the latter estimate. Note that neither the values of $\Sigma$ nor that of $K$ have been used in adjusting the parameters. These values result from fixing  parameters such as $\alpha_3$ and $\alpha_4$ using constraints discussed earlier in this section.

\subsubsection{Final choices of parameters}

%\begin{figure}[t]
%	\centering
%	\includegraphics[scale=1.0]{parameter_bands_PD_single_91}
%	\includegraphics[scale=1.0]{parameter_bands_PD_single_93}
%	\includegraphics[scale=1.0]{parameter_bands_PD_single_95}
%	\caption{Parameter bands for the parity-doublet model in the $m_{\sigma}$-$m_{0}$-plane
%	matching experimental data (within errorbars) of Table \ref{tab:experiment}. Shown is
%	the result for the Landau effective mass $0.91 \le M_{\ast}/M_{N} \le 0.94$ in steps of
%	$0.01$ [in consecutive subfigures (a) to (d)]. The legend of subfigure (a) applies to 
%	all panels. The gray-shaded area indicates parameter sets leading to $T_{c} > 30\ \mathrm{MeV}$ 
%	for the liquid-gas transition; these sets are excluded from the detailed parameter analysis. }
%	\label{fig:parameter_bands_PD}
%\end{figure}

\begin{figure}[t]
	\centering
	\includegraphics[scale=1.0]{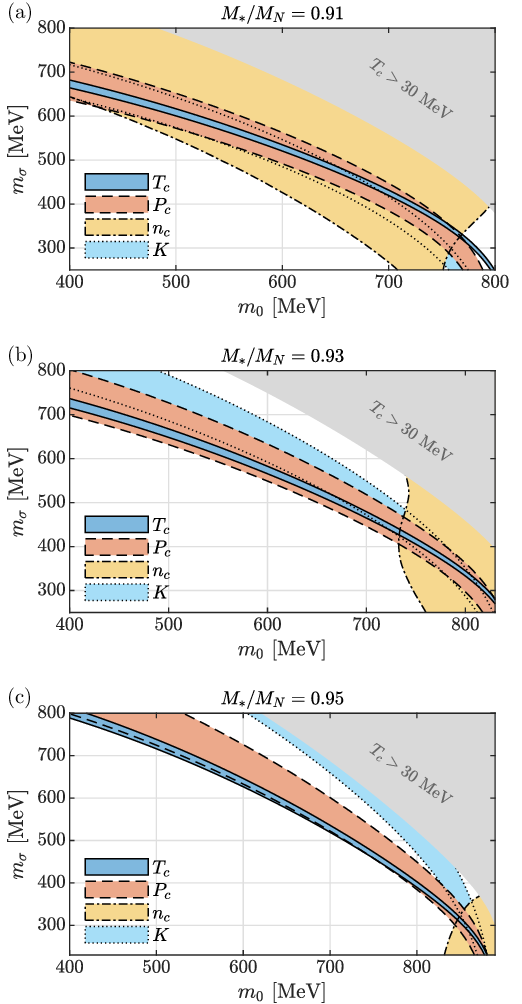}
	\caption{Parameter bands for the parity-doublet model in the $m_{\sigma}$-$m_{0}$-plane
	matching experimental data (within errorbars) of Table \ref{tab:experiment} (and including the
	compression modulus drawn here as the band $K = 230\pm 20\ \mathrm{MeV}$). Shown is the result for the effective mass $M_{\ast}/M_{N} = 0.91$ 
	[panel (a)], $0.93$ [panel (b)], and $0.95$ [panel (c)]. The gray-shaded area indicates parameter 
	sets leading to $T_{c} > 30\ \mathrm{MeV}$ for the liquid-gas transition.}
	\label{fig:parameter_bands_PD_new}
\end{figure}

Figure \ref{fig:parameter_bands_PD_new} demonstrates the capability of the parity-doublet model
to reproduce empirical data. Because of the sensitivity of the results to the value of $M_*$ we display  parameter bands for three different ratios
$M_*/M_{N}$ in the $m_{\sigma}$-$m_{0}$-plane for which the respective values listed in 
Table \ref{tab:experiment} are matched (within indicated errorbars). There is a clear correlation between $m_\sigma$ and $m_0$, suggesting that an increase of $m_0$ can be compensated by a decrease in $m_\sigma$. Although this correlation is the result of a complicated balance between several effects, involving the interplay of many parameters, the following remark could make it more intuitive. This is based on the dependence of the equilibrium value $\sigma_0$ of the sigma field.  We note that an increase of $m_\sigma$ naturally leads to an increase of  the value of $\sigma_0$ (i.e.\ it gives a value of $\sigma_0$ closer to $f_\pi$). Similarly, by noticing that $M_+(\sigma_0; m_0)$ is fixed by the value of $M_*$ once the density is fixed, we see that by decreasing $m_0$, one increases the difference $M_+(\sigma_0; m_0)-m_0$, which can be compensated again by increasing the value of $\sigma_0$ (given that the Yukawa couplings are fixed by Eq.~(\ref{eq:YukawaMN})). Note that the correlations discussed here are similar to those observed in previous studies of the parity-doublet model (see e.g.\ Refs.~\cite{Zschiesche:2006zj,sasaki_thermodynamics_2010,minamikawa_chiral_2023,Motohiro:2015taa,mun2023nuclear}).

Choices of parameters corresponding to maximally overlapping 
zones are of course preferred, since in these zones  more properties of nuclear matter are reproduced. In fact, 
for  $M_{\ast}/M_{N} = 0.93$, one finds a zone  in the regime 
of $300\ \mathrm{MeV} \lesssim m_{\sigma} \lesssim 450\ \mathrm{MeV}$ and $730\ \mathrm{MeV} 
\lesssim m_{0} \lesssim 820\ \mathrm{MeV}$ in which all data ranges of Table \ref{tab:experiment} 
are matched (even for the compression modulus $K$). It is in this region that we have fixed our  parameters. These are listed in Table \ref{tab:parameter_choice_PD}.  This is of course not a unique choice, and there exist other acceptable regions of the parameter space, involving in particular smaller values of $m_0$. We have however several reasons to prefer  a choice of a relatively small $\sigma$ mass in combination with a large 
value of $m_{0}$. We have shown in a previous paper that such a combination  successfully reproduces pion-pion scattering lengths \cite{Eser:2021ivo}.
  Studies of nuclear properties within the parity-doublet model also favor a large $m_0$ (see e.g.\ Ref.~\cite{mun2023nuclear}). Furthermore, a look at Fig.~\ref{fig:omega_vacuum} in the Appendix~\ref{sec:vacuum} reveals that a small value of $m_0$ implies large cancellations of quantities of order  2 to 3 GeV  for the range of values of $\sigma$ relevant for nuclear matter, which looks unnatural. We note in addition that a large value of $m_0$ seems to be favored by recent studies of the parity-doublet model (see e.g.\ Ref.~\cite{koch2023fluctuations}), and by lattice calculations about the composition of the proton mass assigning only a minor fraction to quark scalar condensates \cite{Yang:2018nqn}. 
 
\begin{figure}[h]
	\centering
	\includegraphics[scale=1.0]{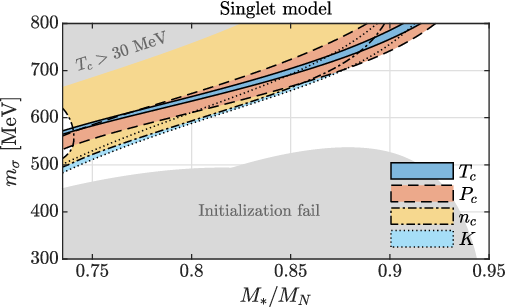}
	\caption{Parameter bands for the singlet model in the $m_{\sigma}$-$M_*/M_{N}$-plane
	matching experimental data (within errorbars) of Table \ref{tab:experiment} (and
	including the compression modulus $K = 230\pm 20\ \mathrm{MeV}$). The
	region called ``initialization fail'' consists of parameters for which the 
	initialization conditions were not met, e.g.\ the solution of the gap equation
	(\ref{eq:MF_eq_sigma}) corresponds to a local maximum, not a minimum.}
	\label{fig:parameter_bands_WT_new}
\end{figure}

When further decreasing or increasing the Landau effective mass, we observe that these complete overlap
zones either shrink and move to even larger $m_{0}$ and smaller $m_{\sigma}$ (see Fig.~\ref{fig:parameter_bands_PD_new}(c)), or move away from the preferred region of large $m_{0}$ and small $m_{\sigma}$ (see Fig.~\ref{fig:parameter_bands_PD_new}(a)), both of which we do not favor for the reasons mentioned above.\footnote{Nevertheless, we could equally have chosen e.g.\ $M_{\ast}/M_{N} = 0.94$ (not shown) with slightly different values for $m_{0}$ and $m_{\sigma}$, but we do not expect drastic changes in the results.} The values that we
eventually obtain for the critical end point of the liquid-gas transition (with the chosen parameter
set) are also listed in Table~\ref{tab:experiment}.

A similar analysis can be made for the singlet model. The results are illustrated in Fig.~\ref{fig:parameter_bands_WT_new}, which reveals a clear correlation between the effective mass (or the vector coupling) and the sigma mass.  This is to be expected. Indeed, an increase in the effective mass implies a decrease of the effective vector coupling $G_v$, hence a reduction of the repulsion between the nucleons. This can be compensated by a reduction of the attraction, controlled by $G_s=y^2/m_\sigma^2$, hence by an increase of $m_\sigma^2$ since the Yukawa coupling $y$ is fixed to the value $M_N/f_\pi$.  The overlap region would suggest a value for $M_* \gtrsim 0.85$ (respecting again $K$ for the moment), but at the cost of $m_{\sigma} \gtrsim 700\ \mathrm{MeV}$. We therefore choose $M_*/M_N=0.8$ corresponding to a not too large $\sigma$ mass, similar to those values quoted in Ref.\ \cite{brandes_fluctuations_2021}, and accepting that this yields a compression modulus which is too high (as reported earlier). Comparing the corresponding parameters given in Table~\ref{tab:parameter_choice_WT} with those of the doublet model in Table~\ref{tab:parameter_choice_PD}, one observes that one gets a larger $\sigma_0$, smaller $\alpha_3$ and $\alpha_4$ (in absolute values), and larger $G_v$. The larger value of $G_v$ reflects the strongest attraction mechanism of the singlet model so that saturation requires more repulsion.  The larger value of $\sigma_0$ is correlated to the larger value of the $\sigma$ mass in the singlet model, as already mentioned.

\subsubsection{The liquid-gas phase transition}

\begin{table}	
	\caption{\label{tab:experiment}Experimental values for the critical end point 
	of the nuclear liquid-gas transition \cite{Elliott:2013pna}, and the values obtained within the doublet and singlet models (including also the respective critical baryon-chemical potential).}
	\begin{ruledtabular}
		\begin{tabular}{lr}
		Observable & Numerical value\\
		\colrule\\[-0.25cm]
		Temperature $T_{c}\ [\mathrm{MeV}]$ & $17.9 \pm 0.4$\\[0.05cm]
		Pressure $P_{c}\ [\mathrm{MeV}\,\mathrm{fm}^{-3}]$ & $0.31 \pm 0.07$ \\[0.05cm]
		Baryon density $n_{c}\ [\mathrm{fm}^{-3}]$ & $0.06 \pm 0.01$ \\[0.05cm]
		%\colrule\\[-0.25cm]
		%Nuclear surface tension $\Sigma\ [\mathrm{MeV}\,\mathrm{fm}^{-2}]$ & $1.08 \pm 0.06$\\[0.05cm]
		%Compression modulus $K\ [\mathrm{MeV}]$ & $240 \pm 20$
		\colrule\\[-0.25cm]
		Parity-doublet model: & \\[0.05cm]
		$T_{c}\ [\mathrm{MeV}]$ & 18.0 \\[0.05cm]
		$P_{c}\ [\mathrm{MeV}\,\mathrm{fm}^{-3}]$ & 0.33 \\[0.05cm]
		$n_{c}\ [\mathrm{fm}^{-3}]$ & 0.06 \\[0.05cm]
		$\mu_{c}\ [\mathrm{MeV}]$ & 905\\[0.05cm]
		\colrule\\[-0.25cm]
		Singlet model: & \\[0.05cm]
		$T_{c}\ [\mathrm{MeV}]$ & 17.9 \\[0.05cm]
		$P_{c}\ [\mathrm{MeV}\,\mathrm{fm}^{-3}]$ & 0.34 \\[0.05cm]
		$n_{c}\ [\mathrm{fm}^{-3}]$ & 0.06\\[0.05cm]
		$\mu_{c}\ [\mathrm{MeV}]$ & 907
		\end{tabular}
	\end{ruledtabular}
\end{table}

\begin{figure}[h]
	\centering
	\includegraphics[scale=1.0]{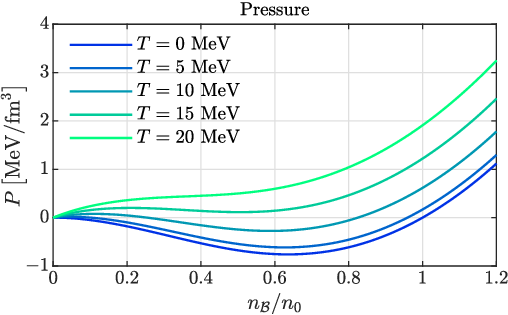}
	\caption{Equation of state (pressure $P$ as function of baryon density $n_{\mathcal{B}}$) 
	for various temperatures.}
 	\label{fig:pressure}
\end{figure}

The liquid-gas transition is an important property of nuclear matter. The transition occurs at densities lower than the saturation density and at a temperature of the order of the binding energy or lower. The transition occurs as the result of the competition between entropy effects and binding energy effects, and the properties of this transition are directly related to the ground state properties of nuclear matter (binding energy, compressibility, effective mass). The various isothermal curves in  Fig.~\ref{fig:pressure} indicate how this transition occurs. These curves are obtained by following continuously the solution of the gap equation as the density increases. As already mentioned, such solutions are found to exist even in regions of negative pressure, or regions between the spinodal points (where the derivative of the pressure vanishes) in which the compressibility is negative and the system is a priori unstable. When the temperature increases, nuclear matter continues to exist as a (metastable) self-bound system of zero pressure. For a slightly positive pressure, it can coexist with a low-density vapor. As the temperature continues to increase the self-bound system ceases to exist, the pressure becoming positive for all values of the density, while coexistence between two phases is still possible. This phase coexistence remains possible until the critical  temperature is reached, above which the pressure becomes a monotonously increasing function of $n$.

The values of the thermodynamic variables at the critical point  can be extracted from nucleus-nucleus collisions \cite{Elliott:2013pna}. They are listed in the Table~\ref{tab:experiment}, together with the corresponding values obtained in the two models for the chosen values of the parameters. One sees that both models, with the present choices of parameters, account rather well for these characteristic properties of the liquid-gas transition.

\section{The nucleon sigma term}

\label{sec:sigma_term}

\subsection{Definitions}

The pion-nucleon sigma term $\sigma_N$ is defined as the matrix element\footnote{The nucleon states are normalized so that  $\bra{\p',s'} \p,s\rangle =(2\pi)^3 \delta^{(3)}(\p'-\p) \delta_{s's}$, with $\p$ the three-momentum and $s$ the spin projection.}
\beq
\sigma_{N}=\bar m\left[ \bra{N}\bar q q\ket{N}-\langle\bar q q\rangle_0\right].
\eeq
In this equation, $\bar q q=\bar{ u} u+\bar{d} d$, $\bar m=(m_u+m_d)/2$, and we ignore  isospin symmetry breaking (i.e., $m_u=m_d$). Furthermore, $\bar q q$ stands for the spatial integral $\int_\x \bar q(\x) q(\x)$. The sigma term  provides a measure of the scalar density within the nucleon and of the direct contribution of light quarks to the nucleon mass (for reviews on the sigma term, see Refs.~\cite{sainio_pion-nucleon_2001,Alarcon:2021dlz}). 

The Feynman-Hellmann theorem allows us to relate the expectation value of $\bar q q$ to that of the symmetry breaking part of the QCD Hamiltonian \cite{Cohen:1991nk} 
\beq\label{eq:FHtheorem}
\bar m \frac{\rmd }{\rmd \bar m}\langle {\cal H}\rangle=\bar m \left\langle \frac{\rmd }{\rmd \bar m} {\cal H}_{\rm SB}\right\rangle=\bar m \langle \bar{ q} q \rangle,
\eeq
where the expectation value of the Hamiltonian ${\cal H}$  is taken in a ground state or with the statistical density operator at finite temperature and density (see later where the same strategy is formulated in terms of the gap equation).  The important point is that only the explicit symmetry breaking term in the Hamiltonian, denoted ${\cal H}_{\rm SB}$, contributes to the derivative with respect to $\bar m$.  By using Eq.~(\ref{eq:FHtheorem}) one can write the following expression for the sigma term
\beq
\sigma_{N}=\bar m  \frac{\rmd M_N}{\rmd \bar m}\,,
\eeq
where $M_N=\bra{N} {\cal H}\ket{N}-\bra{0} {\cal H}\ket{0}$ is the nucleon mass. 

Experimentally, the sigma term is  deduced from the measurement of the $\pi N$ scattering amplitudes, including constraints from pionic atoms. Most recent determinations yield a value $\sigma_N\simeq 60$ MeV \cite{Hoferichter:2015hva,friedman_pion-nucleon_2019}, somewhat larger than the  traditionally accepted value $\sigma\simeq 45$ MeV \cite{Gasser:1990ce}, and definitely larger than the value obtained in lattice calculations, $\sigma\simeq 40$ MeV (see e.g.\ Ref.~\cite{Agadjanov:2023jha} and references therein).  For a discussion of this discrepancy between experimental determinations and lattice calculations see Refs.~\cite{Hoferichter:2016ocj,Hoferichter:2023ptl}.

In the context of dense-matter studies, the nucleon sigma term is interesting  in that it provides information on the ``resistance'' to chiral symmetry restoration, by measuring the change in hadronic properties induced by the small explicit symmetry breaking term proportional to the quark mass.  More broadly, the sigma term informs us on how the presence of (dilute) matter affects the average scalar field which the nucleon mass is sensitive to. We shall later consider larger densities leading eventually to chiral symmetry restoration. In this high-density regime, the sigma term looses its interest since, as we shall see,  density-dependent corrections are large and may not lead to a convergent expansion. 

\begin{figure}[h]
	\centering
	\includegraphics[scale=1.0]{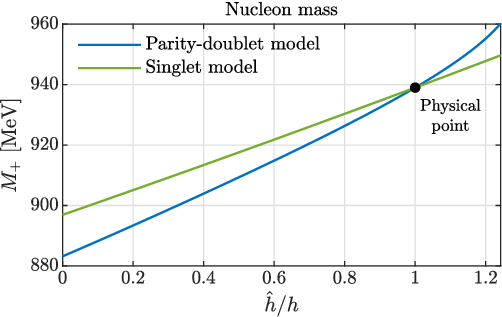}
	\caption{The nucleon mass $M_{+}$, as a function of $\hat{h}$, in both the singlet and doublet models.}
	\label{fig:h_dependenceM}
\end{figure}

In this section, we provide estimates of $\sigma_N$ within the singlet and doublet models. It turns out that  these estimates  depend sensitively on precisely how one does the calculation. This leads to uncertainties, which are particularly important in the parity-doublet model,  and which have  to do with the extrapolation from the physical point to the chiral limit. Thus, by expanding the nucleon mass about its value in the chiral limit, we may define 
\beq\label{sigmadef0}
\sigma_{N}=\bar m\left. \frac{\rmd M_N}{\rmd \bar m} \right|_{\bar m\to 0}\simeq M_N(\bar m)- M_N(\bar m=0), \quad
\eeq
where the limit $\bar m\to 0$ does not concern the first factor $\bar m$ (for which we should use the ``physical'' value of $\bar m$), but only the point where the derivative is evaluated. If the nucleon mass  $M_N$ depended linearly on $\bar m$, all the way from the chiral limit to the physical point, it would not matter whether the derivative is evaluated at the physical point or at the chiral limit. However, in the parity-doublet model, this linearity is only approximately verified (see Fig.~\ref{fig:h_dependenceM}). In the singlet model an almost perfect linearity is observed, and as a result the various estimates of $\sigma_N$ end up being closer to each other (see Table~\ref{tab:sigma_term} below).

\subsection{Remarks on the chiral limit}

At this point, it is useful to recall that chiral symmetry leads to a number of model-independent relations in the vicinity of the chiral limit. To see how these are implemented in the present model, we look at the variation of various parameters as a function of the strength $\hat h$ of the explicit symmetry breaking term. To avoid confusion, we denote by $\hat h$ this parameter, reserving $h$ for its physical value $h=m_\pi^2 f_\pi$ (see Eq.~(\ref{eq:hvalue})). We shall base our analysis on Eq.~(\ref{eq:minVphih}), which is strictly valid only in the vicinity of the physical point. It is of course trivial to proceed to a numerical evaluation involving the complete potential $V(\varphi)$, which we shall do later. The present analysis provides useful  analytical understanding of the results of such numerical calculations.

We start by rewriting the vacuum gap equation (\ref{eq:minVphih}) in the following way
\beq\label{eq:minVphih2}
\frac{1}{\sigma}
\frac{\rmd V(\sigma)}{\rmd \sigma}=\alpha_1+\frac{\alpha_2}{2}(\sigma^2-f_\pi^2)=\frac{\hat h}{\sigma}.
\eeq
The solution for $\hat h=h$ is $\sigma=f_\pi$. 
In the chiral limit, $\hat h=0$, the solution $\sigma_\chi$ is given by
\beq\label{eq:sigmachi}
\sigma_\chi^2-f_\pi^2=-2\frac{\alpha_1}{\alpha_2}=-2 \frac{m_\pi^2 f_\pi^2}{m_\sigma^2-m_\pi^2},
\eeq
where we have kept the parameters $\alpha_1$ and $\alpha_2$ at their ``physical values" as $\hat h\to 0$. The last expression in Eq.~(\ref{eq:sigmachi})  reflects the competition between explicit symmetry breaking (the numerator) and spontaneous symmetry breaking (the denominator). The decrease of the effective value of $f_\pi$ as one moves to the chiral limit is generic (and in qualitative agreement with chiral perturbation theory): with our choice of sign, $h$ is positive, and so is the value  $\sigma_\chi$ at the minimum of the potential, with $ \sigma_\chi< f_\pi$.  Anticipating on the next subsection, we note here that the results of the numerical evaluations indicate  a larger shift in the  doublet model than in the singlet model.  Given that the value of $m_\sigma^2$ in the doublet model is smaller than in the singlet model (see Tables~\ref{tab:parameter_choice_PD} and \ref{tab:parameter_choice_WT}), this result is in qualitative agreement with formula (\ref{eq:sigmachi}). The prediction of chiral perturbation theory is $\Delta f_\pi= \sigma_\chi-f_\pi\approx -6$  MeV, while one finds $\Delta f_\pi \approx- 4\ \mathrm{MeV}$ and $\Delta f_\pi\approx -13\ \mathrm{MeV}$ in the singlet and doublet models, respectively.\footnote{The comparison with chiral perturbation theory is of course only indicative, since in the present approximation, the chiral logarithms are not included in our model estimates. A thorough discussion of the relation between the nucleon and the pion masses can be found in Ref.~\cite{Hoferichter:2015hva}.}

\begin{figure}[h]
	\centering
	\includegraphics[scale=1.0]{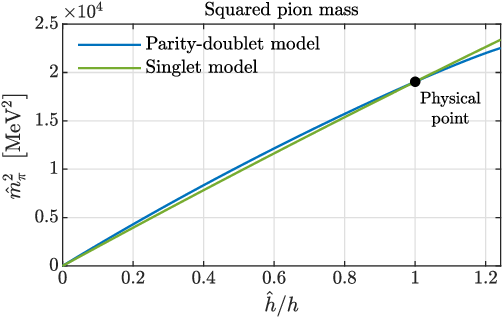}
	\caption{The squared pion mass $\hat{m}_\pi^2$ as a function of $\hat h$, illustrating its non linear behavior in the doublet model.}
	\label{fig:h_dependenceMpi}
\end{figure}

Let us now consider an arbitrary small value of $\hat h$, and expand the solution $\sigma_{\hat h}$ around its chiral limit $\sigma_\chi$. By combining Eq.~(\ref{eq:minVphih2}) with the corresponding equation for $\hat h=0$, one gets
\beq\label{eq:sigmahath}
\frac{\alpha_2}{2}(\sigma_{\hat h}^2-\sigma_\chi^2)=\frac{\hat h}{\sigma_{\hat h}}.
\eeq 
Assuming the difference between $\sigma_{\hat h}$ and $\sigma_\chi$ to be small,   we get
\beq
\sigma_{\hat h}-\sigma_\chi\simeq \frac{\hat h}{\alpha_{2}\sigma_\chi^2},
\eeq
which is indeed  linear in $\hat h$ at small $\hat h$, with a quadratic correction.
The value $\hat m_\pi$  of the pion mass for an arbitrary value of $\hat h$ is given by 
\beq\label{eq:hatmpi}
\hat m_\pi^2=\alpha_1+\frac{\alpha_2}{2}(\sigma_{\hat{h}}^2-f_\pi^2)=\frac{\hat h}{\sigma_{\hat{h}}},
\eeq
where the first equality follows from taking the second derivative of the potential $V(\varphi)$ in Eq.~(\ref{Vvarphi}), and the second one from Eq.~(\ref{eq:minVphih2}). 
In the chiral limit $\hat m_\pi$ vanishes as it should. For the physical value   $\hat h=h$ we get instead $\hat m_\pi^2=\alpha_1=m_\pi^2$. More generally $\hat m_\pi^2=\hat h/\sigma_{\hat h}$ where $\sigma_{\hat h}$ is given by the solution of the equation (\ref{eq:sigmahath}) above. Thus the leading-order expression of $\hat m_\pi^2$ is linear in $\hat h$, with a correction quadratic in $\hat h$.  A plot of $\hat m_\pi^2$ as a function of $\hat h$ is given in Fig.~\ref{fig:h_dependenceMpi}. A small non linearity is visible for the parity-doublet model (the relation for the singlet model is nearly perfectly linear). As we shall see, this non linearity impacts the value of $\sigma_N$. 
Similarly, the sigma mass is given by 
\beq\label{eq:hatmsigma}
\hat m_\sigma^2&=&\alpha_1 +\alpha_2 \sigma^2 +\frac{\alpha_2}{2} (\sigma^2-f_\pi^2)\nn
&=&  m_\sigma^2-3 m_\pi^2 +3\hat m_\pi^2.
\eeq
Thus the sigma mass squared  is reduced to the value $m_\sigma^2-3 m_\pi^2$ in the chiral limit.

\subsection{Various estimates of $\sigma_N$}

We now return to the numerical evaluation of the sigma term. All estimates are done with the parameters determined in the previous section. No attempt is made to use the (uncertain) experimental values of $\sigma_N$ to improve the choice of parameters.  We assume that the effect of the quark masses is captured by the symmetry breaking term proportional to $\hat h$, i.e.,  that  the small variations in $\bar m$ are  proportionally to small variations in $\hat h$.   It is then straightforward to obtain an estimate of $\sigma_{N}$, since from the assumption just mentioned, we can write
 \beq\label{sigmaterm2}
\sigma_N^{(1)}= \bar m\left. \frac{\rmd M_N}{\rmd \bar m} \right|_{\bar m\to 0}= h\left. \frac{\rmd M_N}{\rmd \hat h} \right|_{\hat h\to 0},
\eeq
where the upperscript $(1)$ is meant to specify the particular estimate of $\sigma_N$ considered at this point. 
It follows from Eq.~(\ref{sigmaterm2}) that 
\beq
\frac{\sigma_N^{(1)}}{h}=\,y_+(\sigma_\chi) \left. \frac{\rmd \sigma_{\hat h}}{\rmd \hat h} \right|_{\hat h\to 0},
\eeq
%
%
%
%$m_{q}$ (i.e.\ degenerate up and down mass). With the Gell-Mann-Oakes-Renner relation,
%denoting the quark condensate $\braket{\bar{q}q}$,
%\begin{equation}
%	m_{\pi}^{2} \simeq m_{q} \left| \frac{\braket{\bar{q}q}}{f_{\pi}^{2}} 
%	\right|_{m_{q}\, \rightarrow\, 0},
%\end{equation}
%we are able to rewrite $\sigma_{N}$ as
%\begin{IEEEeqnarray}{rCl}
%	\sigma_{N} & \simeq & m_{\pi}^{2} \left. \frac{\mathrm{d}M_{N}}
%	{\mathrm{d}m_{\pi}^{2}} \right|_{m_{\pi}^{2}\, \rightarrow\, 0} 
%	= m_{\pi}^{2}\;\! y_{\sigma}^{++}(\sigma_{\chi})
%	\left. \frac{\mathrm{d}\sigma_{0}}{\mathrm{d}m_{\pi}^{2}} 
%	\right|_{m_{\pi}^{2}\, \rightarrow\, 0} \nonumber\\[0.1cm]
%	& = & m_{\pi}^{2}(\sigma_{0})\;\! y_{\sigma}^{++}(\sigma_{\chi})\;\! \sigma_{\chi}
%	\left. \frac{\mathrm{d}\sigma_{0}}{\mathrm{d}h_{\mathrm{ESB}}} 
%	\right|_{h_{\mathrm{ESB}}\, \rightarrow\, 0} \nonumber\\[0.1cm]
%	& \simeq & y_{\sigma}^{++}(\sigma_{\chi}) \Delta\sigma 
%	\label{eq:sigma_term}
%\end{IEEEeqnarray}
%due to the leading-order linearity of $m_{\pi}^{2}$ in $m_{q}$. The
where $\sigma_{\hat h}$ denotes the solution of Eq.~(\ref{eq:sigmahath}), while  $\sigma_{\chi}$ is the corresponding solution  in the chiral limit. On the left-hand side of Eq.~(\ref{sigmaterm2}) is  $h=m_\pi^2 f_\pi$. The derivative $\rmd \sigma_{\hat h}/\rmd \hat h|_{\hat h\to 0}$ is  the  chiral susceptibility in the chiral limit 
\beq\label{eq:dsigmadh}
\left. \frac{\rmd \sigma_{\hat h}}{\rmd \hat h} \right|_{{\hat h}\to 0}=\frac{1}{\hat m_\sigma^2 (\sigma_\chi)}.
\eeq
 Note that, in the parity-doublet model, it is much enhanced as compared to its value $1/m_\sigma^2$ in the physical point (the simple formula above, Eq.~(\ref{eq:hatmsigma}), yields $\hat m_\sigma^2(\sigma_\chi)={m_\sigma^2-3m_\pi^2}$). 
We then end up with the following formula for our first estimate of $\sigma_N$:
\beq\label{sigma1}
\sigma_N^{(1)}=m_\pi^2 f_\pi  \frac{y_+ (\sigma_\chi)}{\hat m_\sigma^2 (\sigma_\chi)},
\eeq
which is valid in the linear order in the symmetry breaking parameter $h\propto \bar m$. This formula is consistent with Refs.~\cite{Birse:1992uxe, *Birse:1993hx,
Delorme:1996cc, Chanfray:1998hr, Dmitrasinovic:1999mf}. It has the expected qualitative structure. The first term $m_\pi^2 f_\pi$ measures the deviation of the physical point from the chiral limit, the quantity $y_+(\sigma_\chi)$ measures the response of the nucleon mass to a change in the sigma field near the chiral limit, and the factor $1/\hat m_\sigma^2 (\sigma_\chi)$ is the chiral susceptibilty. The numerical evaluation of Eq.~(\ref{sigma1}) yields $\sigma_N^{(1)}=50.2\ \mathrm{MeV}$. 

Note that we could also estimate directly $\sigma_{N}$ from Eq.~(\ref{sigmadef0}), taking the finite difference rather than the derivative. Because $M(\hat h)$ is not strictly linear in $\hat h$, one gets a slightly larger value, $\sigma_N^{(0)}= 55.8$ MeV. 

%\begin{figure}[h]
%	\centering
%	\includegraphics[scale=1.0]{sigma_term_PD}
%	\caption{Nucleon sigma term (\ref{eq:sigma_term}) as function of $m_{0}$ and $m_{\sigma}$
%	for $M_{\ast}/M_{N} = 0.93$. Thick black contours refer to constant lines of $\sigma_{N}$;
%	they are given every $5\ \mathrm{MeV}$ and steadily increase from ``left'' to ``right.'' 
%	The gray parameter bands in the background are adopted from Fig.~\ref{fig:parameter_bands_PD}. }
%	\label{fig:sigma_term_PD}
%\end{figure}

An alternative to the calculation presented above consists, as often done,  in replacing the derivative with respect to $\hat h$ or $\bar m$ by a derivative with respect to $m_\pi^2$, exploiting the expectably proportional relation between the two quantities. In the present case, this brings a difference though, because the relation is not strictly linear all the way to the physical point, in particular in the doublet model. Indeed the pion mass satisfies the relation (\ref{eq:hatmpi}),  $\hat m_\pi^2 =\hat h/\sigma_{\hat h} $, with $\sigma_{\hat h}$ solution of Eq.~(\ref{eq:sigmahath}).  Thus the relation between $\hat m_\pi^2$ and $\hat h$ may deviate from a linear behavior as one approaches the physical point (see Fig.~\ref{fig:h_dependenceMpi}). Ignoring this non linear correction  one finds the following estimate 
\beq\label{sigma2}
\sigma_N^{(2)}= m_\pi^2 \left.\frac{\rmd M_N}{\rmd \hat m_\pi^2}\right|_{\hat m_\pi^2\to 0}=m_\pi^2 \sigma_\chi \frac{y_+ (\sigma_\chi)}{\hat m_\sigma^2 (\sigma_\chi)}, 
\eeq
which differs from Eq.~(\ref{sigma1}) by the sole substitution of $f_\pi$ by $\sigma_\chi$. Because, as we have seen,  $\sigma_\chi$ differs much from $f_\pi$ in the parity-doublet model, the estimate of (\ref{sigma2}) is much lower than $\sigma_N^{(1)}$, $\sigma_N^{(2)}= 43.1\ \mathrm{MeV}$.

Finally, we consider a fourth estimate, which is relevant to the forthcoming discussion related to the effect of the baryon density. We note that we could read Eq.~(\ref{sigmadef0}) in two ways. Either as the expansion of the nucleon mass around the chiral limit, which we have done in the estimates above. Or as an expansion around the physical point toward the chiral limit, in which case the derivative should be evaluated at the physical point rather than in the chiral limit. Because $M_N(\hat h)$ is not a strictly linear function of $\hat h$ (see Fig.~\ref{fig:h_dependenceM})   the derivative $\rmd M_N/\rmd \bar m$ takes different values depending on where it is evaluated.  By evaluating the derivative at the physical point, one gets 
\beq\label{eq:sigmaterm3}
\sigma_N^{(3)}=h \left. \frac{\rmd M_N}{\rmd \hat h} \right|_{\hat h= h}=m_\pi^2 f_\pi  \frac{y_+ }{m_\sigma^2}=68.6 \:{\rm MeV}, \quad
\eeq
where $y_+$ is evaluated at $\sigma=f_\pi$. 

\begin{table}
	\caption{\label{tab:sigma_term}Summary of estimates of the
	nucleon sigma term.}
	\begin{ruledtabular}
		\begin{tabular}{lcc}
		Estimate & Parity-doublet model & Singlet model\\
		\colrule\\[-0.25cm]
		(0) [MeV] & $55.8$ & $42.1$ \\[0.05cm]
		(1) [MeV] & $50.2$ & $40.6$ \\[0.05cm]
		(2) [MeV] & $43.1$ & $38.8$ \\[0.05cm]
		(3) [MeV] & $68.6$ & $43.7$
		\end{tabular}
	\end{ruledtabular}
\end{table}
The various estimates that we have discussed in this section are summarized in Table~\ref{tab:sigma_term}. The spread in the values obtained with the parity-doublet model is larger than with the singlet model. This, as we have argued at several places, is due to the enhanced non linearities of the relations connecting the chiral limit to the physical point within the parity-doublet model. 

\subsection{Density dependence of $\sigma_N$}

\begin{figure}[t]
	\centering
	\includegraphics[scale=1.0]{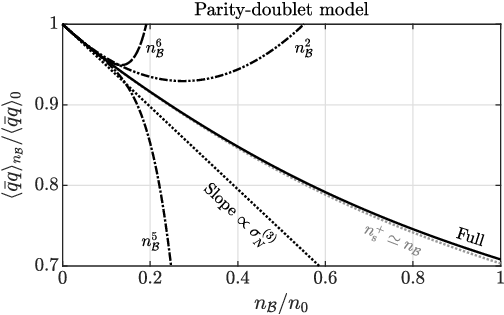}
	\caption{Density dependence of the chiral condensate in the parity-doublet model.
	The full solution to the gap equation (solid line) is successively approximated for small 
	baryon density $n_{\mathcal{B}}$ by a linear function proportional 
	to $\sigma_{N}^{(3)}$ (black dotted line) and higher Taylor 
	polynomials of $\mathcal{O}\!\left(n_{\mathcal{B}}^{2}\right)$ (dash-double-dotted line), 
	$\mathcal{O}\!\left(n_{\mathcal{B}}^{5}\right)$ (dash-dotted line), and 
	$\mathcal{O}\!\left(n_{\mathcal{B}}^{6}\right)$ (dashed line). The gray dotted line 
	indicates the difference to the full solution when approximating $n_{\rm s}^{+}$ by 
    $n_{\mathcal{B}}$, which can be seen to be an excellent approximation in this range of densities. }
	\label{fig:density_dependence_PD}
\end{figure}

\begin{figure}[t]
	\centering
	\includegraphics[scale=1.0]{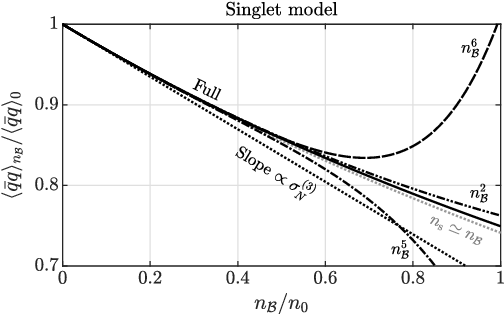}
	\caption{Density dependence of the chiral condensate in the singlet model.
	Analogous plot to Fig.~\ref{fig:density_dependence_PD}.}
	\label{fig:density_dependence_WT}
\end{figure}

The nucleon sigma term provides a measure of the scalar density (or quark condensate) inside a nucleon (as compared to the vacuum). It may also be used, more broadly, to estimate   how the presence of baryonic matter modifies the quark condensate.
This is most easily seen by using the Feynman-Hellmann theorem, which yields
\beq\label{eq:kinetic}
\bar m\braket{\bar{q} q}_{n_\mathcal{B}}=\bar m\braket{\bar{q}q}_0+\bar m\frac{\rmd }{\rmd \bar m} \left[{\cal E}(n_{\mathcal{B}}) - {\cal E}(0)\right], \quad
\eeq
where $\braket{\bar{q}q}_{0}$ denotes the condensate in vacuum.
As a first orientation, and repeating a standard argument \cite{Cohen:1991nk},  we consider a low-density gas of independent nucleons, for which the energy density is given by
\begin{equation}
	{\cal E}(n_{\mathcal{B}}) - {\cal E}(0) = 4 \int_{|p|\, <\, p_{F}} 
	\frac{\mathrm{d}^{3}p}{(2\pi)^{3}}\, \sqrt{p^{2} + M_{N}^2}
\end{equation}
and $p_{F} = (3\pi^{2} n_{\mathcal{B}}/2)^{1/3}$. By differentiating this expression with respect to $\bar m$ at fixed baryon density, one gets
\beq\label{eq:GORrelation}
\bar m\braket{\bar{q} q}_{n_\mathcal{B}}=\bar m \braket{\bar{q}q}_0+n_{\rm s} \sigma_N,
\eeq
where $n_{\rm s}$ is the average scalar density  and the factor $\sigma_N$ originates from the derivative of the nucleon mass with respect to $\bar m$. On the other hand, by applying the Feynman-Hellmann theorem to the vacuum state, and using for the vacuum the present model, one gets
\beq\label{eq:GORrelation2}
\bar m \langle \bar{ q} q\rangle_0= -h\langle \sigma\rangle_0,
\eeq
where the right-hand side follows from ${\cal E}(0)=U=V(\sigma)-h\sigma$. This equation allows us to identify the quark condensate with the expectation value of the scalar field. 

By using in Eq.~(\ref{eq:GORrelation2})  the physical value of $h$, $h=m_\pi^2 f_\pi$, and the vacuum expectation value of the $\sigma$ field $\langle \sigma\rangle_0=f_\pi$, one gets the Gell-Mann Oakes Renner relation\footnote{Note that the Feynman-Hellmann theorem holds for any value of  $\bar m$, so that this is, strictly speaking, a generalisation of the GOR relation, whose original derivation from current algebra invokes the chiral limit.}
\beq
\bar m \langle \bar q q\rangle_0=-m_\pi^2 f_\pi^2.
\eeq
By combining these results, it follows that 
\begin{equation}
	\frac{\langle\sigma\rangle_{n_{\mathcal{B}}}}{\langle\sigma\rangle_0}=\frac{\braket{\bar{q}q}_{n_{\mathcal{B}}}}{\braket{\bar{q}q}_{0}} \simeq 
	1 - \frac{\sigma_{N}^{(3)} n_{\rm s}}{f_\pi^{2} m_{\pi}^{2}} ,
	\label{eq:condensate_density}
\end{equation}
% This model-independent 
%formula derives from the quark mass dependence of the energy density $\epsilon$, that is,
%\begin{equation}
%	\frac{\braket{\bar{q}q}(n_{\mathcal{B}})}{\braket{\bar{q}q}_{0}} =  
%	1 + \frac{1}{\braket{\bar{q}q}_{0}} 
%	\left.\frac{\mathrm{d}}{\mathrm{d}m_{q}}\right|_{m_{q}\, \rightarrow\, 0}
%	\left[\epsilon(n_{\mathcal{B}}) - \epsilon(0)\right] ,
%\end{equation}
%where 
%\begin{equation}
%	\epsilon(n_{\mathcal{B}}) - \epsilon(0) = 4 \int_{|p|\, <\, p_{F}} 
%	\frac{\mathrm{d}^{3}p}{(2\pi)^{3}}\, \sqrt{p^{2} + M_{N}}
%\end{equation}
%and $p_{F} = (3\pi^{2} n_{\mathcal{B}}/2)^{1/3}$.
%, the leading term in the low-density expansion of the scalar density,
%\begin{equation}
%	n_{\rm s} = n_{\mathcal{B}} \left(1 - \frac{3 p_{F}^{2}}{10 M_{N}^{2}}
%	+ \frac{9 p_{F}^{4}}{56 M_{N}^{4}} 
%	- \frac{5 p_{F}^{6}}{48 M_{N}^{6}} + \cdots \right) .
%	\label{eq:ns_expansion}
%\end{equation}
%This expansion however does not exhibit good convergence properties.
where, for small enough baryon density,  we can set $n_{\rm s}\simeq n_{\mathcal{B}}$. 
What the relation (\ref{eq:condensate_density}) then says is that,  in the low baryon density regime, each additional nucleon occupies a region initially filled with vacuum. Since the quark condensate is lower in the nucleon than in the vacuum, the presence of the new nucleon decreases the average value of the quark condensate, or equivalently, of the average $\sigma$ field.  
The formula (\ref{eq:condensate_density}) predicts a linear decrease of the quark condensate with increasing baryon density. It  suggests a  reduction of the condensate in normal nuclear matter by about 1/3 of its vacuum value, as well as a restoration of chiral symmetry at about three times nuclear matter density. However this linear estimate neglects the effects of the interaction, which we now consider.

In order to make contact with the previous literature 
\cite{Birse:1992uxe, Birse:1993hx, Kaiser:2007nv}, we generalise the formula (\ref{eq:condensate_density}) as follows 
\begin{equation}\label{eq:sigmabarN}
	\frac{\braket{\bar{q} q}_{n_\mathcal{B}}}{\braket{\bar{q}q}_{0}} \simeq
	1 - \frac{\bar{\sigma}_{N}(n_{\mathcal{B}})\;\! n_{\mathcal{B}}}{f_{\pi}^{2} m_{\pi}^{2}},
\end{equation}
where we have set $n_{\rm s}\approx n_{\mathcal{B}}$, an approximation which remains valid for a baryon density up to normal nuclear matter density. The effect of interactions is thus considered as a 
(model-dependent)  modification of the nucleon sigma term, $\sigma_{N} \mapsto
\bar{\sigma}_{N}(n_{\mathcal{B}})$. 

To take the interactions into account, we use the complete expression of the energy density (see Eq.~(\ref{eq:OmegacalEt0})), and determine $\sigma$ by solving the gap equation 
\beq
\frac{\rmd V(\sigma)}{\rmd \sigma}=h-y_{+} n_{\rm s}^{+}\simeq h-y_{+} n_\mathcal{B}. 
\eeq
The last term, in which we have approximated the scalar density by the baryon density, represents the matter contribution. We see that this contribution opposes that of the explicit symmetry breaking term $h$. It tends to drive the system to a chirally symmetric state, that is, it leads to a reduction of the value of $\sigma$. In order to solve the gap equation for small baryon densities, we assume the following expansion for the solution
\beq
\sigma(n_{\mathcal{B}})=\sigma^{(0)}+\sigma^{(1)} n_{\mathcal{B}}+\frac{1}{2} \sigma^{(2)} n_{\mathcal{B}}^2+\cdots
\eeq
with $\sigma^{(0)}=f_\pi$, and we expand $y_{+}$ and $\rmd U/\rmd\sigma$. The coefficients $\sigma^{(1)}, \sigma^{(2)},\ldots$ are determined by solving the gap equation to the required order. One obtains then, e.g.\ to order $n_{\mathcal{B}}^{2}$,
%\begin{equation}
%	\left.\frac{\partial U}{\partial\sigma}
%	\right|_{\sigma(n_{\mathcal{B}})} + y_{+}(\sigma(n_{\mathcal{B}})) 
%	\;\! n_{\mathcal{B}} = 0, \label{eq:gap_eq_simplified}
%\end{equation}
%where $n_{\rm s}^{-} = 0$ in the low-density approximation. At next-to-leading order, which 
%corresponds to first corrections of $\mathcal{O}(n_{\mathcal{B}}^{2})$ to the linear density 
%approximation (\ref{eq:condensate_density}), we find
\begin{IEEEeqnarray}{rCl}\label{sigmanB}
	{\sigma(n_{\mathcal{B}})} & = & {f_{\pi}}
	- \frac{y_{+}(f_{\pi})}{m_{\sigma}^{2}} \;\! n_{\mathcal{B}}
	+y_{+}(f_{\pi}) \;\! \Bigg\lbrace \frac{1}{m_{\sigma}^{4}} 
	\left.\frac{\mathrm{d} y_{+}}{\mathrm{d}\sigma}\right|_{f_{\pi}}
	\nonumber\\[0.2cm] & & \qquad
	-\, \frac{y_{+}(f_{\pi})}{2 m_{\sigma}^{6}} 
	\left. \frac{\partial^{3} U}{\partial\sigma^{3}}\right|_{f_{\pi}} 
	\Bigg\rbrace \; n_{\mathcal{B}}^{2} .
\end{IEEEeqnarray}
The density-dependent nucleon sigma term, as defined in Eq.~(\ref{eq:sigmabarN}), is then given by
\begin{IEEEeqnarray}{rCl}
	\bar{\sigma}_{N}(n_{\mathcal{B}}) & = & \sigma_{N}^{(3)}
	- \left.\frac{\mathrm{d} y_{+}}{\mathrm{d}\sigma}\right|_{f_{\pi}}
	\frac{\big(\sigma_{N}^{(3)}\big)^{2} n_{\mathcal{B}}}{m_{\pi}^{2} f_{\pi} y_{+}(f_{\pi})} \nonumber\\[0.2cm]
	& & +\, \frac{\big(\sigma_{N}^{(3)}\big)^{3} n_{\mathcal{B}}}{2 m_{\pi}^{4} f_{\pi}^{2} y_{+}(f_{\pi})} 
	\left. \frac{\partial^{3} U}{\partial\sigma^{3}}\right|_{f_{\pi}}.
	\label{eq:density_modification}
\end{IEEEeqnarray}
We have recognized in the first correction the expression (\ref{eq:sigmaterm3}) of the sigma term. The first term of $\mathcal{O}(n_{\mathcal{B}})$ in Eq.~(\ref{eq:density_modification}) reduces 
the nucleon sigma term, as $y_{+}(f_{\pi})$ as well as its first derivative are positive, while 
the second term enhances or reduces $\sigma_{N}$ (depending on the sign of the derivative of $U$). The  effect of these corrections, together with higher order ones,  is illustrated in Fig.~\ref{fig:density_dependence_PD}. As suggested by the plots in this figure, the systematic expansion in powers of the density does not converge well. The full solution of the gap equation, obtained numerically,  has a smooth behavior and yields a reduction of about $30\%$  of the sigma field at nuclear matter density, in agreement with the value of $\sigma_0$ quoted in Table~\ref{tab:parameter_choice_PD}.

It is straightforward to obtain the corresponding corrections for the singlet model, 
in which case the first contribution in Eq.~(\ref{eq:density_modification}) vanishes since
the Yukawa coupling carries no $\sigma$ dependence. As the potential $U$
contains high powers of the $\sigma$ field, the expression (\ref{eq:density_modification}) may be seen as the generalization of 
the enhancement of $\sigma_{N}$ discussed in Ref.~\cite{Birse:1992uxe} for a bosonic 
potential of quartic order in $\sigma$. The results of the corresponding analysis are plotted in Fig.~\ref{fig:density_dependence_WT}.

\begin{figure}[t]
	\centering
	\includegraphics[scale=1]{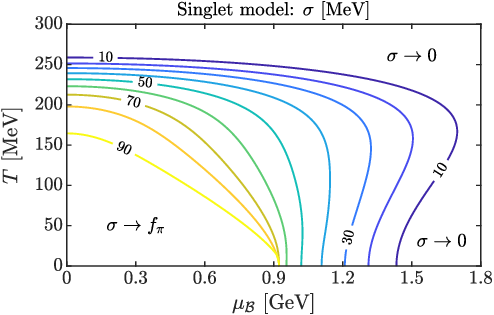}
	\caption{Phase diagram of the singlet model: Isoscalar condensate $\sigma$ as function of 
	temperature $T$ and baryon-chemical potential $\mu_{\mathcal{B}}$. Contours of constant 
	$\sigma$ are given every $10\ \mathrm{MeV}$.}
	\label{fig:phase_diagram_WT}
\end{figure}

\begin{figure}[t]
	\centering
	\includegraphics[scale=1.0]{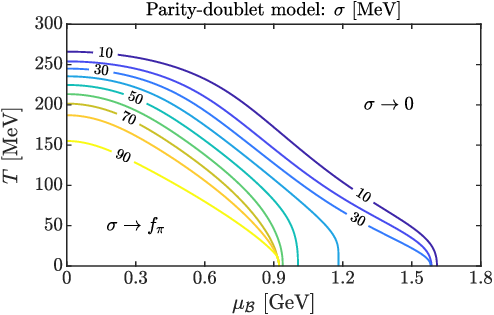}
	\caption{Phase diagram of the doublet model: Isoscalar condensate $\sigma$ as function of 
	temperature $T$ and baryon-chemical potential $\mu_{\mathcal{B}}$. Contours of constant 
	$\sigma$ are given every $10\ \mathrm{MeV}$.}
	\label{fig:phase_diagram_PD}
\end{figure}

\section{Chiral symmetry restoration}\label{sec:chiraltrans}

The analysis of the $\sigma$ term in the previous section reveals that non linear effects play an increasingly important role as the density increases, and that an expansion of the value of the $\sigma$ field in powers of the density has a limited range of validity. In this section,  we turn to a more thorough study of the dependence of the sigma field on the baryon density, and we extend our analysis to finite temperature.

We start by considering the general behavior of the scalar field in the presence of matter at finite temperature and baryon density. Figures \ref{fig:phase_diagram_WT} and \ref{fig:phase_diagram_PD} display the contour plots of the magnitude of the sigma field in the $T$-$\mu_{\mathcal{B}}$-plane, for the singlet and doublet models. There are similarities and differences between the two models that are clearly visible on these plots. The first similarity concerns the regime of low density-low temperature, where the two models exhibit remarkably similar behaviors: this is the regime of nuclear matter, with the well identified first-order liquid-gas transition at low temperature. That the two models behave in the same way in this regime should not come as a surprise since their respective parameters are precisely adjusted to reproduce nuclear matter properties, which they do as we have seen in Sec.~\ref{sec:results}. 

A further similarity is visible as one moves up along the temperature axis at low baryon density. In this case, the dominant degrees of freedom are nucleons and anti\-nucleons, the parity partners becoming to be significantly populated only at larger temperatures.  As we shall see, at vanishing baryon density, the chiral transition is a second-order one in both models in the chiral limit, and as suggested by the contours in these figures, it occurs within the same temperature range (in fact, at nearly the same temperature). These second-order transitions are smeared out to smooth crossovers for physical pion mass.

Things are different at zero temperature. Indeed, Fig.~\ref{fig:phase_diagram_PD} suggests a first-order chiral transition in the parity-doublet model, and this is indeed so. In this case, the structure of the parity-doublet model plays an important role, as we shall discuss in detail in the later part of this section. In the singlet model the chiral transition at $T=0$ is a second-order transition in the chiral limit and a mere crossover for finite pion mass. 

\begin{figure}[h]
	\centering
	\includegraphics[scale=1.0]{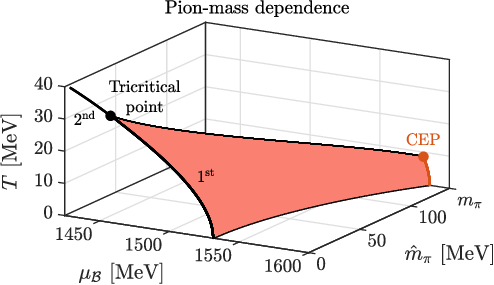}
	\caption{First-order chiral transition in the parity-doublet model, as a function of $\mu_{\mathcal{B}}$ 
	and $\hat m_{\pi}$. The pion mass $\hat m_\pi$ varies continuously between the chiral limit ($\hat m_{\pi} = 0$) 
	and the physical value $m_{\pi} = 138\ \mathrm{MeV}$. In the chiral limit, 
	the first-order transition line (black) ends in a tricritical point (black dot) at 
	$T_{\mathrm{tri}} \simeq 33\ \mathrm{MeV}$ and $\mu_{\mathrm{tri}} \simeq 1460\ \mathrm{MeV}$.
	Above $T_{\mathrm{tri}}$, the transition line is of second order. Values of $(T,\mu_{\mathcal{B}})$ on the red surface correspond to a first-order transition. The first-order transition line (red) for physical pion mass terminates in the critical end point (``CEP'') at
$(T_c, \mu_c) \simeq (8.5\ \mathrm{MeV}, 1580\ \mathrm{MeV})$.}
	\label{fig:pion_dependence}
\end{figure}

A summary of the phase diagram of the parity-doublet model in the $T$-$\mu_{\mathcal{B}}$-plane, featuring the chiral transition, is given in Fig.~\ref{fig:pion_dependence}. There, we also indicate the dependence on the pion mass, which varies from $\hat m_\pi=0$ (chiral limit) to its physical value $\hat m_\pi=m_\pi$. In the chiral limit, the chiral transition is first order at small temperature, and turns into a second-order transition at the critical point as the temperature increases. The critical point depends on the pion mass, and decreases towards the physical point. Its location as a function of $\hat m_\pi $ delineates the region of first-order transition (indicated by the red surface in Fig.~\ref{fig:pion_dependence}), and terminates in the critical end point denoted ``CEP'' in Fig.~\ref{fig:pion_dependence} when $\hat m_\pi=m_\pi$. 

\begin{figure}[t]
	\centering
	\includegraphics[scale=1.0]{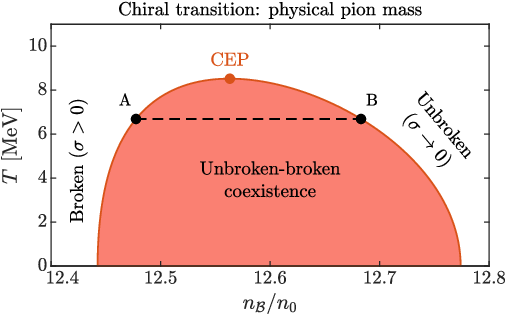}
	\caption{Phase coexistence in the $T$-$n_{\mathcal{B}}$-plane (for physical pion mass), where the density is measured in unit of the nuclear matter density $n_0$. The points $\mathrm{A}$ and $\mathrm{B}$ represent 
	the two coexisting phases with identical 
	pressure and baryon-chemical potential at a given temperature. In the low-density phase ($\mathrm{A}$), chiral symmetry is still
	broken ($\sigma > 0$), while in the high-density phase ($\mathrm{B}$) chiral symmetry is restored ($\sigma$ approaches zero). With increasing temperature, the critical chemical potential decreases (see Fig.~\ref{fig:pion_dependence}) and the coexistence region terminates at a critical end point (``CEP'') characterized by $(T_{c},\, n_{c}) \simeq 
	(8.5\ \mathrm{MeV},\, 2.01\ \mathrm{fm}^{-3})$. }
	\label{fig:chiral_transition1}
\end{figure}

\begin{figure}[h]
	\centering
	\includegraphics[scale=1.0]{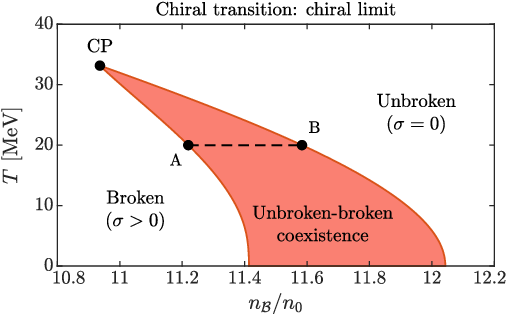}
	\caption{Phase coexistence in the $T$-$n_{\mathcal{B}}$-plane (in the chiral limit). Analogous plot to Fig.~\ref{fig:chiral_transition1}. The black dot denoted ``CP'' corresponds to the tricritical point in Fig.~\ref{fig:pion_dependence}.}
	\label{fig:chiral_transition1_chiral_limit}
\end{figure}

Figures~\ref{fig:chiral_transition1} and \ref{fig:chiral_transition1_chiral_limit} display respectively the coexistence region of the chiral transition in the $T$-$n_{\mathcal{B}}$-plane for physical pion mass as well as in the chiral limit.
At a temperature below  the temperature $T_c$ of the critical end point,  two different phases of matter with different densities coexist. These two phases, labelled $\mathrm{A}$ and $\mathrm{B}$,  have the  same thermodynamic pressure and
baryon-chemical potential. While the density of phase A increases with increasing $T$ in the case of the physical pion mass, it decreases in the chiral limit, whereas the density of phase B decreases in both cases.  The composition of the two phases A and B with respect to nucleons and their chiral partners will be further discussed at the end of this section, both for physical pion mass as well as in the chiral limit.

In the rest of this section, we analyze further the transition in the two cases of zero temperature or zero chemical potential. 

\subsection{The chiral transition in the singlet model}

Since only the positive-parity baryons (the nucleons) are involved in the singlet model, we omit the subscript + on all quantities ($M_+\mapsto M, y_+\mapsto y$, etc). 

\subsubsection{The transition at $T=0$}

\begin{figure}[h]
	\centering
	\includegraphics[scale=1.0]{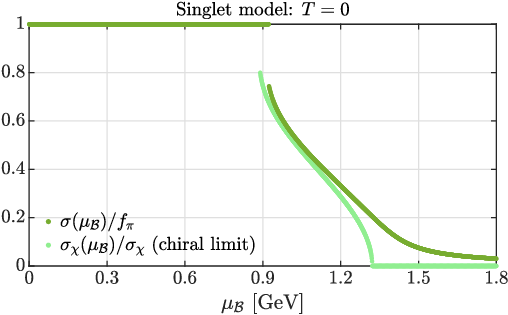}
	\caption{The average $\sigma$ field at zero temperature as a function of the baryon-chemical 
	potential $\mu_{\mathcal{B}}$ in the singlet model (dark green: physical pion mass; light green:
	chiral limit).}
	\label{fig:phase_diagram_chempot_WT}
\end{figure}

The variation of the $\sigma$ field as a function of the baryon-chemical potential is displayed in Fig.~\ref{fig:phase_diagram_chempot_WT}. The first-order liquid-gas transition is clearly visible for a value of the chemical potential of order $0.9\ \mathrm{GeV}$. We are concerned here with the chiral transition that takes place for a larger chemical potential. As suggested by the plots, this transition is in fact a simple crossover for the physical pion mass. However, in the chiral limit, it becomes a continuous second-order  transition, as we shall verify in this section.

The value of the sigma field   is obtained as a function  of  the baryon density by solving the gap equation (\ref{eq:gap}). For the singlet model this equation reads simply
\beq
\frac{\rmd U}{\rmd\sigma}= -y n_{\rm s}, 
\eeq
where the zero-temperature scalar density is given by ($p_F$ is the Fermi momentum)
\beq\label{eq:nsint}
n_{\rm s}=4M\int_{|\p| \,\le\, p_F} \frac{1}{\sqrt{\p^2+M^2}}.
\eeq
The integral can be calculated analytically as a function of $x\equiv M/p_F$, which yields
\beq\label{eq:nsanal}
\frac{n_{\rm s}}{n_{\mathcal{B}}}= \frac{3x}{2} \left[\sqrt{1+x^2}-x^2 \sinh^{-1}\left(\frac{1}{x}\right) \right],
   \eeq
 a monotonously increasing function of $x$, going from $0$ to $1$ as $x$ runs from $0$ to $\infty$.
  %\begin{figure}[h]
%	\centering	\includegraphics[scale=1.0]{ns_nb}
%	\caption{Plot of $ns/n_{\mathcal{B}}$ as a function of $M/p_F$. \jpb{This plot should be continued all the way to zero.}}
%	\label{fig:nsplotT0}
%\end{figure}
This function  has two simple limits. 
The first one is that of low density or large-mass limit, $M\gg p_F$ or $x\to \infty$.  This limit is a non relativistic limit. It  can be obtained simply by expanding the denominator of the integrand in Eq.~(\ref{eq:nsint})  in powers of $p^2/M^{2}$. The leading order gives $
n_{\rm s}\approx n_{\mathcal{B}}$, independent of $\sigma$. By including the first correction, we get 
\beq
\frac{n_{\rm s}}{n_{\mathcal{B}}}\simeq 1-\frac{3}{10 x^2}+\mathcal{O}\left(\frac{1}{x^4}\right),\qquad(x\to \infty). \quad
\eeq

The second  limit corresponds to the small mass or large-density limit, relevant for the chiral transition. We have, for $x\to 0$, 
   \beq\label{nssmallx}
  \frac{n_{\rm s}}{n_{\mathcal{B}}}\simeq\frac{3 x}{2}+\frac{3}{4} x^3 \left[2 \log (x/2)+1\right]+ \mathcal{O}\left(x^4\right).   
  \eeq
Again the limiting behavior can be easily obtained by noticing that, when $p_F\gg M$, the mass in the denominator in Eq.~(\ref{eq:nsint}) can be ignored. One then gets $n_{\rm s}\approx Mp_F^2/\pi^2$, in agreement with the formula (\ref{nssmallx}) above. 
The logarithmic correction in Eq.~(\ref{nssmallx}) originates from a potential infrared logarithmic divergence, as $M\to 0$, of the integral involved in the derivative of $n_{\rm s}$ with respect to $M$. The same logarithmic contribution arises in the expansion of the kinetic contribution to the energy density, which reads (with $z\equiv p_F/M$)
\beq\label{eq:energydenskin}
&& \!\!\!\!\!\!\!\!\!\!\!\!\!\!\!\!\!\!\!\!4  \int_{|\p|\, \le\, p_{F}} 
	 \sqrt{\p^{2} + M^2}\qquad\qquad\nn[0.2cm] \qquad&=&\frac{M^4}{4\pi^2} \left[z \sqrt{z^2+1} \left(2 z^2+1\right)-\sinh ^{-1}z\right]  \nn[0.2cm]
	&&\!\!\!\!\!\simeq \frac{M^4}{2\pi^2} \left\lbrace z^4+z^2+\frac{1}{8} \left[1 -4 \log (2z)\right]\right\rbrace,    \quad
   \eeq
where the last line provides the large-$z$ expansion up to the logarithmic correction. We shall verify shortly that the logarithmic contributions $\propto \sigma^4 \ln \sigma$ cancel out when solving the gap equation.

In order to study the vicinity of the  chiral transition we assume that, in the chiral limit,  the potential $U(\sigma)$ has  the following form near $\sigma=0$:
\beq\label{UBsigapprox}
U(\sigma)\simeq U_0-\frac{r}{2}\sigma^2+\frac{u}{4} \sigma^4-l\sigma^4\ln\frac{\sigma}{f_\pi},
\eeq
where $U_0$ is a constant and $l$ is obtained from Eq.~(\ref{Ubosons}), $l=y^4/(4\pi^2)$. The value of $r$  can also be obtained from Eq.~(\ref{Ubosons}), $r = (1452\ \mathrm{MeV})^{2}$. However, the presence of the logarithm makes the fourth derivative ill defined. The value of $u$ is then obtained from a numerical fit. One gets $u \approx 140$.

It follows  from Eq.~(\ref{UBsigapprox}) that 
\beq\label{UBsigapproxder}
\frac{1}{\sigma}\frac{\rmd U}{\rmd \sigma}\simeq -r+(u-l)\sigma^2-\frac{y^4\sigma^2}{\pi^2} \ln\frac{\sigma}{f_\pi}.
\eeq
On the other hand, by using the expansion (\ref{nssmallx}) above for $n_{\rm s}$ one gets
\beq\label{eq:gappFsinglet}
\frac{y n_{\rm s}}{\sigma} 
	\simeq   \frac{y^{2}}{\pi^{2}} \left(p_{F}^{2} + \frac{y^{2} \sigma^{2}}{2}
	- y^{2}\sigma^{2} \ln\frac{2p_{F}}{y \sigma} \right).  
	\eeq
It follows that for small $\sigma$, and leaving aside the trivial solution corresponding to the maximum of $U$, one can write the gap equation in the following  form
\beq\label{eq:gappFsinglet1}
-r+\frac{y^{2}p_{F}^{2}}{\pi^{2}}+\left(u+\frac{y^4}{4\pi^2}-\frac{y^4}{\pi^2} \ln\frac{2p_F}{yf_\pi}\right)\sigma^2= 0. \qquad
	\eeq
	Note that the logarithmic terms proportional to $  \sigma^2\log\sigma$ have cancelled, as anticipated, leaving their trace in a simple renormalisation of the parameter $u$. 
		
\begin{figure}[t]
	\centering
	\includegraphics[scale=1.0]{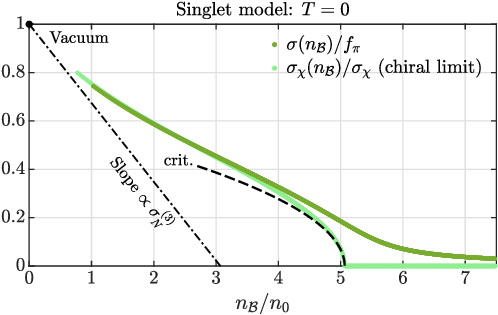}
	\caption{Condensate $\sigma$ at zero temperature as a function of the baryon density 
	$n_{\mathcal{B}}$ in the singlet model. Results for physical pion mass are given in dark green, 
	those of the chiral limit in light green. The black dashed line (``crit.'') represents the analytic solution (\ref{eq:sigmanc}) valid near $n_c$. The dash-dotted line shows the linear
	density approximation proportional to the sigma term $\sigma_{N}^{(3)}$. See also 
	Fig.~\ref{fig:phase_diagram_density_PD}.}
	\label{fig:phase_diagram_density_WT}
\end{figure}

The critical density is obtained from the solution of the gap equation (\ref{eq:gappFsinglet1}) corresponding to  $\sigma =0$. This yields the critical value  of the Fermi momentum $p_c^2=\pi^2 r/y^2$, corresponding to a critical density 
\beq
 n_{c} = \frac{2\pi}{3} \left(\frac{r}{y^{2}}\right)^{3/2} \approx  5.06 n_{0}.
\eeq
A simple calculation shows that in the vicinity of $n_c$, $\sigma$ behaves as a function of $n_c-n$ as 
\begin{equation}\label{eq:sigmanc}
    \sigma \simeq \left(\frac{y^{6}}{\pi^{2}r\tilde{u}^{2}}\right)^{1/4} \sqrt{n_{c} - n},
\end{equation}
with 
\beq
 \tilde{u} = u + \frac{y^{4}}{4\pi^{2}} \left(1 - 2 \ln \frac{4\pi^{2}r}{y^{4}f_{\pi}^{2}}\right).
 \eeq
The square root behavior is characteristic of mean-field theory, and the formula above reproduces accurately the behavior of $\sigma$ as can be seen in Fig.~\ref{fig:phase_diagram_density_WT}.

Note that the critical density $n_c$ depends only on the value of $r$, which itself depends crucially on the renormalisation of the coefficients of $(\sigma^2-f_\pi^2)$ and $(\sigma^2-f_\pi^2)^2$, as well as on the Taylor coefficients $\alpha_3$ and $\alpha_4$. These quantities have been adjusted in order to fit nuclear matter properties. This is an illustration of the strong correlation that exists in this model between the nuclear matter properties near its ground state, and the chiral transition.

The graphical solution of the gap equation provides insight on how the phase transition proceeds (see Fig.~\ref{fig:gap_equation_singlet_1}). The graphical solution indicates that as $n_{\mathcal{B}}$ increases starting from values of the order of $n_0$, the intersection point moves to smaller and smaller values of $\sigma$, while the corresponding scalar density increases. The  scalar density reaches a maximum value when the intersection point coincides with the point where $\rmd^2 U/\rmd \sigma^2=0$, as we shall verify shortly. This occurs for $n\approx 4n_0$. From that point on,   the scalar density rapidly decreases with further increase of the baryon density and eventually vanishes at the transition. For small values of $\sigma$, $n_{\rm s}(\sigma) $ is linear in $\sigma$ (see Eq.~(\ref{nssmallx})), and  the transition takes place when the corresponding slope matches that of $\rmd U/\rmd\sigma$ at $\sigma=0$, that is when $y\partial n_{\rm s}/\partial\sigma\simeq y^2p_F^2/\pi^2=r$, which is the critical value of the Fermi momentum determined above. Note that, at the transition, $m_\sigma^2\equiv\rmd^2 U/\rmd\sigma^2+y\partial n_{\rm s}/\partial\sigma |_{n_{\mathcal{B}}}=0$.\footnote{This definition of the sigma mass follows directly from the second derivative of the energy density with respect to $\sigma$, at fixed baryon density. In the present context it is consistent with the general considerations of Ref.~\cite{haensch_medium_2023} about the chiral critical mode in this class of sigma models.}

%\begin{figure}[h]
%	\centering
%	\includegraphics[scale=0.75]{gap_equation_singlet_3}
%	\caption{Graphical solution of the gap equation. The blue curve represents $\rmd U/\rmd \sigma$ as a function of $\sigma$. Note that this curve is independent of $n_{\mathcal{B}}$.  The orange curves give the scalar density as a function of $\sigma$ for given values of  $n_{\mathcal{B}}$. The intersections points determine the value of $\sigma$ solution of the gap equation as a function of $n_{\mathcal{B}}$.  \jpb{We should draw several curves in order to illustrate the solution as a function of $n_{\mathcal{B}}$.} \question{This figure is not needed.}}
%	\label{fig:gap_equation_singlet_3}
%\end{figure}

To determine the maximum of $n_{\rm s}$ as a function of $n_{\mathcal{B}}$, we note that 
\beq\label{eq:nsmax1}
\frac{\rmd n_{\rm s}}{\rmd n_{\mathcal{B}}}=\left.\frac{\del n_{\rm s}}{\del n_{\mathcal{B}}}\right|_\sigma+\left.\frac{\del n_{\rm s}}{\del \sigma}\right|_{n_{\mathcal{B}}} \frac{\rmd \sigma}{\rmd n_{\mathcal{B}}},
\eeq
so that the requirement that  $n_{\rm s}$ be a maximum as a function of $n_{\mathcal{B}}$ yields the relation
\beq\label{eq:nsmax2}
\left.\frac{\del n_{\rm s}}{\del n_{\mathcal{B}}}\right|_\sigma=-\left.\frac{\del n_{\rm s}}{\del \sigma}\right|_{n_{\mathcal{B}}} \frac{\rmd \sigma}{\rmd n_{\mathcal{B}}}.
\eeq
By taking the derivative of the solution of the gap equation  with respect to $n_{\mathcal{B}}$ (cf.\ again Eq.~(\ref{eq:disgmadn0})), one obtains 
\beq\label{eq:nsmax3}
\frac{\rmd\sigma}{\rmd n_{\mathcal{B}}}=-\frac{y}{m_\sigma^2}\left.\frac{\del n_{\rm s}}{\del n_{\mathcal{B}}}\right|_\sigma ,
\eeq
which can be used to simplify the condition (\ref{eq:nsmax2}) above into
\beq\label{eq:nsmax4}
m_\sigma^2=y\left.\frac{\del n_{\rm s}}{\del \sigma}\right|_{n_{\mathcal{B}}}.
\eeq
On the other hand, 
\beq\label{eq:nsms2}
m_\sigma^2=\frac{\rmd^2 U}{\rmd \sigma^2}+y\left.\frac{\del n_{\rm s}}{\del \sigma}\right|_{n_{\mathcal{B}}}.
\eeq
By comparing the two expressions above for $m_\sigma^2$, one concludes that  $n_{\rm s}$ reaches its maximum value when the second derivative of $U$ vanishes.

The square of the $\sigma$ mass, Eq.~(\ref{eq:nsms2}), 
informs us on the stability of the solution. In fact it is easy to verify that the solution is always stable. For $n<n_c$, $m_\sigma^2$ is positive, and it vanishes at the transition, as we have seen. Above the transition, $\sigma$ remains equal to $0$, and the dependence of $m_\sigma^2$ on the density is dictated by the corresponding dependence of $\partial n_{\rm s}/\partial\sigma=yp_F^2/\pi^2$ on the density.  It follows that, above  the transition, we have 
\beq
m_\sigma^2 =m_\pi^2\simeq\frac{y^2}{\pi^2}\left(p_F^2-p_c^2\right), 
\eeq
where the equality $m_\sigma^2 =m_\pi^2$ reflects the fact that chiral symmetry is restored ($\sigma=0$). Above the transition, these degenerate masses increase with increasing density.

So far, we have examined the transition in the chiral limit. It is interesting to consider also what happens for the physical value of the pion mass, that is when the linear term $h\sigma$ is present in the potential. This term not only affects the behavior of the potential near $\sigma=f_\pi$ but also near $\sigma=0$. With $h$ non vanishing, the potential (\ref{UBsigapprox}) reads (ignoring the logarithmic contribution)
\beq\label{UBsigapprox2}
U(\sigma)\simeq U_0-\frac{r}{2}\sigma^2+\frac{u}{4} \sigma^4- h\sigma.
\eeq
The presence of the term $h\sigma$ entails a modification of the gap equation in the vicinity of $\sigma=0$ which is easily understood with the graphical solution of Fig.~\ref{fig:gap_equation_singlet_1}. Since $\rmd U/\rmd\sigma|_{\sigma=0}=-h$,  the corresponding green curve in Fig.~\ref{fig:gap_equation_singlet_1} is shifted downwards by $h$. It follows that  when the density reaches the value  corresponding to the critical density for $h=0$, that is when $r=y^2 p_F^2/\pi^2$, the intersection of the (shifted) green line with a black line occurs at finite $\sigma$, rather than at $\sigma=0$. Then, as the density increases beyond that point, $\sigma$ continues to decrease with increasing density. This behavior can also be deduced from the explicit form of the gap equation, which for  the potential (\ref{UBsigapprox2}), is given by 
\beq
\left(-r+\frac{y^2 p_F^2}{\pi^2}\right)\sigma+u\sigma^3= h.
\eeq
This shows  in particular that the value $\sigma=0$ is reached only asymptotically, i.e.\ when $p_F\to\infty$.  One can also use this equation to verify that  $m_\sigma^2$  never vanishes.

As an alternative to solving the gap equation, one may obtain $\sigma$ as a function of $n_{\mathcal{B}}$ by solving a simple differential equation. Consider indeed  Eq.~(\ref{eq:nsmax3}). After noticing that  $\del n_{\rm s}/\del n_{\mathcal{B}}|_\sigma=M/M_*$, we may rewrite this equation as follows
\beq\label{eq:nsmax3b}
\frac{\rmd\sigma}{\rmd n_{\mathcal{B}}}=-\frac{y}{m_\sigma^2} \frac{M}{M_*} .
\eeq
The right-hand side of this equation is a known function of $\sigma$ and $n_{\mathcal{B}}$. It can then be integrated from some initial condition all the way to the chiral transition, and beyond. Since this is a first-order differential equation, the solution is determined by the initial condition, i.e.\ by $\sigma(n_{\mathcal{B}}=0)$. This depends on the value of $\hat h$:   $\sigma(n_{\mathcal{B}}=0)=\sigma_\chi$ for $\hat h=0$, and   $\sigma(n_{\mathcal{B}}=0)=f_\pi$ for $\hat h=h$. Note that the integration does not depend on $\hat h$: the only place where $\hat h$ enters is the potential $U$ and only its second derivative enters the expression of $m_\sigma^2$ in Eq.~(\ref{eq:nsms2}). Thus $\hat h$ enters only the initial condition. It implies that whether the transition is second order or a mere crossover is entirely dictated by the initial condition for the differential equation (\ref{eq:nsmax3b}), a rather remarkable feature.

%\begin{figure}[t]
%	\centering
%	\includegraphics[scale=0.75]{derivative_gap}
%	\caption{The mass squared}
%	\label{fig:mass2}
%\end{figure}

\subsubsection{The transition at $\mu_{\mathcal{B}}=0$}

\begin{figure}[h]
	\centering
	\includegraphics[scale=1.0]{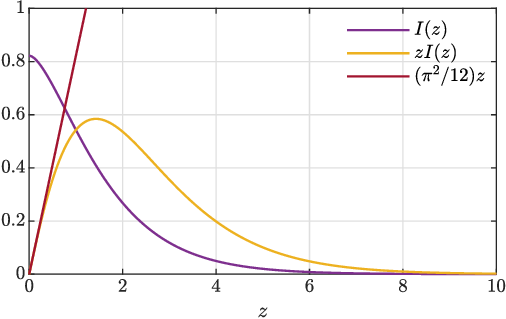}
	\caption{The integral $I(z)$ defined in Eq.~(\ref{eq:Iintegral}) and
	related functions.}
	\label{fig:nsT}
\end{figure}

\begin{figure}[t]
	\centering
	\includegraphics[scale=1.0]{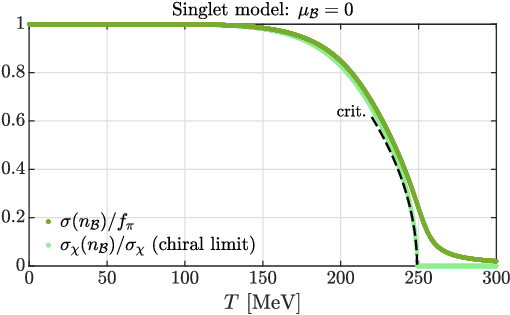}
	\caption{Condensate $\sigma$ at zero baryon-chemical potential
	as function of temperature $T$ in the singlet model. The dashed
	line (``crit.'') shows the critical behavior (\ref{eq:crit_sigma_mu_WT}).}
	\label{fig:phase_diagram_temperature_WT}
\end{figure}

As is the case at finite density, the increase of the fermion scalar density is the driving term for the decrease of $\sigma$. For vanishing baryon density and finite temperature, the scalar density is given by \beq
n_{\rm s}(T)=8 M\int_\p   \frac{n_\mathrm{F}(\varepsilon_\p)}{\varepsilon_\p}  ,
\eeq
where the factor 8 accounts for the antiparticle contributions in addition to spin and isospin of the nucleons. 
 It can be written as the following integral (with $ z\equiv M/T$)
\beq\label{eq:IntegralIz}
n_{\rm s}(T)=\frac{4}{\pi^2} M T^2 I(z)=\frac{4T^3}{\pi^2} z I(z) ,
\eeq
where
\beq \label{eq:Iintegral}
I(z)= \int_0^\infty \rmd x\, \frac{x^2}{\sqrt{x^2+z^2}}\frac{1}{\rme^{\sqrt{x^2+z^2}}+1}.
\eeq
The integral can be calculated analytically when $z=0$: $I(0)=\pi^2/12$. It is a rapidly decreasing function of $z$ (see Fig.~\ref{fig:nsT}). It starts to be non negligible when $z\lesssim 6$. 
This is in line with what we see in Fig.~\ref{fig:phase_diagram_temperature_WT} where the scalar field starts to drop when $T$ is of the order of 150 MeV ($\approx 940/6\ \mathrm{MeV}$).  This is the temperature at which the scalar density starts to increase significantly.  Until that temperature is reached, the sigma field remains at its vacuum value and so do the masses of the nucleons and antinucleons. As the temperature increases above $150\ \mathrm{MeV}$,  the scalar density increases, which entails the decrease of $\sigma$ and of the nucleon mass.

% decreases and so does the nucleon mass.   One then reaches a point where the integral becomes maximal. As is the case at $T=0$ discussed above, this maximum occurs at the value of $\sigma$ which corresponds to the  inflexion point in the potential $U$. Indeed relations analogous to Eqs.~(\ref{eq:nsmax1}) to (\ref{eq:nsms2}) are easily seen to hold with derivatives with respect to $n_B$ are replaced by derivatives with respect to $T$.
%\question{Is this really correct?} \jpb{I guess this is ok, but we need to discuss about the definition of $m_\sigma^2$. }

In the vicinity of the transition, and in the chiral limit,  we may set $M=0$ in the integral that multiplies $M$ in the expression of $n_{\rm s}$ given above. We get
\beq
n_{\rm s}(T)\simeq \frac{M}{3} T^2.
\eeq
 By using the same approximate expression for $U$ as in Eq.~(\ref{UBsigapprox}), one easily solves the gap equation and gets
\beq
T_{c} = \sqrt{\frac{3r}{y^{2}}} \approx 249\ \mathrm{MeV}. 
\eeq
To obtain the behavior of $\sigma(T)$ in the vicinity of $T_c$ requires more work. There are logarithmic contributions both in the function $I(z)$ and in the potential that cancel each other, as is the case at finite density. The logarithmic contribution from the fermion loop is obtained from the well known high-temperature expansion \cite{kapusta2007finite}, which leads to 
    \beq
    n_{\rm s}& \simeq & M\frac{T^{2}}{3} + \frac{M^3}{\pi^{2}}
    \left(\gamma_{\mathrm{E}} - \frac{1}{2} + \ln\frac{M}{\pi T}\right),
\eeq
with the Euler-Mascheroni constant $\gamma_{\mathrm{E}}$.
By examining Eq.~(\ref{UBsigapproxder}) one verifies easily that the logarithmic contribution cancels out, as announced. By following the same steps as at finite density one then obtains 
the critical behavior  
\beq \label{eq:crit_sigma_mu_WT}
    \sigma \simeq \left(\frac{4 y^{2} r}{3 \tilde{u}^{2}}\right)^{1/4} \sqrt{T_{c} - T},
    \eeq
 where    
\beq    \tilde{u} = u + \frac{y^{4}}{\pi^{2}} \left(\gamma_{\mathrm{E}} - \frac{3}{4}
    - \frac{1}{2} \ln\frac{3 \pi^{2} r}{y^{4} f_{\pi}^{2}}\right).
\eeq

The effect of the physical pion mass is  to smear out the second-order transition 
found within the chiral limit, turning it into a smooth crossover where $\sigma$ only asymptotically approaches zero
(cf.\ again the dark green line in Fig.~\ref{fig:phase_diagram_temperature_WT}).

\subsection{The chiral transition in the doublet model}

We start with the transition at $\mu_{\mathcal{B}}=0$. This  is a second-order transition which bears strong similarity with the corresponding one in the singlet model. Then we discuss the transition at $T=0$, which is a first-order transition with very specific features.

\subsubsection{The chiral transition at $\mu_{\mathcal{B}}=0$}

In the parity-doublet model, the gap equation reads (see Eq.~(\ref{eq:gap}))
\beq
\frac{\rmd U}{\rmd \sigma}= -y_{+} n_{\rm s}^+-y_{-} n_{\rm s}^-,
\eeq
where the scalar densities $n_{\rm s}^\pm$ are given by expressions similar to that of the singlet model (Eq.~(\ref{eq:nsanal})), in which one substitutes $M\mapsto M_\pm$, respectively. It follows that the solution of the gap equation for the sigma field exhibits similar behavior as that of the singlet model. In particular, $\sigma$ starts to decrease only after a temperature of the order of $150\ \mathrm{MeV}$. Then $n_{\rm s}^+$ increases as $T$ increases. When the temperature reaches a value $\sim 200\ \mathrm{MeV}$, the density of the negative-parity baryons $n_{\rm s}^-$ starts to increase [this is lower than the anticipated value $\sim 1500/6\ \mathrm{MeV}$ (see the discussion after Eq.~(\ref{eq:IntegralIz})) because the mass of the negative-parity baryons has already decreased when they start to be populated]. Eventually, the masses of the two parity states become (approximately) equal, and the chiral symmetry is restored. 

\begin{figure}[h]
	\centering
	\includegraphics[scale=1.0]{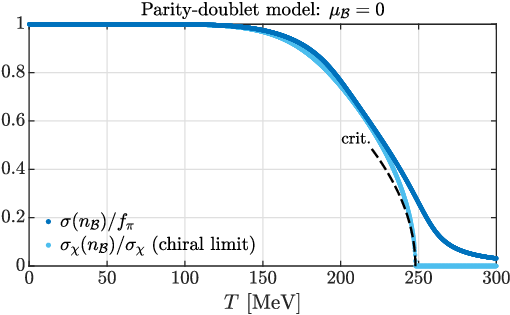}
	\caption{Condensate $\sigma$ at zero baryon-chemical potential
	as function of temperature $T$ in the parity-doublet model (dark blue: physical pion mass; light blue: chiral limit). The dashed line (``crit.'') represents the approximate solution (\ref{eq:sigmacT}).}
	\label{fig:phase_diagram_temperature_PD}
\end{figure}

As is the case in the singlet model,  shortly before the transition occurs, the density $n_{\rm s}^+$ starts to decrease rapidly with increasing temperature. In the vicinity of the transition and in the chiral limit, we have
\beq
n_{\rm s}^\pm(T) = \frac{4}{\pi^2} M_\pm T^2 I(z_\pm),\qquad z_\pm\equiv \frac{M_\pm}{T},
\eeq
where $I(z)$ is the integral (\ref{eq:Iintegral}). To linear order the masses $M_\pm$ are given by (see Eq.~(\ref{eq:massphys}))
\beq
M_\pm\simeq m_0\pm \frac{\sigma}{2} (y_a-y_b).
\eeq
The expansion of $n_{\rm s}^\pm(T)$ for small $\sigma$ reads then
\begin{multline}
n_{\rm s}^\pm(T)\simeq \frac{4T^2}{\pi^2} \left[m_0\pm \frac{\sigma}{2} (y_a-y_b)\right] \\[0.1cm]
\times \left[ I(z_0)\pm \frac{\sigma}{2T} (y_a-y_b)I'(z_0)\right].
\end{multline}
We have also, again to linear order in $\sigma$ (see Eq.~(\ref{eq:yukawa_phys})), 
\beq
y_\pm\simeq\frac{1}{2} \left[\pm(y_a-y_b) +(y_a+y_b)^2\frac{\sigma}{2 m_0}   \right].
\eeq
It follows that the right-hand side of the gap equation can be written as
\begin{multline}
-y_{+} n_{\rm s}^+-y_{-}n_{\rm s}^- \simeq \\[0.2cm] -\sigma \frac{4T^2}{\pi^2} \left[\left(y_a^2+y_b^2\right) I(z_0)
+ \frac{z_0 I'(z_0)}{2} (y_a - y_b)^2 \right],
\end{multline}
where we have used Eq.~(\ref{eq:jurgenformula}).
The critical temperature is then determined by the equation
\beq
r= \frac{4T^2}{\pi^2}\left[\left(y_a^2+y_b^2\right) I_0+\frac{z_0 I'_0}{2} (y_a-y_b)^2\right],
\eeq
where $I_0=I(z_0)$, $z_0=m_0/T$, and $I'(z)=\rmd I(z)/\rmd z$. We eventually find
\begin{equation}
	T_{c} \approx 248\ \mathrm{MeV}
\end{equation}
with $r \approx (634\ \mathrm{MeV})^{2}$, which we obtained in the same fashion as before.

The determination of the critical behavior with respect to temperature requires more work, but proceeds as in the case of the singlet model. We just quote the result 
\begin{equation}\label{eq:sigmacT}
    \sigma \simeq \sqrt{\frac{\tilde{r}}{\tilde{u}}} \sqrt{T_{c} - T},
\end{equation}
where
\begin{widetext}
\begin{IEEEeqnarray}{rCl}
    \tilde{r} & = & \frac{4}{\pi^{2}} \left[2 T_{c}\left(y_{a}^{2} + y_{b}^{2}\right)
    I_{c} - \frac{m_{0}}{2} \left(y_{a} + y_{b}\right)^{2} I_{c}'
    - \frac{m_{0} z_{c}}{2} \left(y_{a} - y_{b}\right)^{2} I_{c}''\right], \\[0.2cm]
    \tilde{u} & = & u + \frac{1}{4 T_{c} \pi^{2}} \bigg\lbrace \frac{T_{c}}{z_{c}} \left(y_{a} + y_{b}\right)^{2}
    \left[4\left(y_{a} - y_{b}\right)^{2} + \left(y_{a} + y_{b}\right)^{2}\right] I_{c}' 
    + T_{c} \left(y_{a} - y_{b}\right)^{2} \left[2\left(y_{a} + y_{b}\right)^{2} 
    + \left(y_{a} - y_{b}\right)^{2}\right] I_{c}'' \nonumber\\[0.2cm]
    & & \qquad +\, \frac{m_{0}}{3} \left(y_{a} - y_{b}\right)^{4} I_{c}'''\bigg\rbrace ,
\end{IEEEeqnarray}
\end{widetext}
with $z_{c} = m_{0}/T_{c}$, $I_{c} = I(z_{c})$. The coupling $u$ is given by the fourth 
derivative of the bosonic potential at the origin,
\begin{equation}
    u = \frac{1}{6}\left.\frac{\partial^{4} U}
    {\partial\sigma^{4}}\right|_{\sigma\, =\, 0} \approx 207,
\end{equation}
which does not exhibit a divergence at $\sigma = 0$ (contrary to the singlet model).

\subsubsection{Transition at $T=0$}

\begin{figure}[t]
	\centering
	\includegraphics[scale=1.0]{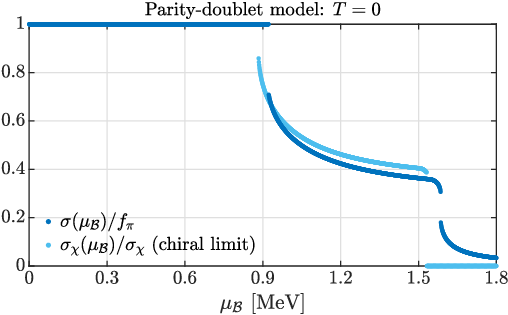}
	\caption{Condensate $\sigma$ at zero temperature as function of baryon-chemical 
	potential $\mu_{\mathcal{B}}$ in the parity-doublet model (dark blue: physical pion mass; light blue: chiral limit).}
	\label{fig:phase_diagram_chempot_PD}
\end{figure}

In contrast to what happens in the singlet model, in the doublet model, and  for our choice of parameters,  the zero-temperature chiral transition is  discontinuous both in the chiral limit, and for the physical pion mass. This is illustrated in Fig.~\ref{fig:phase_diagram_chempot_PD} which displays the value of the $\sigma$ field as a function of the baryon-chemical potential. Both the liquid-gas transition and the chiral transition manifest themselves as jumps in the value of $\sigma$ for $\mu_{\mathcal{B}}\simeq 0.9 $ GeV (liquid-gas transition) and $\mu_{\mathcal{B}}\simeq 1.55$ GeV (chiral transition).  

\begin{figure}[h]
	\centering
	\includegraphics[scale=1.0]{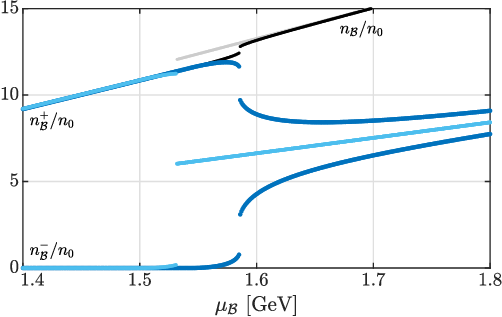}
	\caption{The densities $n_{\mathcal{B}}^+$ and $n_{\mathcal{B}}^-$ as functions of the baryon-chemical potential.
	The coloring (physical pion mass/chiral limit) coincides with the one of 
	Fig.~\ref{fig:phase_diagram_chempot_PD}. The respective total baryon density 
	$n_{\mathcal{B}} = n_{\mathcal{B}}^{+} + n_{\mathcal{B}}^{-}$ is given 
	in black/gray.}
	\label{fig:chiral_transition_den_mu}
\end{figure}

To start understanding the mechanisms at work in the chiral transition, it is instructive to look first at the variations of the populations of positive ($\mathcal{B}^+$) and negative ($\mathcal{B}^-$) parity baryons as the chemical potential increases. This is illustrated in Fig.~\ref{fig:chiral_transition_den_mu}. One sees that the $\mathcal{B}^-$ density remains negligible until the immediate vicinity of the transition, where it increases very rapidly (albeit by a small amount). At the same time the rate of increase of the  $\mathcal{B}^+$ density diminishes, the  $\mathcal{B}^+$ density  eventually decreasing for the physical pion mass. The total baryon density continues to increase, roughly linearly with increasing $\mu_{\mathcal{B}}$, until the transition where it makes a small positive jump. Above the transition the $\mathcal{B}^+$ and $\mathcal{B}^-$ populations are equal (in the chiral limit), as expected once chiral symmetry is restored ($\sigma=0$) and $M_+=M_-$. For the physical pion mass, the chiral symmetry is restored only asymptotically, and the transition, while remaining of first order, is smoother: the $\mathcal{B}^+$ density starts to drop at the transition, and continues to do so after, until it begins to grow and eventually merges with the growing $\mathcal{B}^-$ density at asymptotically large chemical potential or density.

\begin{figure}[t]
	\centering
	\includegraphics[scale=1.0]{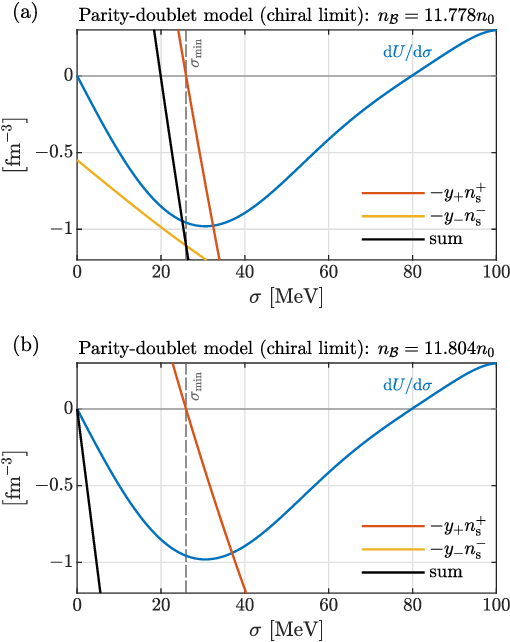}
	\caption{Gap equation in the parity-doublet model. Note that given the large value of $m_0$, the scalar densities coincide with the respective baryon density, that is, $n_{\rm s}^\pm\simeq n_{\mathcal{B}}^\pm$.}
	\label{fig:gap_equation_doublet_1}
\end{figure}

The  solution of  the gap equation provides the value of $\sigma$ as a function of the baryon density. The graphical solution (in the chiral limit) is illustrated in Fig.~\ref{fig:gap_equation_doublet_1} for two values of the density that are close to that corresponding to the chiral transition. The dashed vertical line indicates the value $\sigma_{\rm min}$ of $\sigma$ that corresponds to the minimum of $M_+$. For  $\sigma=\sigma_{\rm min}$,  $y_+=0$ so that the line $-y_+ n_{\rm s}^+$ crosses the dashed line on the horizontal axis. At small and intermediate density, there are no $\mathcal{B}^-$  present in the system and the line corresponding to $-y_- n_{\rm s}^-$ coincides with the horizontal axis while the black line coincides with the red one. The situation is then similar to that represented in Fig.~\ref{fig:gap_equation_singlet_1} for the singlet model. The solution of the gap equation is given by the intersection of the red curve with the blue curve and for a large interval of densities, the corresponding value of $\sigma$ decreases very slowly as the density increases. We shall explain shortly this specific behavior. As the density gets closer to the transition value,  the  $\mathcal{B}^-$ start to appear, and their contribution is indicated by the orange curve (which crosses the vertical dashed line at the same place  as the black line since, as we have seen,  for $\sigma=\sigma_{\rm min}$, $ y_{+} n_{\rm s}^+=0$).  When the density of $\mathcal{B}^-$  starts to increase, things develop rapidly, leading eventually to a jump of the black curve so that it intercepts the blue curve only at $\sigma=0$. This jump in $\sigma$ corresponds of course to the first-order transition.

The (absolute) maximum of the right-hand side of the gap equation can be determined as in the singlet model. We note that 
\begin{IEEEeqnarray}{rCl}\label{eq:maxnspm1}
\frac{\rmd (y_\pm n_{\rm s}^\pm)}{\rmd n_{\mathcal{B}}} & = & y_\pm \left.\frac{\del n_{\rm s}^\pm}{\del n_{\mathcal{B}}}\right|_\sigma+y_\pm  \left.\frac{\del n_{\rm s}^\pm}{\del \sigma}\right|_{n_{\mathcal{B}}}\frac{\rmd \sigma}{\rmd n_{\mathcal{B}}} \nn[0.2cm]
& & +\, n_{\rm s}^{\pm} \frac{\rmd^{2} M_\pm}{\rmd\sigma^{2}} \frac{\rmd \sigma}{\rmd n_{\mathcal{B}}},
\end{IEEEeqnarray}
%\beq\label{eq:maxnspm1}
%y^\pm \frac{\rmd n_{\rm s}^\pm}{\rmd n_{\mathcal{B}}}=y^\pm \left.\frac{\del n_{\rm s}^\pm}{\del n_{\mathcal{B}}}\right|_\sigma+y^\pm  \left.\frac{\del n_{\rm s}^\pm}{\del \sigma}\right|_{n_{\mathcal{B}}}\frac{\rmd \sigma}{\rmd n_{\mathcal{B}}},
%\eeq
where we use the concise notation $a_\pm b_\pm=a_+ b_+ +a_- b_-$. Requiring that $y_\pm n_{\rm s}^\pm$ be a maximum as a function of $n_{\mathcal{B}}$, one gets
\beq\label{eq:maxnspm2}  
y_\pm \left.\frac{\del n_{\rm s}^\pm}{\del n_{\mathcal{B}}}\right|_\sigma=- \left(y_\pm  \left.\frac{\del n_{\rm s}^\pm}{\del \sigma}\right|_{n_{\mathcal{B}}}   + n_{\rm s}^{\pm} \frac{\rmd^{2} M_{\pm}}{\rmd\sigma^{2}}\right)\frac{\rmd \sigma}{\rmd n_{\mathcal{B}}}. \qquad
\eeq
On the other side, by taking the derivative of the gap equation with respect to $n_{\mathcal{B}}$, one gets
\beq\label{eq:maxnspm3}
\frac{\rmd \sigma}{\rmd n_{\mathcal{B}}}=-\frac{y_\pm}{m_\sigma^2} \left.\frac{\del n_{\rm s}^\pm}{\del n_{\mathcal{B}}}\right|_\sigma,
\eeq
with 
\beq\label{eq:maxnspm4}
m_\sigma^{2}= \frac{\rmd^2 U}{\rmd \sigma^2}+n_{\rm s}^\pm \frac{\rmd^2 M_\pm}{\rmd \sigma^2}+y_\pm\left.\frac{\del n_{\rm s}^\pm}{\del \sigma}\right|_{n_{\mathcal{B}}}.
\eeq
The combination of Eqs.~(\ref{eq:maxnspm2}) and (\ref{eq:maxnspm3}) yields the condition
\beq\label{eq:maxnspm5}
m_\sigma^2=y_\pm \left.\frac{\del n_{\rm s}^\pm}{\del \sigma}\right|_{n_{\mathcal{B}}}
+ n_{\rm s}^{\pm} \frac{\rmd^{2} M_{\pm}}{\rmd\sigma^{2}}.
\eeq
By comparing  this expression of $m_\sigma^2$ with the general formula (\ref{eq:maxnspm4}) above, one concludes that the maximum of $y_\pm n_{\rm s}^\pm$ occurs when
\beq\label{eq:maxnspm6}
\frac{\rmd^2 U}{\rmd \sigma^2}=0,
\eeq
which matches the corresponding finding within the singlet model. Note however that the field dependence of the Yukawa couplings induces a slight displacement of the location of the maximum of the scalar density $n_{\rm s}^{+}$ with respect to the inflexion point of the potential, in contrast to the singlet model.

%\begin{figure}[t]
%	\centering
%	%\includegraphics[scale=0.25]{Gap_674}
%	\includegraphics[scale=0.25]{Gap_778}
%	\includegraphics[scale=0.25]{Gap_804}
%	\caption{Gap equation in the parity-doublet model. Note that given the large value of $m_0$, the scalar densities coincide with the respective baryon density, that is, $n_{\rm s}^\pm\simeq n_{\mathcal{B}}^\pm$.}
%	\label{fig:gap_equation_doublet_1}
%\end{figure}

%\begin{figure}[t]
%	\centering
%	\includegraphics[scale=0.65]{dsigmadnB}
%	\caption{The derivative $\rmd\sigma/\rmd n_{\mathcal{B}}$ showing a rapid decrease beyond the threshold for $\mathcal{B}^-$.}
%	\label{fig:dsigmadnB}
%\end{figure}

The analysis of the gap equation reveals interesting features of the transition: (i) The slow increase of $\sigma$ over a wide range of densities. (ii) The  role of the value $\sigma_{\rm min}$  which controls the overall ``topology'' of the graphical solution of the gap equation, as we have just discussed. (iii) The rapid growth of the $\mathcal{B}^-$ population above some threshold.  Note also that from the point of view of the baryon density, the transition is weakly first order, meaning that the relative jump in the baryon density is small at the transition. 
We shall now analyze these various features.

\begin{figure}[t]
	\centering
	\includegraphics[scale=1.0]{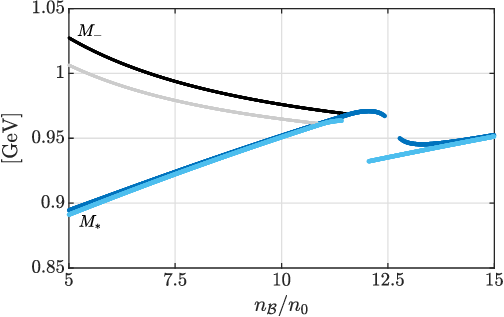}
	\caption{Effective mass $M_{\ast}$ as function of density (dark blue: physical pion mass; 
	light blue: chiral limit). The black and gray lines represent the mass $M_-$ of the 
	negative-parity baryons (up to threshold), for physical pion mass and in the chiral limit, respectively.}
	\label{fig:effective mass}
\end{figure}

\begin{figure}[t]
	\centering
	\includegraphics[scale=1.0]{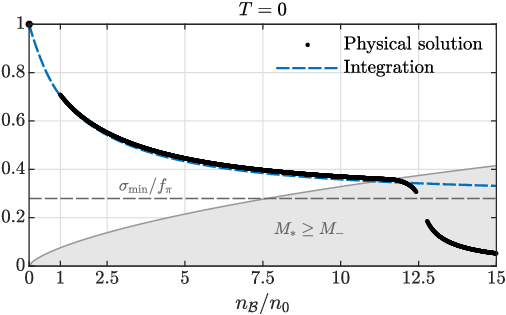}
	\caption{Condensate $\sigma$ at zero temperature as function of baryon density 
	$n_{\mathcal{B}}$ in the parity-doublet model (for physical pion mass). The blue 
	dashed line represents the result obtained by continuously integrating the 
	differential equation (\ref{eq:DEnBsigma}).}
	\label{fig:sigma)DE}
\end{figure}

We start with the threshold for the population of negative-parity baryons (point (iii)). In Fig.~\ref{fig:effective mass} is plotted the effective mass $M_{\ast}$ as a function of density. One sees that between $n_{\mathcal{B}}=5n_0$ and $11n_0$, the increase of $M_{\ast}$ with increasing density  is nearly linear. This can be understood in the following way. We have $M_{\ast}=\sqrt{p_{+}^2+M_+^2}$, where $p_+$ is the Fermi momentum of the $\mathcal{B}^+$ baryons. In this range of densities, $M_+$ depends weakly on $\sigma$, which furthermore does not vary much with density, as we shall see. It is then easily verified that in the range $2n_{0}$ to $11n_0$, the function $M_{\ast}=\sqrt{p_{+}^2+M_+^2}$ for constant $M_+$ is nearly linear as a function of $n_{\mathcal{B}}\sim p_{+}^{3}$. The black line in Fig.~\ref{fig:effective mass} represents the variation of $M_-$ with density. In contrast to $M_+$, the dependence of $M_-$ on $\sigma$ is much stronger: it drops by nearly a factor 2 as $\sigma$ drops from $f_\pi$ to zero. The intersection of the black line with the blue line (gray and light blue in the chiral limit, respectively) defines the $\mathcal{B}^-$ threshold (cf.\ also Ref.~\cite{Larionov:2021ycq}):
\beq\label{eq:threshold1}
M_-=\sqrt{p_F^2+M_+^2},
\eeq
where we have substituted $p_+$ by the baryon Fermi momentum $p_F$ since the $\mathcal{B}^-$ density vanishes at threshold. Since both $M_-$ and $M_+$ are functions of $\sigma$, this equation defines a relation between the baryon density $n_{\mathcal{B}}$ and the value of the sigma field $\sigma$ at threshold. By using the explicit expressions of the mass $M_-$ and $M_+$, Eq.~(\ref{eq:massphys}),  we can rewrite this relation as follows 
\beq\label{eq:threshold2}
p_F^2=\sigma(y_b-y_a)\sqrt{\sigma^2(y_a+y_b)^2+4m_0^2}.
\eeq
The relation (\ref{eq:threshold2}) between $\sigma$ and the Fermi momentum $p_F$ (or equivalently the baryon density) marks the boundary of the gray zone in Fig.~\ref{fig:sigma)DE}. 

In order to complete the determination of the $\mathcal{B}^-$ threshold, we need to know the value of $\sigma$ at the threshold, that is the function $\sigma(n_{\mathcal{B}})$ that is obtained by solving  the gap equation. This is what we turn to now, which will give us the opportunity to explain the origin of the very slow increase of $\sigma$ over a wide range of baryon densities (point (i) above). Rather than solving directly the gap equation, we may rely on Eqs.~(\ref{eq:maxnspm3}) and (\ref{eq:maxnspm4}):
%take the derivative of this equation with respect to $n_{\mathcal{B}}$, and get
%\beq
%\frac{\rmd \sigma}{\rmd n_{\mathcal{B}}}=-  \frac{y_\pm}{m_\sigma^2}\left.\frac{\del n^\pm_s}{\del n_{\mathcal{B}}}\right|_\sigma , 
%\eeq
%where 
%\beq\label{eq:msigmanB}
%m_\sigma^2=\frac{\rmd^2 U}{\rmd \sigma^2}+n_{\rm s}^\pm \frac{\rmd^2 M_\pm}{\rmd \sigma^2}+y_\pm\left.\frac{\rmd n_{\rm s}^\pm}{\rmd \sigma}\right|_{n_{\mathcal{B}}}.
%\eeq
Before the $\mathcal{B}^-$ threshold, one can ignore $n_{\rm s}^-$. Furthermore, because of the mass $m_0$, $n_{\rm s}^+$ starts to differ from $n_{\mathcal{B}}$ only when $p_{+}\gtrsim m_0$. For $m_0=800$ MeV, this inequality is not satisfied until $n_{\mathcal{B}}\gtrsim 28 n_0$. So in the relevant range of densities, $n_{\rm s}^+\simeq n_{\mathcal{B}}$ is an excellent approximation. Thus, in Eqs.~(\ref{eq:maxnspm3}) and (\ref{eq:maxnspm4}) above, we can substitute $n_{\rm s}$ by $n_{\mathcal{B}}$, set $M_+/M_* = 1$,   and also  ignore  the two derivatives ${\partial n_{\rm s}^+}/{\partial \sigma}|_{n_{\mathcal{B}}}$. We obtain then 
\beq\label{eq:DEnBsigma}
\frac{\rmd \sigma}{\rmd n_{\mathcal{B}}}\simeq -y_{+}\left( \frac{\rmd^2 U}{\rmd \sigma^2}+n_{\mathcal{B}} \frac{\rmd^2 M_+}{\rmd \sigma^2}  \right)^{-1},
\eeq
where ${\rmd^2 M_+}/{\rmd \sigma^2}$ is given explicitly in Eq.~(\ref{eq:d2M+}). The equation above can be considered as a differential equation for $\sigma(n_{\mathcal{B}})$. Integrating this equation from the initial condition $\sigma(n_{\mathcal{B}}=0)=f_\pi$ yields the dashed curve in Fig.~\ref{fig:sigma)DE} which almost perfectly overlaps with  the exact solution all the way to the $\mathcal{B}^-$ threshold. Since the sigma field varies little over a large part of the integration range, we may get a qualitative understanding of the long plateau seen in Fig.~\ref{fig:sigma)DE}. Indeed, in the region where the sigma field is nearly constant, the integrand is of the form $(a+b n_{\mathcal{B}})^{-1}$, with $a$ and $b$ two constants,  and the integral exhibits only a smooth, logarithmic variation with respect to $n_{\mathcal{B}}$. 

In fact, it is interesting to continue the integration by ignoring the negative-parity baryons. As indicated by the blue dashed line in Fig.~\ref{fig:sigma)DE}, the obtained function continues its smooth decrease. Because the coefficient  $y_{+}$ in Eq.~(\ref{eq:DEnBsigma}) vanishes at $\sigma=\sigma_{\rm min}$ this behavior is expected to continue asymptotically until $\sigma$ reaches this value $\sigma_{\rm min}$ where $\rmd \sigma/\rmd n_{\mathcal{B}}\to 0$. In the graphical solution of the gap equation (in the chiral limit), this corresponds to the red line becoming vertical, i.e.\ coinciding with the gray dashed line. One may combine this argument with Eq.~(\ref{eq:threshold2})  for the $\mathcal{B}^-$ onset, by plugging into this formula the value $\sigma=\sigma_{\rm min}$. Since the decrease of $\sigma$ with increasing $n_{\mathcal{B}}$ given by Eq.~(\ref{eq:DEnBsigma}) is only logarithmic this is not a very stringent lower bound. This can be seen from Fig.~\ref{fig:sigma)DE}. The relation (\ref{eq:threshold2}) delineates the gray zone, and its intersection with the dashed line corresponding to $\sigma_{\rm min}$ provides the estimate of the lower bound, about $7.5n_0$.

\begin{figure}[t]
	\centering
	\includegraphics[scale=1.0]{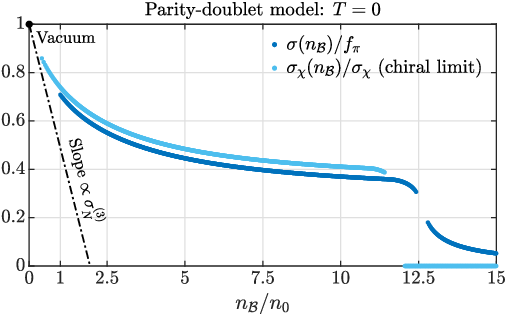}
	\caption{Condensate $\sigma$ at zero temperature as function of baryon density 
	$n_{\mathcal{B}}$ in the parity-doublet model. The solutions to the gap equation 
	were computed in discrete steps of the baryon-chemical potential 
	(between $\mu_{\mathcal{B}} = 0$ and $\mu_{\mathcal{B}} = 1.8\ \mathrm{GeV}$; cf.\ 
	Fig.~\ref{fig:phase_diagram_chempot_PD}). Results for physical pion mass are given in dark blue, 
	those of the chiral limit in light blue. The dash-dotted line shows the linear
	density approximation proportional to the sigma term $\sigma_{N}^{(3)}$.}
	\label{fig:phase_diagram_density_PD}
\end{figure}

Once the threshold is passed, the $\sigma$ field rapidly decreases. This phenomenon amplifies itself. As the population of $\mathcal{B}^-$ increases, so does the quantity $y_{-}n_{\rm s}^-$ in the right-hand side of the gap equation. This leads to a decrease of the magnitude of $\sigma$ which entails a decrease of the mass $M_-$. In turn, this opens further the phase space allowed to the $\mathcal{B}^-$, allowing the Fermi momentum $p_-$ to increase, which in turn contributes to the increase of $y_{-} n_{\rm s}^-$. Figure \ref{fig:phase_diagram_density_PD} summarizes the physical solutions of the gap equations for both the physical pion mass and the chiral limit, together with the initial linear decrease controlled by the nucleon sigma term.

To understand better the underlying mechanism, and in particular what  drives the increase of the $\mathcal{B}^-$ population, let us consider the total energy density. This is  a function of $n_{\mathcal{B}}^+$, $n_{\mathcal{B}}^-$ and $\sigma$:
\begin{IEEEeqnarray}{rCl}
{\cal E}(n_{\mathcal{B}}^+,n_{\mathcal{B}}^-;\sigma) & = & {\cal E}^+_{\rm qp}(n_{\mathcal{B}}^+;\sigma)+{\cal E}^-_{\rm qp}(n_{\mathcal{B}}^-;\sigma) \nn[0.1cm]
& & +\,\frac{1}{2} G_{v} n_{\mathcal{B}}^2+U(\sigma).
\end{IEEEeqnarray}
The equilibrium state is determined by the minimum of ${\cal E}(n_{\mathcal{B}}^+,n_{\mathcal{B}}^-;\sigma)$ with respect to all three variables, subjected to the constraint that $n_{\mathcal{B}}=n_{\mathcal{B}}^++n_{\mathcal{B}}^-$, that is the unconstrained minimum of ${\cal E}(n_{\mathcal{B}}^+,n_{\mathcal{B}}^-;\sigma)-\mu_{\mathcal{B}} (n_{\mathcal{B}}^++n_{\mathcal{B}}^-)$, with $\mu_{\mathcal{B}}$ the baryon-chemical potential. The minimisation with respect to $\sigma$ consistently yields again the gap equation,
\beq\label{eq:gapplusminus}
\left.\frac{\del {\cal E}}{\del\sigma}\right|_{n_{\mathcal{B}}^+,n_{\mathcal{B}}^-} = 0 
= \frac{\rmd U}{\rmd \sigma} + y_{+} n_{\rm s}^+ +y_{-}n_{\rm s}^- .
\eeq
The minimisation with respect to $n_{\mathcal{B}}^+$ and $n_{\mathcal{B}}^-$ yields
\beq\label{eq:minimumplusminus}
\left.\frac{\del {\cal E}}{\del n_{\mathcal{B}}^\pm}\right|_\sigma\equiv \mu^\pm=\mu_{\mathcal{B}}.
\eeq
This equation translates into the equality of the chemical potentials $\mu^+=\mu^-=\mu_{\mathcal{B}}$, with 
\beq
\mu^\pm= \sqrt{p_{\pm}^2+M_{\pm}^2}+G_v n_{\mathcal{B}},
\eeq
or equivalently
\beq\label{eq:Bminusthreshold}
p_+^2-p_-^2=M_-^2-M_+^2.
\eeq
At threshold, $p_-=0$. The condition above says that by turning a $\mathcal{B}^+$ into a $\mathcal{B}^-$, one gains the energy associated with the drop of the Fermi momentum from $p_+$ to $p_-=0$, but at threshold this is just compensated by the mass difference. However, as the density increases, $M_-$ decreases, more than $M_+$ does, so that there is a net energy gain resulting when increasing  the number of $\mathcal{B}^-$ rather than the number of $\mathcal{B}^+$ as one increases $n_{\mathcal{B}}$. In other words, what drives the transition is a form of ``symmetry energy''.

To quantify the effect, we note that the  right-hand side of Eq.~(\ref{eq:Bminusthreshold}) is a known function of $\sigma$. To solve for $p_\pm$ at constant $n_{\mathcal{B}}$, we introduce the Fermi momentum $p_F=(3\pi^2 \,n_{\mathcal{B}}/2)^{1/3}$ and set $p_-/p_F=\alpha^{1/3} $ and $p_+/p_F=(1-\alpha)^{1/3} $. We obtain then 
\beq
f(\alpha)\equiv(1-\alpha)^{2/3}-\alpha^{2/3}=\frac{M_-^2-M_+^2}{p_F^2}.
\eeq
The function $f(\alpha)$ is a decreasing function of $\alpha$, going from $1$ to $-1$ as alpha runs from $0$ to $1$. In fact, since $M_-$ is always bigger than $M_+$, only $f(\alpha)\ge 0$ is relevant, which limits $\alpha$ to values between $0$ and $1/2$.
At threshold, $\alpha=0$, and the right-hand side is unity. As one increases the baryon density, one increases $p_F$ in the denominator, while the numerator is an increasing function of $\sigma$, and $\sigma$ decreases as the density increases. It follows that the right-hand side becomes smaller than unity as $n_{\mathcal{B}}$ increases beyond threshold, and the equation has a solution $\alpha>0$, with $\alpha$ growing with increasing density. This confirms that as $n_{\mathcal{B}}$ increases, the population of $\mathcal{B}^-$ grows. 

In order to see that this growth corresponds to a gain in energy, we consider the variation of  ${\cal E}(n_{\mathcal{B}}^+,n_{\mathcal{B}}^-;\sigma)$ first by restricting the variations  to be such that $n_{\mathcal{B}}=n_{\mathcal{B}}^-+n_{\mathcal{B}}^+$ is constant, as well as $\sigma$. Setting as above 
\beq
\alpha=\frac{n_{\mathcal{B}}^-}{n_{\mathcal{B}}},\qquad \frac{n_{\mathcal{B}}^+}{n_{\mathcal{B}}}=1-\alpha,
\eeq
we note that since $p_F$ and $\sigma$ are kept constant, only ${\cal E}^\pm_{\rm qp}$ depends  on $\alpha$. In other words, the $\alpha$ dependence of the symmetry energy is entirely contained in the sum of the quasiparticle energies. 
We have (see Eq.~(\ref{eq:energydenskin}))
\beq
{\cal E}^\pm_{\rm qp}=\frac{M_\pm^4}{4\pi^2}  \left[z_\pm \sqrt{z_\pm^2+1} \left(2 z_\pm^2+1\right)-\sinh ^{-1}z_\pm\right], \qquad
\eeq
where $z_\pm\equiv p_\pm/M_\pm$. Substituting in this expression, 
\beq
z_-=\alpha^{1/3}\frac{p_F}{M_-} ,\qquad z_+=(1-\alpha)^{1/3}\frac{p_F}{M_+},
\eeq
one finds that ${\cal E}^+_{\rm qp}+{\cal E}^-_{\rm qp}$ exhibits a minimum as a function of $\alpha$. At  threshold, the minimum is at $\alpha=0$, but as one moves closer to the transition, the minimum occurs  for positive $\alpha$, in agreement with the argument above.

\begin{figure}[t]
	\centering
	\includegraphics[scale=1.0]{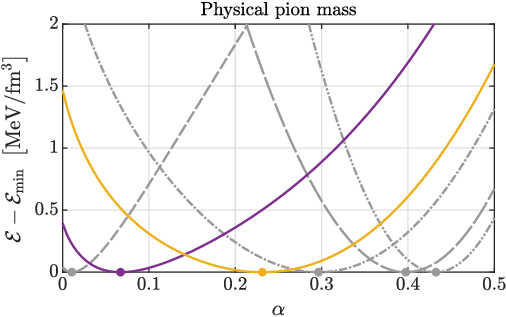}
	\caption{Energy density as function of $\alpha$ for different $n_{\mathcal{B}}$
	(for physical pion mass). Dots indicate the respective minima; gray lines from 
	left to right: $n_{\mathcal{B}} = 12n_{0}$, $13n_{0}$, $14n_{0}$, and $15n_{0}$; 
	magenta line: last value before the transition ($n_{\mathcal{B}} \approx 12.44n_{0}$); 
	orange line: first value after the transition ($n_{\mathcal{B}} \approx 12.77n_{0}$).}
	\label{fig:energy_phys}
\end{figure}

\begin{figure}[t]
	\centering
	\includegraphics[scale=1.0]{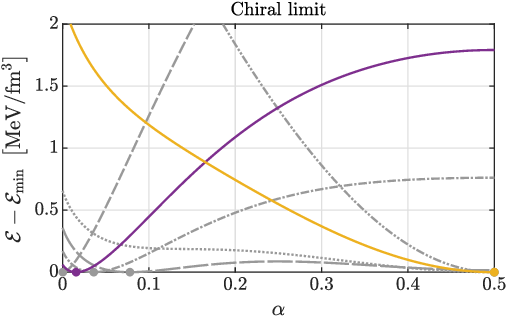}
	\caption{Energy density as function of $\alpha$ for different $n_{\mathcal{B}}$
	(in the chiral limit). Color code equivalent to Fig.~\ref{fig:energy_phys}; gray lines 
	from left to right: $n_{\mathcal{B}} = 11n_{0}$ (dashed), $11.6n_{0}$ (dash-dotted), 
	$11.75n_{0}$ (long-dashed), $11.8n_{0}$ (dotted), and $12.5n_{0}$ (dash-double-dotted); 
	magenta line: $n_{\mathcal{B}} \approx 11.41n_{0}$; orange line: 
	$n_{\mathcal{B}} \approx 12.04n_{0}$. Note that the last two gray dots (for $n_{\mathcal{B}}
	= 11.8n_{0}$ and $12.5n_{0}$) coincide with the orange dot. The lines demonstrate in particular how
	the original minimum at small $\alpha$ becomes a saddle point for increasing density (around $11.8n_{0}$).}
	\label{fig:energy_chiral_limit}
\end{figure}

It is straightforward to calculate numerically the full energy density for a given value of the baryon density, taking into account the effect of the scalar interactions. This amounts to take into account the variation of $\sigma$ as a function of $\alpha$ as given by the gap equation (\ref{eq:gapplusminus}). The results of such a calculation are displayed in Figs.~\ref{fig:energy_phys} and \ref{fig:energy_chiral_limit}, respectively for the physical pion mass and the chiral limit. The minimum at finite value of $\alpha$ and its evolution with increasing density  is clearly visible, in particular in the case of the finite pion mass. In the chiral limit, the evolution before the transition is more restricted. 
%\begin{figure}[t]
%	\centering
%	\includegraphics[scale=0.65]{sym_energy2}
%	\caption{Energy density $\mathcal{E}((1 - x)n_{\mathcal{B}}, x n_{\mathcal{B}})$
%	for $n_{\mathcal{B}} = 13n_{0}$.}
% 	\label{fig:sym_energy1}
%\end{figure}
%\begin{figure}[t]
%	\centering
%	\includegraphics[scale=0.65]{sym_energy_chiral_limit}
%	\caption{Energy density $\mathcal{E}((1 - x)n_{\mathcal{B}}, x n_{\mathcal{B}})$
%	for $n_{\mathcal{B}} = 13n_{0}$.}
% 	\label{fig:sym_energy2}
%\end{figure}

It is instructive to calculate the expansion of the energy density around its minimum, as this will provide insight on the effect of the interactions.  To do so, let us  denote respectively by $\bar\sigma$ and $\bar n_{\mathcal{B}}^\pm$ the values of $\sigma $ and the densities $n_{\mathcal{B}}^\pm$ at the minimum. These are obtained by solving Eqs.~(\ref{eq:gapplusminus}) and (\ref{eq:minimumplusminus}). Keeping $n_{\mathcal{B}}$ fixed, we then expand the energy density in quadratic order in the fluctuation $\delta n_{\mathcal{B}}^-$  ($=-\delta n_{\mathcal{B}}^+$). 
%\begin{figure}[t]
%	\centering
%	\includegraphics[scale=0.8]{sym_energy}
%	\caption{Energy density $\mathcal{E}((1 - x)n_{\mathcal{B}}, x n_{\mathcal{B}})$
%	for $n_{\mathcal{B}} = 13n_{0}$.}
% 	\label{fig:sym_energy}
%\end{figure}
We then obtain the symmetry energy density in the form
\begin{IEEEeqnarray}{rCl}\label{eq:symenergy2}
	\delta^2\mathcal{E} \simeq  \frac{1}{2} 
	\left[\frac{1}{N_{0}^{+}} + \frac{1}{N_{0}^{-}}
	 -\, \frac{\left(y_{+}M_{+} - y_{-}M_{-}\right)^{2}}{m_{\sigma}^{2} M_{\ast}^{2}}
	\right]( \delta n_{\mathcal{B}}^-)^{2},  \nn
\end{IEEEeqnarray}
with $\delta^2 \mathcal{E}$ is shorthand notation for $\mathcal{E}(n_{\mathcal{B}}^-,n_{\mathcal{B}}^+,\sigma)-\mathcal{E}(\bar n_{\mathcal{B}}^-,\bar n_{\mathcal{B}}^+,\bar\sigma)$ expanded to second order in $\delta n_{\mathcal{B}}^-$. In deriving Eq.~(\ref{eq:symenergy2}) we made use of the relation
\begin{equation}
	\frac{\partial\sigma}{\partial n_{\mathcal{B}}^{\pm}} = 
	- \frac{y_{\pm}}{m_{\sigma}^{2}} \frac{M_{\pm}}{M_{\ast}},
\end{equation}
together with  $\mu^{+} = \mu^{-}$ at the minimum. Note that the coefficient of $(\delta n_{\mathcal{B}}^-)^2$ in the expression (\ref{eq:symenergy2}) can vanish. This indeed occurs, as one can see on the plot related to the chiral limit.  As the density increases, a minimum develops at $\alpha=1/2$ corresponding to the symmetric system with $n_{\mathcal{B}}^-=n_{\mathcal{B}}^+$. At the same time the minimum at small $\alpha$ gets  shifted to higher energy, and disappears as a minimum (when the coefficient vanishes) leaving eventually the symmetric minimum as the only stable one. We shall see that the coefficient of $\delta n_{\mathcal{B}}^-$ in the expression (\ref{eq:symenergy2}) enters the expression of  the 
derivatives $\mathrm{d}n_{\mathcal{B}}^{\pm}/\mathrm{d}n_{\mathcal{B}}$ as given below, and its vanishing in the chiral limit is related to a softening of the symmetry energy near the transition.

In order to see that, we need to digress on the compressibility of the system and how it evolves as one approaches the transition. 
The formulae for the compressibility obtained above for a single species of baryon  can be generalized to include the contribution of  the negative-parity partners. We have, with $n_{\mathcal{B}}=n_{\mathcal{B}}^++n_{\mathcal{B}}^-$, 
\beq
\frac{\rmd n_{\mathcal{B}}^\pm}{\rmd\mu_{\mathcal{B}}}&=& 4\int_\p \delta(\mu_{\mathcal{B}}-E^\pm_\p) \left(1-\frac{\rmd E^\pm_\p}{\rmd n_{\mathcal{B}}}\frac{\rmd n_{\mathcal{B}}}{\rmd\mu_{\mathcal{B}}}  \right). \quad
\eeq
In analogy with the calculation of the compression modulus performed in Sec.~\ref{sec:results} (see Eq.~(\ref{eq:K3})), we set
\beq
f_0^\pm =\frac{\rmd E^\pm_\p}{\rmd n_{\mathcal{B}}}=G_v +y_\pm \frac{M_\pm}{M_{\ast}} \frac{\rmd\sigma}{\rmd n_{\mathcal{B}}}, 
\eeq
 so that\footnote{Note that these derivatives ${\rmd n_{\mathcal{B}}^\pm}/{\rmd \mu_{\mathcal{B}}}$ are closely related to the susceptibilities studied in Refs.~\cite{Marczenko_2023,koch2023fluctuations}. Since the densities $n^+_{\mathcal{B}}$ and $n^-_{\mathcal{B}}$ are internal variables of a coupled system, we refrain here to attribute these derivatives  a physical meaning beyond that of indicating the rates of change of the $\mathcal{B}^+$ and $\mathcal{B}^-$ populations as the system approaches the chiral transition.  }
 \beq
\frac{\rmd n_{\mathcal{B}}^\pm}{\rmd \mu_{\mathcal{B}}}= N_0^\pm \left( 1-f_0^\pm  \frac{\rmd n_{\mathcal{B}}}{\rmd \mu_{\mathcal{B}}}\right),\quad N_0^\pm=\frac{2 p_\pm M_{\ast}}{\pi^2}. \qquad
\eeq
By using $n_{\mathcal{B}}=n_{\mathcal{B}}^++n_{\mathcal{B}}^-$, we obtain from the formulae above
\beq
\frac{\rmd n_{\mathcal{B}}}{\rmd \mu_{\mathcal{B}}}\left( 1+F_0  \right)=N_0, 
\eeq
with
$
N_0=N_0^++N_0^-$ and    $F_0=N_0^+f_0^++N_0^- f_0^-, 
$
thereby recovering the expression for the full compression modulus (see Eq.~(\ref{eq:compressibility0})).
\begin{figure}[t]
	\centering
	\includegraphics[scale=1.0]{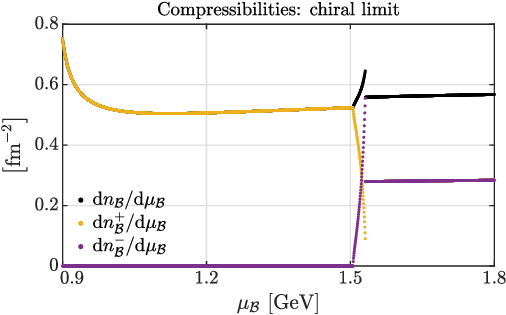}
	\caption{Compressibilities in the chiral limit of the parity-doublet model.}
	\label{fig:compressibilities}
\end{figure}

As can be seen in Fig.~\ref{fig:compressibilities},   the total compressibility (in the chiral limit), proportional to $\rmd n_{\mathcal{B}}/\rmd \mu_{\mathcal{B}}$, increases rapidly before the transition, as the growing population of the negative-parity baryons starts to contribute significantly to the total density. At the transition, the compressibility exhibits a negative jump and returns to a value slightly larger than before the transition region. We may separate the contributions of $\mathcal{B}^-$ and $\mathcal{B}^+$ to the compressibility. These contributions,  $\rmd n_{\mathcal{B}}^\pm /\rmd\mu_{\mathcal{B}}$, are also plotted in Fig.~\ref{fig:compressibilities}. They exhibit opposite behaviors close to the transition,  the $\mathcal{B}^-$ contribution increasing with increasing $\mu_{\mathcal{B}}$ while the $\mathcal{B}^+$ contribution decreases at nearly an identical rate.  At the transition, $\rmd n_{\mathcal{B}}^+ /\rmd\mu_{\mathcal{B}}$ almost vanishes.

From the formulae above, we can obtain the variation of the $\mathcal{B}^+$ population. We get
\beq\label{eq:nBplusnB1}
\frac{\rmd n_{\mathcal{B}}^+}{\rmd n_{\mathcal{B}}}= \frac{   \frac{1}{N_0^-}+f_0^- -f_0^+  }{  \frac{1}{N_0^+}+\frac{1}{N_0^-}  },
\eeq
and a similar formula  for  $\rmd n_{\mathcal{B}}^-/\rmd n_{\mathcal{B}}$ obtained by exchanging plus and minus. 
The difference $f_0^- -f_0^+$ is given by 
\beq
f_0^--f_0^+=\frac{y_- M_- -y_+ M_+}{M_*}\frac{\rmd \sigma}{\rmd n_{\mathcal{B}}}.
\eeq
Since $y_- M_->y_+ M_+$ and $\rmd \sigma/\rmd n_{\mathcal{B}}<0$, $f_0^--f_0^+<0$. It follows that 
\beq\label{eq:boundsnpm}
\frac{\rmd n_{\mathcal{B}}^+}{\rmd n_{\mathcal{B}}}<\frac{p_+}{p_++p_-},\qquad \frac{\rmd n_{\mathcal{B}}^-}{\rmd n_{\mathcal{B}}}>\frac{p_-}{p_++p_-}.
\eeq
Note that these bounds correspond to the values of the derivatives obtained by neglecting the interactions, that is, by ignoring the contribution of $f_0^--f_0^+$ in Eq.~(\ref{eq:nBplusnB1}).
\begin{figure}[h]
	\centering
	\includegraphics[scale=1.0]{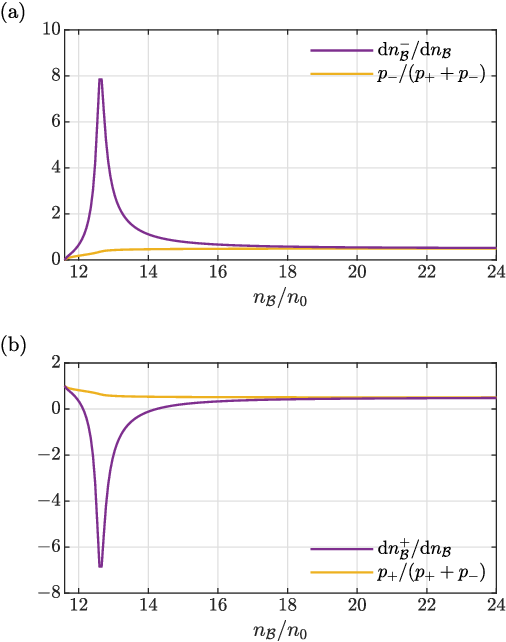}
	\caption{The derivatives $\rmd n_{\mathcal{B}}^\pm/\rmd n_{\mathcal{B}}$ as functions of $n_{\mathcal{B}}$ for physical pion mass, together with the bounds (\ref{eq:boundsnpm}). The derivatives are plotted with $n_{\mathcal{B}}$ starting at the $\mathcal{B}^-$ threshold.}
	\label{fig:dnpndnB}
\end{figure}
A plot of $ \rmd n_{\mathcal{B}}^\pm/\rmd n_{\mathcal{B}}$ as a function of $n_{\mathcal{B}}$ is given in Fig.~\ref{fig:dnpndnB} for the case of the physical pion mass, together with the bounds given in the formula above. To understand the origin of the large peaks in these derivatives, we recall that  
\beq
\frac{\rmd \sigma}{\rmd n_{\mathcal{B}}}=-\frac{1}{m_\sigma^2}\left( y_{+} \left.\frac{\del n_{\rm s}^+}{\del n_{\mathcal{B}}}\right|_\sigma+y_{-} \left.\frac{\del n_{\rm s}^-}{\del n_{\mathcal{B}}}\right|_\sigma  \right).
\eeq
We have
\beq
\left.\frac{\del n_{\rm s}^\pm}{\del n_{\mathcal{B}}^\pm}\right|_\sigma=\frac{M_\pm}{M_*}. 
\eeq
By using this relation together with $n_{\mathcal{B}}=n_{\mathcal{B}}^+ +n_{\mathcal{B}}^-$, we obtain
\beq\label{eq:dsigmadnB2}
\frac{\rmd \sigma}{\rmd n_{\mathcal{B}}}&=&-\frac{1}{m_\sigma^2 M_*} \left( y_{+} M_+\frac{\rmd n_{\mathcal{B}}^+}{\rmd n_{\mathcal{B}}} + y_{-} M_- \frac{\rmd n_{\mathcal{B}}^-}{\rmd n_{\mathcal{B}}}\right)\nn[0.2cm]
&=&-\frac{y_{+} M_+ - y_{-} M_- }{m_\sigma^2 M_*} \frac{\rmd n_{\mathcal{B}}^+}{\rmd n_{\mathcal{B}}}-\frac{y_{-} M_-}{m_\sigma^2 M_*}.
\eeq
We may then combine Eq.~(\ref{eq:dsigmadnB2}) with Eq.~(\ref{eq:nBplusnB1}) in order to get an equation which controls the variation of $n_{\mathcal{B}}^\pm$ as a function of $n_{\mathcal{B}}$. We get
\beq\label{eq:nBplusnB2}
\frac{\rmd n_{\mathcal{B}}^+}{\rmd n_{\mathcal{B}}}=\frac{   \frac{1}{N_0^-}+ \frac{y_{-} M_-}{m_\sigma^2 M_*^2} \left( y_{+} M_+ - y_{-} M_- \right) }{  \frac{1}{N_0^+}+\frac{1}{N_0^-} -\frac{1}{m_\sigma^2 M_*^2} \left( y_{+} M_+ - y_{-} M_- \right)^2 }, \qquad
\eeq
and similarly for $n_{\mathcal{B}}^-$. Note that $\rmd n_{\mathcal{B}}^+/\rmd n_{\mathcal{B}}=1- \rmd n_{\mathcal{B}}^-/\rmd n_{\mathcal{B}}$.  Consider first the derivative $\rmd n_{\mathcal{B}}^-/\rmd n_{\mathcal{B}}$. At the $\mathcal{B}^-$ threshold, $N_0^-\propto p_-$ vanishes, so that the derivative goes to $0$. Since $ y_{-} M_- -y_{+} M_+$ remains positive, the derivative is positive. However, the denominator is the difference of two positive numbers and it can potentially vanish. In fact, we recognize in this denominator the curvature of the symmetry energy obtained in Eq.~(\ref{eq:symenergy2}). There is thus a direct correlation between the softening of the symmetry energy and the peak in the derivative $\rmd n_{\mathcal{B}}^-/\rmd n_{\mathcal{B}}$ as a function of $n_{\mathcal{B}}$. 

The corresponding derivative $\rmd n_{\mathcal{B}}^+/\rmd n_{\mathcal{B}}$ can be analyzed similarly. Its initial value is $1$, for $p_{-} \to 0$. The main difference with the $\mathcal{B}^-$ case is that the numerator in Eq.~(\ref{eq:nBplusnB2}) may change sign. It indeed does so for the physical pion mass, as shown in Fig.~\ref{fig:dnpndnB}: thus, as the $\mathcal{B}^-$ population starts to grow, the $\mathcal{B}^+$ population first continues to grow and then starts to decrease with increasing $n_{\mathcal{B}}$.  In the chiral limit, the (negative) peak is replaced by an infinite slope approaching the corresponding density from below. This is connected with the fact that the $\mathcal{B}^+$ contribution to the compressibility never vanishes in the chiral limit (the chiral transition occurs before).
%\begin{figure}
%	\centering
%	\includegraphics[scale=0.75]{dnminusdnB}
%	\includegraphics[scale=0.75]{dnplusdnB}
%	\caption{The derivatives $\rmd n_{\mathcal{B}}^\pm$ as a function of $n_{\mathcal{B}}$ for the physical pion mass, together with the bounds (\ref{eq:boundsnpm}). The derivatives are plotted with $n_{\mathcal{B}}$ starting at the $\mathcal{B}^-$ threshold. }
%	\label{fig:dnpndnB}
%\end{figure}
%
%\begin{figure}[t]
%	\centering
%	\includegraphics[scale=0.75]{diffsigma}
%	\includegraphics[scale=0.75]{diffnplus}
%	\caption{Flow lines $\sigma(n_{\mathcal{B}})$ (top panel) and $n_{\mathcal{B}}^+(n_{\mathcal{B}})$ obtained by solving the coupled equations (\ref{eq:dsigmadnB3}) and (\ref{eq:nBplusnB2}). The horizontal linesindicate the value $\sigma_{\rm min}$. The gray line corresponds to the $\mathcal{B}^-$ threshold, $M_*=M_-$. Note that for the bottom panel, this is just a straight line since at the $\mathcal{B}^-$ threshold, $\rmd n_{\mathcal{B}}^+/\rmd n_{\mathcal{B}}=1$. }
%	\label{fig:diffsigma}
%\end{figure}

\begin{figure}[t]
	\centering
	\includegraphics[scale=1.0]{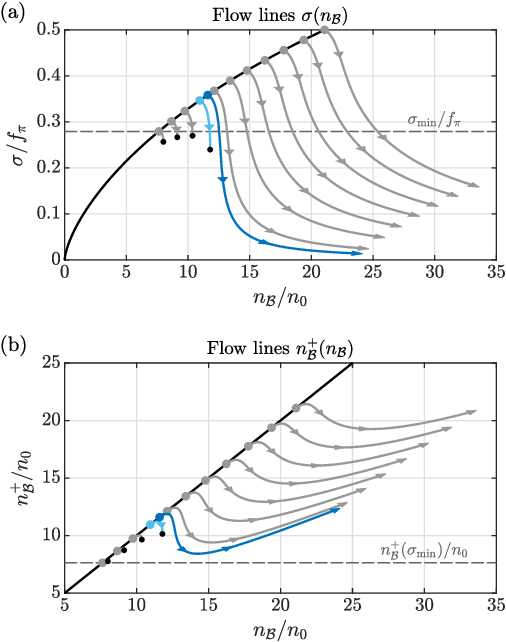}
	\caption{Flow lines $\sigma(n_{\mathcal{B}})$ [panel (a)] and $n_{\mathcal{B}}^+(n_{\mathcal{B}})$ [panel (b)] obtained by solving the coupled equations (\ref{eq:dsigmadnB2}) and (\ref{eq:nBplusnB2}). The horizontal dashed line in panel (a) indicates the value $\sigma_{\rm min}$. The black line corresponds to the $\mathcal{B}^-$ threshold, $M_*=M_-$. Note that for the bottom panel, this is just a straight line since at the $\mathcal{B}^-$ threshold, $\rmd n_{\mathcal{B}}^+/\rmd n_{\mathcal{B}}=1$. }
	\label{fig:diffsigma}
\end{figure}

The previous discussion relies on the solution of the gap equation. We have already mentioned that an alternative to solving the gap equation is to solve a differential equation for the function $\sigma(n_{\mathcal{B}})$. This is easily done for densities below the $\mathcal{B}^-$ threshold, where  
 Eq.~(\ref{eq:dsigmadnB2}) reduces to (see also Eq.~(\ref{eq:DEnBsigma}))
\beq\label{eq:dsigmadnB3}
\frac{\rmd \sigma}{\rmd n_{\mathcal{B}}}=-\frac{y_{+}}{m_\sigma^2} \frac{M_+}{M_*}.
\eeq
However, as soon as the $\mathcal{B}^-$ appear, we need an additional equation that controls the composition of matter as a function of $n_{\mathcal{B}}$. This additional equation is provided by the equation (\ref{eq:nBplusnB2}) for $\rmd n_{\mathcal{B}}^+/\rmd n_{\mathcal{B}}$. The right-hand side of these equations, where $m_\sigma^2$ is given by Eq.~(\ref{eq:maxnspm4}), are known functions of $\sigma$ and $n_{\mathcal{B}}$. They can be integrated together in order to obtain $\sigma$ and $n_{\mathcal{B}}^{+}$ as functions of $n_{\mathcal{B}}$. In Fig.~\ref{fig:diffsigma} we plot resulting flow lines obtained by fixing the initial condition on the line that corresponds to the $\mathcal{B}^-$ threshold (and initialize Eq.~(\ref{eq:nBplusnB2}) with a tiny density for the negative-parity baryons, $n_{\mathcal{B}}^{-} > 0$, in the sense of a perturbation). The integration of Eq.~(\ref{eq:dsigmadnB3}) from some initial condition at $n_{\mathcal{B}}=0$ will bring the system on a point of the black line. Typical initial conditions depend on the value of the pion mass, with the dark blue dot corresponding to the physical pion mass, and the light blue dot to the chiral limit. Note that the solution emanating from the light blue dot terminates at some finite value of $\sigma$ (indicated by a small black dot). This is where $\sigma$ drops to zero, and the system of differential equations runs into a divergence, corresponding to the vanishing symmetry energy (\ref{eq:symenergy2}). In contrast, the solution emanating from the dark blue dot reaches $\sigma=0$ only asymptotically. Interestingly, these flow lines indicate the expected behavior as one increases the pion mass beyond its physical value. This we cannot do in the standard approach, since this would amount to explore regions where the potential $U(\sigma)$ is not bounded from below. As the plots suggest, the variations of $\sigma$ with $n_{\mathcal{B}}$ becomes smoother as the pion mass increases. For sufficiently large values one may expect the transition to turn into a mere crossover. 

Figure~\ref{fig:diffsigma} furthermore demonstrates how the population of the $\mathcal{B}^{+}$ changes once the system passed the $\mathcal{B}^{-}$ threshold, where the horizontal dashed line shows the density $n_{\mathcal{B}}^{+}$ that pairs the value $\sigma_{\mathrm{min}}$ by virtue of Eq.~(\ref{eq:threshold1}), and $\sigma_{\mathrm{min}}$ marks the lowest value on the threshold line that can be reached by integration of Eq.~(\ref{eq:dsigmadnB3}). Finally, the flow lines $n_{\mathcal{B}}^{+}(n_{\mathcal{B}})$ indicate where to expect a reduction of the $\mathcal{B}^{+}$ in favor of the $\mathcal{B}^{-}$ population.

\begin{figure}[t]
	\centering
	\includegraphics[scale=1.0]{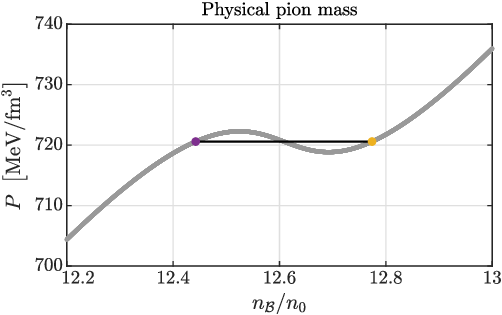}
	\caption{Zero-temperature pressure in the parity-doublet model as function of 
	$n_{\mathcal{B}}$ (for physical pion mass). The black horizontal line represents 
	the Maxwell construction corresponding to the chiral transition (the colored dots correspond to the densities of the colored lines in Fig.~\ref{fig:energy_phys}).}
	\label{fig:pressure_maxwell_phys}
\end{figure}

\begin{figure}[t]
	\centering
	\includegraphics[scale=1.0]{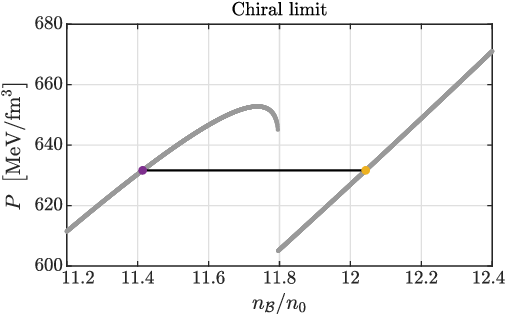}
	\caption{Zero-temperature pressure in the parity-doublet model as function of 
	$n_{\mathcal{B}}$ (in the chiral limit). The black horizontal line represents 
	the Maxwell construction corresponding to the chiral transition (the colored dots correspond to the densities of the colored lines in Fig.~\ref{fig:energy_chiral_limit}). Approaching the transition from below, one can follow continuously the pressure as a function of increasing density up to the point where the sigma fields jumps to zero.}
	\label{fig:pressure_maxwell_chiral_limit}
\end{figure}

The advantage of the approach based on the differential equations is that it allows us to follow continuously the evolution of the system even in regions where the corresponding phases are not stable. In the chiral limit, it also leads us to expect a discontinuous behavior characteristic of a first-order transition. However, the approach does not allow us to determine precisely the location and nature of the transition. To do so, we need to calculate the pressure as a function of the chemical potential, or the pressure as a function of density and rely on the Maxwell construction. The latter is presented in Figs.~\ref{fig:pressure_maxwell_phys} and \ref{fig:pressure_maxwell_chiral_limit}, for both the physical pion mass and the chiral limit. For the case of the physical pion mass, one can follow the pressure continuously as the baryon density increases, even in the unphysical region. In the chiral limit, one is prevented to do so because $\sigma$ exhibits a jump from a finite value to zero.

%\begin{figure}[t]
%	\centering
%	\includegraphics[scale=0.75]{maxwell}
%	\includegraphics[scale=0.75]{maxwell-chiral}
%	\caption{The zero temperature pressure in  the parity-doublet model. Top panel as a function of $x$ and physical pion mass. Lower panel, as a function of $n_{\mathcal{B}}$ and chiral limit.  \jpb{ Should not we plot the physical pion mass also versus $n_{\mathcal{B}}$?} The hozontal lines represent the Maxwell construction.  }
%	\label{fig:maxwell}
%\end{figure}

\begin{figure*}[t]
	\centering
	\includegraphics[scale=1.0]{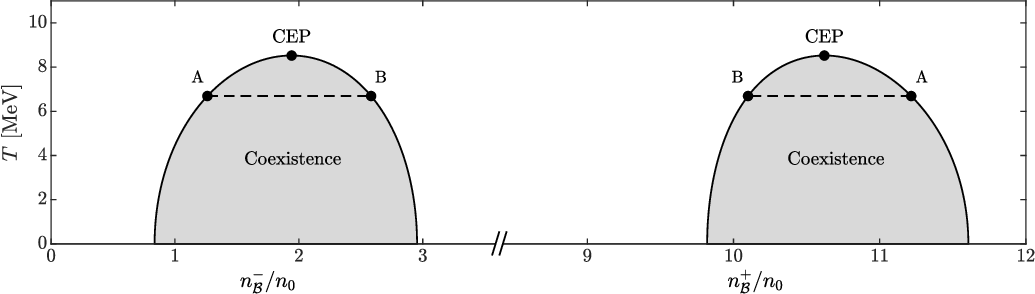}
	\caption{ Illustration of the
	composition of coexisting phases. The points $\mathrm{A}$ and $\mathrm{B}$ correspond to the points $\mathrm{A}$ and $\mathrm{B}$ of Fig.~\ref{fig:chiral_transition1}. The concentration of negative-parity baryons is greater in Phase $\mathrm{B}$ (chirally symmetric) than in Phase 
	$\mathrm{A}$. The x-labels correspond respectively to the two gray ``blobs'', i.e.\ $n_{\mathcal{B}}^{-}/n_{0}$ to the left one and $n_{\mathcal{B}}^{+}/n_{0}$ to the right one, both data drawn on a single axis. This means that e.g.\ the densities $n_{\mathcal{B}}^{+}$ and $n_{\mathcal{B}}^{-}$ of phase A add up to the total baryon density $n_{\mathcal{B}} = n_{\mathcal{B}}^{+} + n_{\mathcal{B}}^{-}$ of point A given in Fig.~\ref{fig:chiral_transition1}.}
	\label{fig:chiral_transition2}
\end{figure*}

\begin{figure}[h]
	\centering
	\includegraphics[scale=1.0]{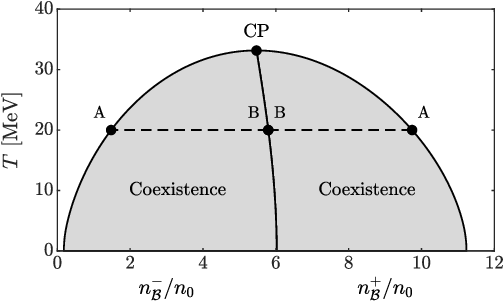}
	\caption{Same as Fig.~\ref{fig:chiral_transition2} in the chiral limit, and corresponding to Fig.~\ref{fig:chiral_transition1_chiral_limit}. The two formerly separate gray blobs merge and coincide in the black middle line, as the populations of $\mathcal{B}^{+}$ and $\mathcal{B}^{-}$ are identical in phase B.}
	\label{fig:chiral_transition2_chiral_limit}
\end{figure}

Finally, let us comment on the structure of the phases at the chiral transition. The transition being first order,  at the critical chemical potential, there is phase coexistence. This was already shown in Figs.~\ref{fig:chiral_transition1} (for physical pion mass) and \ref{fig:chiral_transition1_chiral_limit} (in the chiral limit) as a function of temperature and baryon density. Let us focus on the zero-temperature case. At the phase transition, a chirally symmetric phase develops, whose density is slightly above that of the broken phase (this corresponds to the small jumps in density observed in Fig.~\ref{fig:phase_diagram_density_PD} for both data sets). When analyzed in terms of the populations $n_{\mathcal{B}}^+$ and $n_{\mathcal{B}}^-$, the coexistence regions for physical pion mass split as indicated in Fig.~\ref{fig:chiral_transition2}. It is
clearly visible that the $\mathcal{B}^{-}$ become populated already before the transition, $n_{\mathcal{B}}^{-} > 0$. The two coexistence regions
are roughly mirror-imaged to each other, with the fraction $n_{\mathcal{B}}^{-}/n_{\mathcal{B}}^{+}$ being closer to $1$ in the large-density phase (where chiral symmetry is approximately restored). Regarding the same illustration in the chiral limit, see Fig.~\ref{fig:chiral_transition2_chiral_limit}, one finds that the two corresponding coexistence regions merge, in the sense that they coincide in the black middle line. This line contains the phase B with now identical populations of the $\mathcal{B}^{+}$ and $\mathcal{B}^{-}$, i.e.\ $n_{\mathcal{B}}^{+} = n_{\mathcal{B}}^{-}$, whereas in phase A we still have $n_{\mathcal{B}}^{-}/n_{\mathcal{B}}^{+} < 1$. The two densities to be equal in phase B is of course induced by the restoration of chiral symmetry ($\sigma = 0$), hence the masses of the $\mathcal{B}^{+}$ and $\mathcal{B}^{-}$ being degenerate. We finally note that the middle line between the two coexistence regions is not vertical but slightly distorted to the left with increasing $T$ (towards smaller densities), which finds its explanation in the fact that the total baryon density of phase B is larger than the one of phase A (recall the positive jumps in $n_{\mathcal{B}} = n_{\mathcal{B}}^{+} + n_{\mathcal{B}}^{-}$ occurring at the first-order transitions, as shown in Fig.~\ref{fig:phase_diagram_density_PD}).

\section{Conclusions}

We presented a detailed discussion of the thermodynamics of the parity-doublet model for
isospin-symmetric matter in a mean-field approach, basically covering the entire phase diagram in the 
$T$-$\mu_{\mathcal{B}}$-plane. This model features the parity doublet of the positive-parity nucleon and 
its negative-parity chiral partner with the chiral-invariant mass $m_{0}$ that confers the doublet a nonzero 
bulk mass in the high-energy regime, where chiral symmetry is restored. We 
gave special emphasis to the zero-temperature phase structure and the chiral transition, thereby 
extrapolating from low-density nuclear matter to the large-density regime of several times the 
nuclear saturation density $n_{0}$. The model parameters were adjusted so that empirical data of nuclear 
matter in its ground state and the critical end point of the nuclear liquid-gas transition were 
reproduced.

In order to provide a rather complete picture of the chiral transition occurring at large baryon 
densities and vanishing temperature, and to understand and identify the underlying mechanisms at work, 
we systematically contrasted the findings within the doublet model to the corresponding results within the singlet 
model. The singlet model was obtained by ignoring the influence of the chiral partner as a whole,
such that the bulk mass $m_{0}$ is lost and the nucleon mass is exclusively generated by chiral
symmetry breaking at low energies (i.e.\ the nucleon becomes massless in the chiral-restored
phase).

Regarding the zero-temperature chiral transition, the effect of the mass $m_{0}$ is that it locks the scalar 
density to the baryon density throughout a large range of densities relevant for the discussion of the
phase structure. This is a direct consequence of the fact that the fraction of the nucleon mass resulting 
from chiral symmetry breaking is substantially reduced as compared to the singlet model,
where the mass is proportional to the $\sigma$ condensate. The increase of the scalar density with
increasing $\mu_{\mathcal{B}}$ is necessary to drive the ``melting'' of the $\sigma$ condensate
in order to restore chiral symmetry at large $\mu_{\mathcal{B}}$. Another important effect is the 
splitting of the mass spectrum of the opposite parity baryons, which creates a nontrivial minimum of 
the nucleon mass at $\sigma_{\mathrm{min}}$, with $0 < \sigma_{\mathrm{min}} < f_{\pi}$. It turned out 
that these two effects decelerate the melting of the $\sigma$ condensate when approaching $\sigma_{\mathrm{min}}$
from above, starting from its initial value of $\sigma = f_{\pi}$ in vacuum. In consequence, the chiral transition 
within the doublet model, where $\sigma \to 0$, gets significantly delayed. The opening of the phase space for
populating the negative-parity baryons, which is suppressed for small baryon densities, is then necessary to
cross the value of $\sigma_{\mathrm{min}}$ and eventually bring down the condensate to (approximately) zero.
For the chosen set of parameters, the chiral transition in the singlet model occurs at $n_{\mathcal{B}} 
\approx 5n_{0}$, whereas in the doublet model it only occurs at about $12n_{0}$, thus at densities 
more than twice as high as in the singlet model.

Concerning the order of the chiral phase transition at zero temperature, we found a first-order transition within 
the doublet model for physical pion mass as well as in the chiral limit (where the pion mass vanishes). These 
first-order transitions are accompanied by a jump in density, and by the coexistence of two phases with the same 
pressure and baryon-chemical potential. In the large-density phase of these two coexisting phases, the chiral 
symmetry is (approximately) restored, and in the low-density phase, it is still broken. In contrast, in the 
singlet model, the transition is of second order in the chiral limit, which is smeared out to a crossover 
for physical pion mass. Its continuous character in the singlet model even allowed us to determine the critical 
behavior in the vicinity of the transition. For zero chemical potential and finite temperature, the chiral
transition is of second order in the chiral limit in both the doublet and singlet models, and a smooth crossover
in the case of physical pion mass.

The comparison of results obtained in the chiral limit with those obtained for a finite pion mass was very instructive. 
Results are ``sharper'' in the chiral limit. For example, in the singlet model, the chiral transition is of second 
order in the chiral limit, while it becomes a continuous crossover for the finite pion mass, as mentioned above. 
Moreover, we combined the study of the chiral transition at large densities with a detailed discussion 
of the nucleon sigma term $\sigma_{N}$ at low baryon densities, which describes the initial decrease 
of the average $\sigma$ field as soon as the vacuum gets populated with dilute matter. This discussion was then 
completed by the computation of respective density modifications of $\sigma_{N}$. Also, by calculating various 
estimates for the sigma term, we found that its value can be much affected by non linear effects in the relation 
between the chiral limit and the physical point. As a result, the value obtained in the doublet model is somewhat 
larger than in the singlet model.

We have seen that the presence of the massive baryon doublet affects the nature of the chiral phase 
transition, and its dynamics, in a significant way. In particular, the dynamics of the phase transition 
at zero temperature and finite baryon density turned out to be very interesting. We have identified the 
role of a kind of symmetry energy as the driving force for the chiral transition in order to equilibrate 
the respective baryon densities of the two opposite parity baryons, once the phase space for populating 
the chiral partner is opened. This symmetry energy dictates the composition of matter at large baryon
densities around the chiral transition, i.e.\ the fraction of positive to negative-parity baryons. 
We have furthermore investigated in great detail the overall ``topology'' of the corresponding gap equations 
that provide the physical solutions of the system, and exploited the character of their density derivatives 
as first-order differential equations to get an alternative view on the important ingredients for the chiral transition. 

In summary, the doublet model features a rich phase structure that offered an interesting playground for many 
detailed calculations.
It provides an interesting perspective on how chiral symmetry may be restored, allowing for the presence of massive parity doublets in the symmetric phase, for which there is some evidence from lattice calculations at finite temperature \cite{datta_nucleons_2013,aarts_light_2017}. 

We have also identified some limitations of the model, and there are also limitations of the calculations presented in this paper:
There are uncertainties in the phenomenological parametrization of the effective potential for the mesonic degrees of freedom. We have used an expansion in powers of the chiral field, as commonly done, and we have used the simplest form, ignoring for instance self-interactions of the vector field. The parameters of the potential are adjusted so as to reproduce nuclear matter properties, but the same potential is used all the way to the region of vanishing $\sigma$ field when discussing the chiral transition. This is clearly an uncertain extrapolation, for which we have very little control. One particular consequence of this extrapolation is the strong correlation that it induces  between the chiral transition and the liquid-gas transition, which one could consider a priori as two distinct physical phenomena. 

We have taken into account the (one-loop) fermionic fluctuations, but we have ignored the mesonic ones. Fermionic fluctuations contribute significantly to the scalar densities, which play an essential role in the chiral transition. Mesonic fluctuations presumably do not play much of a role at vanishing temperature, and some of them have been taken into account in functional renormalisation group calculations \cite{Weyrich_2015, Tripolt_2021}. However, mesonic fluctuations are certainly important at vanishing baryon density and finite temperature. Since they are ignored in the present paper, one should keep in mind that the results that we have obtained at finite temperature and small baryon density may change quantitatively (and perhaps even qualitatively) once these fluctuations are taken into account. 

In spite of these shortcomings, there is interest to further analyze the physical content of the model, exploring for instance its predictions for neutron matter (and isospin-asymmetric matter in general \cite{nextpaper}). Finally, more input from QCD computations would be helpful to further constrain the phase structure, especially at large densities and low temperatures, although this currently remains an intricate problem.

\begin{acknowledgments}
We acknowledge interesting exchanges with G. Col\`o, U. Meissner, F. Rennecke, and S. Shlomo on  several topics discussed in this paper. 
JE acknowledges funding by the German National Academy of Sciences Leopoldina (through the scholarship 2020-06). JE thanks 
the IPhT in Saclay for hospitality. JPB thanks ECT* (the European Center for Theoretical Studies in Nuclear Physics and 
Related Areas) in Trento for hospitality. 
\end{acknowledgments}

%%%%%%%%%%%%%%%%%%%%%%%%%%%%%%%%%%%%%%%%%%%%%%%%%%%%%%%%
\appendix
%%%%%%%%%%%%%%%%%%%%%%%%%%%%%%%%%%%%%%%%%%%%%%%%%%%%%%%%

\section{Renormalised vacuum fluctuations}
\label{sec:vacuum}

The renormalised fermionic vacuum contribution to the thermodynamic potential can be written as a function of the baryon masses as follows (see e.g.\ Ref.~\cite{Skokov:2010sf})
\begin{equation}
	\Omega_{\ln}(\sigma) = - \sum_{i\, =\, \pm} \frac{M_{i}(\sigma)^{4}}{4\pi^{2}} 
	\ln\frac{M_{i}(\sigma)}{M_{+}(f_{\pi})}.
\end{equation}
The subtraction point has been chosen to be the nucleon mass in vacuum, 
$M_{+}(f_{\pi}) \equiv M_{N}$. The other renormalisation condition (related to mass and coupling constant divergences) are conveniently implemented by    subtracting the second-order polynomial $\delta\Omega_{\ln}$,
\begin{equation}
	\delta\Omega_{\ln} = \sum_{n\, =\, 0}^{2}
	\frac{1}{n!} \left. \frac{\partial^{n}\Omega_{\ln}}
	{(\partial\sigma^{2})^{n}} \right|_{f_{\pi}^{2}}
	\left(\sigma^{2} - f_{\pi}^{2}\right)^{n} ,
\end{equation}
such that the total renormalised vacuum contribution then reads
\begin{equation}
	\Omega_{\mathrm{vacuum}} = \Omega_{\ln} - \delta\Omega_{\ln} .
\end{equation}

The above subtraction guarantees that the vacuum contribution from the baryon loop vanishes for $\sigma=f_\pi$ as well as its first and second-order derivatives with respect to $\sigma^2$. This entails a modification of  the initial values  $\alpha_{1}=m_\pi^2$ and $\alpha_{2}=(m_\sigma^2-m_\pi^2)/f_\pi^2$ of the first two Taylor coefficients of the classical potential $V(\varphi)$.
The full bosonic potential $U$ then reads:
\begin{widetext}
%\begin{IEEEeqnarray}{rCl}
%	U_{\rm{B}} & = & \left(\frac{m_{\pi}^{2}}{2} 
%	+ \frac{1}{8\pi^{2}f_{\pi}}\left\lbrace M_{+}(f_{\pi})^{3} y_{\sigma}^{++}(f_{\pi})
%	+ \left[1 + 4 \ln\frac{M_{-}(f_{\pi})}{M_{+}(f_{\pi})}\right]
%	M_{-}(f_{\pi})^{3} y_{\sigma}^{--}(f_{\pi}) \right\rbrace\right)
%	\left(\sigma^{2} - f_{\pi}^{2}\right) \nonumber\\[0.2cm]
%	& & +\, \bigg(\frac{m_{\sigma}^{2} - m_{\pi}^{2}}{8 f_{\pi}^{2}} 
%	+ \frac{1}{32\pi^{2}f_{\pi}^{3}}\, \bigg\lbrace 7 f_{\pi} M_{+}(f_{\pi})^{2} y_{\sigma}^{++}(f_{\pi})^{2}
%	+ f_{\pi} \left[7 + 12 \ln\frac{M_{-}(f_{\pi})}{M_{+}(f_{\pi})}\right]
%	M_{-}(f_{\pi})^{2} y_{\sigma}^{--}(f_{\pi})^{2} \nonumber\\[0.2cm] 
%	& & \hspace{4cm} +\, M_{+}(f_{\pi})^{3}\left[\frac{2 f_{\pi} y_{\sigma}^{+-}(f_{\pi})^{2}}
%	{M_{+}(f_{\pi}) + M_{-}(f_{\pi})} - y_{\sigma}^{++}(f_{\pi})\right] \nonumber\\[0.2cm] 
%	& & \hspace{4cm} + \left[1 + 4 \ln \frac{M_{-}(f_{\pi})}{M_{+}(f_{\pi})}\right] 
%	M_{-}(f_{\pi})^{3}\left[\frac{2 f_{\pi} y_{\sigma}^{+-}(f_{\pi})^{2}}
%	{M_{+}(f_{\pi}) + M_{-}(f_{\pi})} - y_{\sigma}^{--}(f_{\pi})\right] \bigg\rbrace \bigg)
%	\left(\sigma^{2} - f_{\pi}^{2}\right)^{2} \nonumber\\[0.2cm]
%	& & + \sum_{n\, =\, 3}^{4} \frac{\alpha_{n}}{2^{n} n!} 
%	\left(\sigma^{2} - f_{\pi}^{2}\right)^{n}
%	- m_{\pi}^{2}f_{\pi} (\sigma - f_{\pi})
%	- \frac{1}{2}\left(m_{v}^{2}v^{2} + m_{w}^{2}w^{2}\right) \nonumber\\[0.2cm]
%	& & - \sum_{i\, =\, \pm} \frac{M_{i}(\sigma)^{4}}{4\pi^{2}} 
%	\ln\frac{M_{i}(\sigma)}{M_{+}(f_{\pi})}
%	+ \frac{M_{-}(f_{\pi})^{4}}{4\pi^{2}} \ln\frac{M_{-}(f_{\pi})}{M_{+}(f_{\pi})},
%\end{IEEEeqnarray}
\begin{IEEEeqnarray}{rCl}\label{Ubosons}
	U & = & \left\lbrace\frac{m_{\pi}^{2}}{2} 
	+ \frac{1}{8\pi^{2}f_{\pi}}\left[ M_{+}^{3} \,y_{+}
	+ \left(1 + 4 \ln\frac{M_{-} }{M_{+} }\right)
	M_{-}^{3}\, y_{-}\right]\right\rbrace
	\left(\sigma^{2} - f_{\pi}^{2}\right) \nonumber\\[0.2cm]
	& & +\, \bigg\lbrace\frac{m_{\sigma}^{2} - m_{\pi}^{2}}{8 f_{\pi}^{2}} 
	+ \frac{1}{32\pi^{2}f_{\pi}^{3}}\, \bigg[ 7 f_{\pi} M_{+}^{2}\, y_{+}^{2}
	+ f_{\pi} \left(7 + 12 \ln\frac{M_{-}}{M_{+}}\right)
	M_{-}^{2}  \, y_{-}^{2}  +\, M_{+}^{3}\left(\frac{2 f_{\pi} y_{+-}^{2}}
	{M_{+} + M_{-}} - y_{+}\right) \nonumber\\[0.2cm] 
	& &\qquad \qquad\hspace{4cm} + \left(1 + 4 \ln \frac{M_{-}}{M_{+}}\right) 
	M_{-}^{3}\left(\frac{2 f_{\pi} \,y_{+-}^{2}}
	{M_{+}+ M_{-}} - y_{-}\right) \bigg] \bigg\rbrace
	\left(\sigma^{2} - f_{\pi}^{2}\right)^{2} \nonumber\\[0.2cm]
	& & + \sum_{n\, =\, 3}^{4} \frac{\alpha_{n}}{2^{n} n!} 
	\left(\sigma^{2} - f_{\pi}^{2}\right)^{n}
	- m_{\pi}^{2}f_{\pi} (\sigma - f_{\pi})
	- \frac{1}{2}m_{\omega}^{2}\omega^{2}\ \nonumber\\[0.2cm]
	& & - \sum_{i\, =\, \pm} \frac{M_{i}(\sigma)^{4}}{4\pi^{2}} 
	\ln\frac{M_{i}(\sigma)}{M_{+}}
	+ \frac{M_{-}^{4}}{4\pi^{2}} \ln\frac{M_{-}}{M_{+}},
\end{IEEEeqnarray}
\end{widetext}
where the quantities $M_+, M_-, y_+, y_- $, and $y_{+-}$ are evaluated at $\sigma=f_\pi$ (e.g.\ $M_+=M_+(f_\pi)=M_N$). 
We have used the relation
\begin{equation}\label{eq:d2M+}
	\frac{\mathrm{d}^{2}M_{\pm}(\sigma)}{\mathrm{d}\sigma^{2}} = 
	\frac{2 y_{+- }^{2}}{M_{+} + M_{-}},
\end{equation}
with
\begin{equation}
	y_{+-} = \frac{m_{0}(y_{a} + y_{b})}
	{\sqrt{\rule[0ex]{0ex}{2ex}\sigma^{2}(y_{a} + y_{b})^{2} + 4 m_{0}^{2}}}.
\end{equation}

\begin{figure}
	\centering
	\includegraphics[scale=1.0]{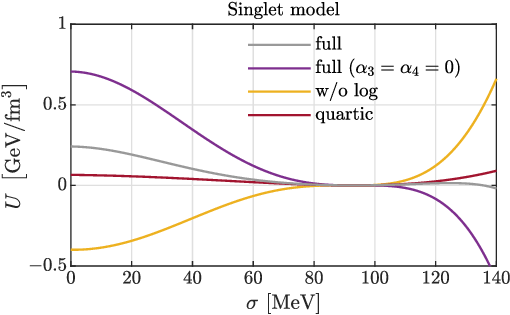}
	\caption{Bosonic potential in vacuum $\Omega = U$ showing the magnitude of
	its various contributions. Orange: the potential $V(\sigma) - h(\sigma - f_{\pi})$,
	without the fermion loop contribution; magenta: the potential including the renormalised
	fermion loop, but with $\alpha_{3} = \alpha_{4} = 0$; gray: the full potential $U(\sigma)$;
	dark red: the quartic approximation to $V(\sigma) - h(\sigma - f_{\pi})$.}
	\label{fig:UB_WT}
\end{figure}
\begin{figure}
	\centering
	\includegraphics[scale=1.0]{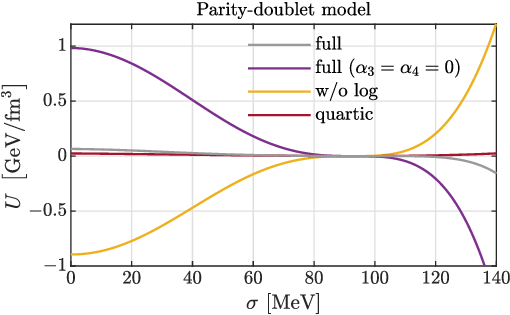}
	\caption{Same as Fig.~\ref{fig:UB_WT} for the parity-doublet model.}
	\label{fig:UB_PD}
\end{figure}

It is interesting to note that, in the parity-doublet model,  the potential logarithmic contributions in $\ln\sigma$ cancel out near $\sigma=0$. In the vicinity of $\sigma=0$, the next-to-last term in Eq.~(\ref{Ubosons}) can  be expanded as a polynomial in $\sigma^2$ with coefficients that depend on $y_a$, $y_b$, $m_0$, and, up to order $\sigma^4$, on $\ln(m_0/M_N)$. Such a cancellation is specific to the parity-doublet model and it does not take place in the singlet model. In both models though, the contribution of $\Omega_{\rm vacuum}$ to the second and fourth derivatives of $U(\sigma=0)$ are large and negative and nearly cancel those of $V(\varphi)$. To appreciate the magnitude of these cancellations, we have plotted various contributions to $U(\sigma)$ in Figs.~\ref{fig:UB_WT} and \ref{fig:UB_PD}, for the singlet and doublet models, respectively. It can be seen that the subtractions, which are mostly constrained by the physics near $\sigma=f_\pi$, affect also significantly the region near $\sigma=0$, hence impacting the properties of the chiral transition. 

At large values of $\sigma$, the logarithmic contribution makes $U$ unbounded from below, a well known feature of the one-loop contribution \cite{Skokov:2010sf}. This phenomenon is delayed to larger values of sigma by the contributions of order $(\sigma^2-f_\pi^2)^n$ in $V(\varphi)$, with $n=3,4$. With the present parameters, the second derivative of $U(\sigma)$ with respect to $\sigma$ remains positive until $\sigma\gtrsim 110$ MeV. We could extend somewhat this region by including higher-order terms in $V(\varphi)$ \cite{minamikawa_chiral_2023}. However, since most of the physics discussed in this paper concerns the range of sigma values between zero and $f_\pi$, we do not find it necessary. 
\begin{figure}[t]
	\centering
	\includegraphics[scale=1.0]{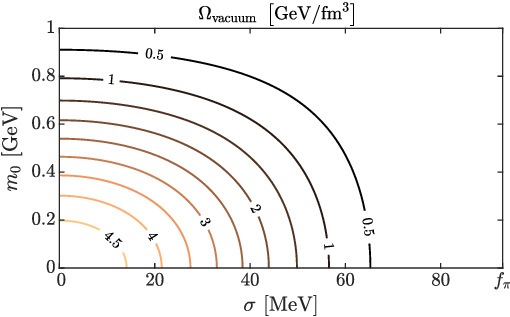}
	\caption{Contour plot of the vacuum contribution $\Omega_{\mathrm{vacuum}}$ to the 
	thermodynamic potential of the parity-doublet model.}
	\label{fig:omega_vacuum}
\end{figure}
The vacuum contribution $\Omega_{\mathrm{vacuum}}$ is
depicted in Fig.~\ref{fig:omega_vacuum} as a function of $\sigma$ and $m_{0}$. This plot shows a large and positive contribution at small values of $\sigma$, the larger the smaller the mass $m_0$. This explains why small values of $m_0$ are not favored for the description of nuclear matter properties, since in the relevant range of $\sigma$ values ($\sigma\sim 60\ \mathrm{MeV}$), this would entail cancellations of contributions of order 1 GeV. In contrast, for $m_0\simeq 800$ MeV, $\Omega_{\rm vacuum}$ is of the order of a few hundreds of MeV for the relevant values of $\sigma$. 

%\begin{figure}[t]
%	\centering
%	\includegraphics[scale=0.7
%	]{potential1.pdf}
%	\caption{Bosonic potential in vacuum 
%	$\Omega \equiv U_{\mathrm{B}}$ showing the maginude of its various contributions. Red: the potential $V(\sigma)-h(\sigma-f_\pi)$, without the fermion loop contribution. Green: the potential including the renormalised fermion loop, but with $\alpha_3=\alpha_4=0$. Blue: the full potential $U(\sigma)$. Orange: the quartic approximation to $V(\sigma)-h(\sigma-f_\pi)$. }
%\end{figure}
%
%\begin{figure}[t]
%	\centering
%	\includegraphics[scale=0.7]{potential_doublet.pdf}
%	\caption{Same as previous figure for the parity-doublet model. }
%\end{figure}

\section{Changing partner}

\begin{figure}[h]
	\centering
	\includegraphics[scale=1.0]{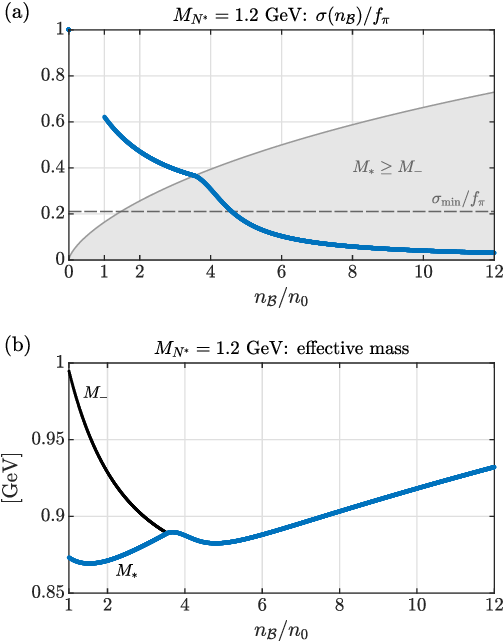}
	\caption{Showing the effect of a reduced mass for the parity partner.}
	\label{fig:M1200}
\end{figure}

There are arguments suggesting that the identification of the $N^*$ as the chiral partner of the nucleon may not be the optimum choice. Hence the authors of Ref.~\cite{Zschiesche:2006zj} considered the possibility of a partner with mass 1200 MeV. They obtained a much lower value for the chiral transition density. We can explain simply this result. To do so consider the expression (\ref{eq:threshold2}) which relates the density to the sigma field at the $\mathcal{B}^-$ onset. Recalling that the Yukawa couplings are fixed by the masses of the parity partners, we can express this relation as a function of $\Delta M=(M_{N^*}-M_N)/2$ and $\bar M=(M_{N^*}+M_N)/2$. One gets, with $\bar\sigma=\sigma/f_\pi$, 
\beq p_{+}^2=4 \bar\sigma\Delta M m_0\sqrt{1+\bar\sigma^2(\bar M^2/m_0^2-1)}.\eeq
As we see from this expression, $p_{+}^2$ is proportional to $\Delta M$. Changing the mass of the partner from $1510\ \mathrm{MeV}$ to $1200\ \mathrm{MeV}$ means $\Delta M$ going from $571\ \mathrm{MeV}$ to $261\ \mathrm{MeV}$. This gives almost a factor 2, which translates by nearly a factor 3 in the density. 
%\begin{figure}[t]
%	\centering
%	\includegraphics[scale=0.5]{M1200_effective_mass}\\
%	\includegraphics[scale=0.40]{M1200_sigma}
%	\caption{ Showing the effect of a reduced mass for the parity partner}
%	\label{fig:M1200}
%\end{figure}
At the same time, it can be verified that $\sigma_{\rm min}$ does not change much. We have
\begin{equation}
    \frac{\sigma_{\rm min}^2}{f_\pi^2} = \frac{m_{0}^{2} (m_{1} - m_{2})^{2}}{(m_{1}m_{2} - m_{0}^{2}) [(m_{1} + m_{2})^{2} - 4 m_{0}^{2}]} ,
\end{equation}
% \beq
% &&\frac{\sigma_{\rm min}^2}{f_\pi^2}=\nn &&\frac{{f_\pi}^2 {m_0}^2 ({m_1}-{m_2})^2}{4 {m_0}^4-{m_0}^2
%    \left({m_1}^2+6 {m_1} {m_2}+{m_2}^2\right)+{m_1} {m_2}
%    ({m_1}+{m_2})^2}\nn
% \eeq
where $m_1=M_N$ and $m_2$ is the mass of the parity partner. It can be verified indeed that $\sigma_{\rm min}$ varies very little in the range of variation of $m_2$ (and beyond), going from $0.28 f_{\pi}$ to $0.21 f_\pi$ as $m_2$ goes from $1510\ \mathrm{MeV}$ to $1200\ \mathrm{MeV}$ (keeping $m_0=800$ MeV). It follows from our estimate of the lower bound given earlier that the value of the sigma field at the $\mathcal{B}^-$ threshold does not change much and that we may expect a factor 3 reduction in the value of the critical density. This is confirmed by Fig.~\ref{fig:M1200}.

Changing the mass of the partner has therefore an important impact on the value of the chiral transition density. Another parameter that can potentially impact this density is the chiral-invariant mass $m_0$. However, the effect is much more modest. Choosing $m_0=500$ MeV for instance leaves the chiral transition density in the vicinity of $10 n_0$. 

%\subsubsection{The dependence on $m_0$}

%One needs to think about the dependence on $m_0$. \jpb{This could be mentioned in words, without a detailed study.}

%\begin{figure}[t]
%	\centering
%	\includegraphics[scale=0.5]{doublet_model_m0_500_1}\\
%	\includegraphics[scale=0.50]{doublet_model_m0_500_2}
%	\caption{ Showing the effect of a reduced $m_0$}
%	\label{fig:M1200}
%\end{figure}

%\subsubsection{Exploring the phase diagram at finite $T$ and finite $\mu_{\mathcal{B}}$}

%Perhaps one or two calculations at finite $T$ and $\mu_{\mathcal{B}}$. 

%\afterpage{\clearpage}

\bibliography{references}

\begin{thebibliography}{74}%
\makeatletter
\providecommand \@ifxundefined [1]{%
 \@ifx{#1\undefined}
}%
\providecommand \@ifnum [1]{%
 \ifnum #1\expandafter \@firstoftwo
 \else \expandafter \@secondoftwo
 \fi
}%
\providecommand \@ifx [1]{%
 \ifx #1\expandafter \@firstoftwo
 \else \expandafter \@secondoftwo
 \fi
}%
\providecommand \natexlab [1]{#1}%
\providecommand \enquote  [1]{``#1''}%
\providecommand \bibnamefont  [1]{#1}%
\providecommand \bibfnamefont [1]{#1}%
\providecommand \citenamefont [1]{#1}%
\providecommand \href@noop [0]{\@secondoftwo}%
\providecommand \href [0]{\begingroup \@sanitize@url \@href}%
\providecommand \@href[1]{\@@startlink{#1}\@@href}%
\providecommand \@@href[1]{\endgroup#1\@@endlink}%
\providecommand \@sanitize@url [0]{\catcode `\\12\catcode `\$12\catcode
  `\&12\catcode `\#12\catcode `\^12\catcode `\_12\catcode `\%12\relax}%
\providecommand \@@startlink[1]{}%
\providecommand \@@endlink[0]{}%
\providecommand \url  [0]{\begingroup\@sanitize@url \@url }%
\providecommand \@url [1]{\endgroup\@href {#1}{\urlprefix }}%
\providecommand \urlprefix  [0]{URL }%
\providecommand \Eprint [0]{\href }%
\providecommand \doibase [0]{http://dx.doi.org/}%
\providecommand \selectlanguage [0]{\@gobble}%
\providecommand \bibinfo  [0]{\@secondoftwo}%
\providecommand \bibfield  [0]{\@secondoftwo}%
\providecommand \translation [1]{[#1]}%
\providecommand \BibitemOpen [0]{}%
\providecommand \bibitemStop [0]{}%
\providecommand \bibitemNoStop [0]{.\EOS\space}%
\providecommand \EOS [0]{\spacefactor3000\relax}%
\providecommand \BibitemShut  [1]{\csname bibitem#1\endcsname}%
\let\auto@bib@innerbib\@empty
%</preamble>
\bibitem [{\citenamefont {Baym}\ \emph {et~al.}(2018)\citenamefont {Baym},
  \citenamefont {Hatsuda}, \citenamefont {Kojo}, \citenamefont {Powell},
  \citenamefont {Song},\ and\ \citenamefont {Takatsuka}}]{Baym:2017whm}%
  \BibitemOpen
  \bibfield  {author} {\bibinfo {author} {\bibfnamefont {G.}~\bibnamefont
  {Baym}}, \bibinfo {author} {\bibfnamefont {T.}~\bibnamefont {Hatsuda}},
  \bibinfo {author} {\bibfnamefont {T.}~\bibnamefont {Kojo}}, \bibinfo {author}
  {\bibfnamefont {P.~D.}\ \bibnamefont {Powell}}, \bibinfo {author}
  {\bibfnamefont {Y.}~\bibnamefont {Song}}, \ and\ \bibinfo {author}
  {\bibfnamefont {T.}~\bibnamefont {Takatsuka}},\ }\href {\doibase
  10.1088/1361-6633/aaae14} {\bibfield  {journal} {\bibinfo  {journal} {Rept.
  Prog. Phys.}\ }\textbf {\bibinfo {volume} {81}},\ \bibinfo {pages} {056902}
  (\bibinfo {year} {2018})},\ \Eprint {http://arxiv.org/abs/1707.04966}
  {arXiv:1707.04966 [astro-ph.HE]} \BibitemShut {NoStop}%
\bibitem [{\citenamefont {DeTar}\ and\ \citenamefont
  {Kunihiro}(1989)}]{detar_linear_1989}%
  \BibitemOpen
  \bibfield  {author} {\bibinfo {author} {\bibfnamefont {C.}~\bibnamefont
  {DeTar}}\ and\ \bibinfo {author} {\bibfnamefont {T.}~\bibnamefont
  {Kunihiro}},\ }\href {\doibase 10.1103/PhysRevD.39.2805} {\bibfield
  {journal} {\bibinfo  {journal} {Physical Review D}\ }\textbf {\bibinfo
  {volume} {39}},\ \bibinfo {pages} {2805} (\bibinfo {year}
  {1989})}\BibitemShut {NoStop}%
\bibitem [{\citenamefont {Datta}\ \emph {et~al.}(2013)\citenamefont {Datta},
  \citenamefont {Gupta}, \citenamefont {Padmanath}, \citenamefont {Mathur},\
  and\ \citenamefont {Maiti}}]{datta_nucleons_2013}%
  \BibitemOpen
  \bibfield  {author} {\bibinfo {author} {\bibfnamefont {S.}~\bibnamefont
  {Datta}}, \bibinfo {author} {\bibfnamefont {S.}~\bibnamefont {Gupta}},
  \bibinfo {author} {\bibfnamefont {M.}~\bibnamefont {Padmanath}}, \bibinfo
  {author} {\bibfnamefont {N.}~\bibnamefont {Mathur}}, \ and\ \bibinfo {author}
  {\bibfnamefont {J.}~\bibnamefont {Maiti}},\ }\href {\doibase
  10.1007/JHEP02(2013)145} {\bibfield  {journal} {\bibinfo  {journal} {Journal
  of High Energy Physics}\ }\textbf {\bibinfo {volume} {2013}},\ \bibinfo
  {pages} {145} (\bibinfo {year} {2013})},\ \bibinfo {note} {arXiv:1212.2927
  [hep-lat, physics:hep-ph, physics:nucl-ex, physics:nucl-th]}\BibitemShut
  {NoStop}%
\bibitem [{\citenamefont {Aarts}\ \emph {et~al.}(2017)\citenamefont {Aarts},
  \citenamefont {Allton}, \citenamefont {De~Boni}, \citenamefont {Hands},
  \citenamefont {J{\"a}ger}, \citenamefont {Praki},\ and\ \citenamefont
  {Skullerud}}]{aarts_light_2017}%
  \BibitemOpen
  \bibfield  {author} {\bibinfo {author} {\bibfnamefont {G.}~\bibnamefont
  {Aarts}}, \bibinfo {author} {\bibfnamefont {C.}~\bibnamefont {Allton}},
  \bibinfo {author} {\bibfnamefont {D.}~\bibnamefont {De~Boni}}, \bibinfo
  {author} {\bibfnamefont {S.}~\bibnamefont {Hands}}, \bibinfo {author}
  {\bibfnamefont {B.}~\bibnamefont {J{\"a}ger}}, \bibinfo {author}
  {\bibfnamefont {C.}~\bibnamefont {Praki}}, \ and\ \bibinfo {author}
  {\bibfnamefont {J.-I.}\ \bibnamefont {Skullerud}},\ }\href {\doibase
  10.1007/JHEP06(2017)034} {\bibfield  {journal} {\bibinfo  {journal} {Journal
  of High Energy Physics}\ }\textbf {\bibinfo {volume} {2017}},\ \bibinfo
  {pages} {34} (\bibinfo {year} {2017})},\ \bibinfo {note} {arXiv:1703.09246
  [hep-lat, physics:hep-ph, physics:nucl-th]}\BibitemShut {NoStop}%
\bibitem [{\citenamefont {Jido}\ \emph {et~al.}(2000)\citenamefont {Jido},
  \citenamefont {Nemoto}, \citenamefont {Oka},\ and\ \citenamefont
  {Hosaka}}]{Jido_2000}%
  \BibitemOpen
  \bibfield  {author} {\bibinfo {author} {\bibfnamefont {D.}~\bibnamefont
  {Jido}}, \bibinfo {author} {\bibfnamefont {Y.}~\bibnamefont {Nemoto}},
  \bibinfo {author} {\bibfnamefont {M.}~\bibnamefont {Oka}}, \ and\ \bibinfo
  {author} {\bibfnamefont {A.}~\bibnamefont {Hosaka}},\ }\href {\doibase
  10.1016/s0375-9474(99)00844-1} {\bibfield  {journal} {\bibinfo  {journal}
  {Nuclear Physics A}\ }\textbf {\bibinfo {volume} {671}},\ \bibinfo {pages}
  {471} (\bibinfo {year} {2000})}\BibitemShut {NoStop}%
\bibitem [{\citenamefont {Jido}\ \emph {et~al.}(2001)\citenamefont {Jido},
  \citenamefont {Oka},\ and\ \citenamefont {Hosaka}}]{jido_chiral_2001}%
  \BibitemOpen
  \bibfield  {author} {\bibinfo {author} {\bibfnamefont {D.}~\bibnamefont
  {Jido}}, \bibinfo {author} {\bibfnamefont {M.}~\bibnamefont {Oka}}, \ and\
  \bibinfo {author} {\bibfnamefont {A.}~\bibnamefont {Hosaka}},\ }\href
  {\doibase 10.1143/PTP.106.873} {\bibfield  {journal} {\bibinfo  {journal}
  {Progress of Theoretical Physics}\ }\textbf {\bibinfo {volume} {106}},\
  \bibinfo {pages} {873} (\bibinfo {year} {2001})},\ \bibinfo {note}
  {arXiv:hep-ph/0110005}\BibitemShut {NoStop}%
\bibitem [{\citenamefont {Zschiesche}\ \emph {et~al.}(2007)\citenamefont
  {Zschiesche}, \citenamefont {Tolos}, \citenamefont {Schaffner-Bielich},\ and\
  \citenamefont {Pisarski}}]{Zschiesche:2006zj}%
  \BibitemOpen
  \bibfield  {author} {\bibinfo {author} {\bibfnamefont {D.}~\bibnamefont
  {Zschiesche}}, \bibinfo {author} {\bibfnamefont {L.}~\bibnamefont {Tolos}},
  \bibinfo {author} {\bibfnamefont {J.}~\bibnamefont {Schaffner-Bielich}}, \
  and\ \bibinfo {author} {\bibfnamefont {R.~D.}\ \bibnamefont {Pisarski}},\
  }\href {\doibase 10.1103/PhysRevC.75.055202} {\bibfield  {journal} {\bibinfo
  {journal} {Phys. Rev. C}\ }\textbf {\bibinfo {volume} {75}},\ \bibinfo
  {pages} {055202} (\bibinfo {year} {2007})},\ \Eprint
  {http://arxiv.org/abs/nucl-th/0608044} {arXiv:nucl-th/0608044} \BibitemShut
  {NoStop}%
\bibitem [{\citenamefont {Walecka}(1974)}]{WALECKA1974491}%
  \BibitemOpen
  \bibfield  {author} {\bibinfo {author} {\bibfnamefont {J.}~\bibnamefont
  {Walecka}},\ }\href {\doibase https://doi.org/10.1016/0003-4916(74)90208-5}
  {\bibfield  {journal} {\bibinfo  {journal} {Annals of Physics}\ }\textbf
  {\bibinfo {volume} {83}},\ \bibinfo {pages} {491} (\bibinfo {year}
  {1974})}\BibitemShut {NoStop}%
\bibitem [{\citenamefont {Eser}\ and\ \citenamefont
  {Blaizot}(2022)}]{Eser:2021ivo}%
  \BibitemOpen
  \bibfield  {author} {\bibinfo {author} {\bibfnamefont {J.}~\bibnamefont
  {Eser}}\ and\ \bibinfo {author} {\bibfnamefont {J.-P.}\ \bibnamefont
  {Blaizot}},\ }\href {\doibase 10.1103/PhysRevD.105.074031} {\bibfield
  {journal} {\bibinfo  {journal} {Phys. Rev. D}\ }\textbf {\bibinfo {volume}
  {105}},\ \bibinfo {pages} {074031} (\bibinfo {year} {2022})},\ \Eprint
  {http://arxiv.org/abs/2112.14579} {arXiv:2112.14579 [hep-ph]} \BibitemShut
  {NoStop}%
\bibitem [{\citenamefont {Hatsuda}\ and\ \citenamefont
  {Prakash}(1989)}]{hatsuda_parity_1989}%
  \BibitemOpen
  \bibfield  {author} {\bibinfo {author} {\bibfnamefont {T.}~\bibnamefont
  {Hatsuda}}\ and\ \bibinfo {author} {\bibfnamefont {M.}~\bibnamefont
  {Prakash}},\ }\href {\doibase 10.1016/0370-2693(89)91040-X} {\bibfield
  {journal} {\bibinfo  {journal} {Physics Letters B}\ }\textbf {\bibinfo
  {volume} {224}},\ \bibinfo {pages} {11} (\bibinfo {year} {1989})}\BibitemShut
  {NoStop}%
\bibitem [{\citenamefont {Dexheimer}\ \emph {et~al.}(2008)\citenamefont
  {Dexheimer}, \citenamefont {Schramm},\ and\ \citenamefont
  {Zschiesche}}]{dexheimer_nuclear_2008}%
  \BibitemOpen
  \bibfield  {author} {\bibinfo {author} {\bibfnamefont {V.}~\bibnamefont
  {Dexheimer}}, \bibinfo {author} {\bibfnamefont {S.}~\bibnamefont {Schramm}},
  \ and\ \bibinfo {author} {\bibfnamefont {D.}~\bibnamefont {Zschiesche}},\
  }\href {\doibase 10.1103/PhysRevC.77.025803} {\bibfield  {journal} {\bibinfo
  {journal} {Phys. Rev. C}\ }\textbf {\bibinfo {volume} {77}},\ \bibinfo
  {pages} {025803} (\bibinfo {year} {2008})},\ \bibinfo {note} {arXiv:0710.4192
  [nucl-th]}\BibitemShut {NoStop}%
\bibitem [{\citenamefont {Weyrich}\ \emph {et~al.}(2015)\citenamefont
  {Weyrich}, \citenamefont {Strodthoff},\ and\ \citenamefont {von
  Smekal}}]{Weyrich_2015}%
  \BibitemOpen
  \bibfield  {author} {\bibinfo {author} {\bibfnamefont {J.}~\bibnamefont
  {Weyrich}}, \bibinfo {author} {\bibfnamefont {N.}~\bibnamefont {Strodthoff}},
  \ and\ \bibinfo {author} {\bibfnamefont {L.}~\bibnamefont {von Smekal}},\
  }\href {\doibase 10.1103/physrevc.92.015214} {\bibfield  {journal} {\bibinfo
  {journal} {Physical Review C}\ }\textbf {\bibinfo {volume} {92}} (\bibinfo
  {year} {2015}),\ 10.1103/physrevc.92.015214}\BibitemShut {NoStop}%
\bibitem [{\citenamefont {Marczenko}\ \emph {et~al.}(2018)\citenamefont
  {Marczenko}, \citenamefont {Blaschke}, \citenamefont {Redlich},\ and\
  \citenamefont {Sasaki}}]{marczenko_chiral_2018}%
  \BibitemOpen
  \bibfield  {author} {\bibinfo {author} {\bibfnamefont {M.}~\bibnamefont
  {Marczenko}}, \bibinfo {author} {\bibfnamefont {D.}~\bibnamefont {Blaschke}},
  \bibinfo {author} {\bibfnamefont {K.}~\bibnamefont {Redlich}}, \ and\
  \bibinfo {author} {\bibfnamefont {C.}~\bibnamefont {Sasaki}},\ }\href
  {\doibase 10.1103/PhysRevD.98.103021} {\bibfield  {journal} {\bibinfo
  {journal} {Phys. Rev. D}\ }\textbf {\bibinfo {volume} {98}},\ \bibinfo
  {pages} {103021} (\bibinfo {year} {2018})},\ \bibinfo {note}
  {arXiv:1805.06886 [astro-ph, physics:hep-ph, physics:nucl-th]}\BibitemShut
  {NoStop}%
\bibitem [{\citenamefont {Yamazaki}\ and\ \citenamefont
  {Harada}(2019)}]{yamazaki_constraint_2019}%
  \BibitemOpen
  \bibfield  {author} {\bibinfo {author} {\bibfnamefont {T.}~\bibnamefont
  {Yamazaki}}\ and\ \bibinfo {author} {\bibfnamefont {M.}~\bibnamefont
  {Harada}},\ }\href {\doibase 10.1103/PhysRevC.100.025205} {\bibfield
  {journal} {\bibinfo  {journal} {Phys. Rev. C}\ }\textbf {\bibinfo {volume}
  {100}},\ \bibinfo {pages} {025205} (\bibinfo {year} {2019})},\ \bibinfo
  {note} {arXiv:1901.02167 [hep-ph, physics:nucl-th]}\BibitemShut {NoStop}%
\bibitem [{\citenamefont {Minamikawa}\ \emph {et~al.}(2021)\citenamefont
  {Minamikawa}, \citenamefont {Kojo},\ and\ \citenamefont
  {Harada}}]{minamikawa_quark-hadron_2021}%
  \BibitemOpen
  \bibfield  {author} {\bibinfo {author} {\bibfnamefont {T.}~\bibnamefont
  {Minamikawa}}, \bibinfo {author} {\bibfnamefont {T.}~\bibnamefont {Kojo}}, \
  and\ \bibinfo {author} {\bibfnamefont {M.}~\bibnamefont {Harada}},\ }\href
  {\doibase 10.1103/PhysRevC.103.045205} {\bibfield  {journal} {\bibinfo
  {journal} {Phys. Rev. C}\ }\textbf {\bibinfo {volume} {103}},\ \bibinfo
  {pages} {045205} (\bibinfo {year} {2021})},\ \bibinfo {note}
  {arXiv:2011.13684 [astro-ph, physics:hep-ph, physics:nucl-th]}\BibitemShut
  {NoStop}%
\bibitem [{\citenamefont {Marczenko}\ \emph
  {et~al.}(2022{\natexlab{a}})\citenamefont {Marczenko}, \citenamefont
  {Redlich},\ and\ \citenamefont {Sasaki}}]{marczenko_chiral_2022}%
  \BibitemOpen
  \bibfield  {author} {\bibinfo {author} {\bibfnamefont {M.}~\bibnamefont
  {Marczenko}}, \bibinfo {author} {\bibfnamefont {K.}~\bibnamefont {Redlich}},
  \ and\ \bibinfo {author} {\bibfnamefont {C.}~\bibnamefont {Sasaki}},\ }\href
  {\doibase 10.1103/PhysRevD.105.103009} {\bibfield  {journal} {\bibinfo
  {journal} {Physical Review D}\ }\textbf {\bibinfo {volume} {105}},\ \bibinfo
  {pages} {103009} (\bibinfo {year} {2022}{\natexlab{a}})},\ \bibinfo {note}
  {arXiv:2203.00269 [astro-ph, physics:nucl-th]}\BibitemShut {NoStop}%
\bibitem [{\citenamefont {Minamikawa}\ \emph
  {et~al.}(2023{\natexlab{a}})\citenamefont {Minamikawa}, \citenamefont {Gao},
  \citenamefont {kojo},\ and\ \citenamefont {Harada}}]{minamikawa_chiral_2023}%
  \BibitemOpen
  \bibfield  {author} {\bibinfo {author} {\bibfnamefont {T.}~\bibnamefont
  {Minamikawa}}, \bibinfo {author} {\bibfnamefont {B.}~\bibnamefont {Gao}},
  \bibinfo {author} {\bibfnamefont {T.}~\bibnamefont {kojo}}, \ and\ \bibinfo
  {author} {\bibfnamefont {M.}~\bibnamefont {Harada}},\ }\href
  {http://arxiv.org/abs/2302.00825} {\enquote {\bibinfo {title} {Chiral
  restoration of nucleons in neutron star matter: studies based on a parity
  doublet model},}\ } (\bibinfo {year} {2023}{\natexlab{a}}),\ \bibinfo {note}
  {arXiv:2302.00825 [nucl-th]}\BibitemShut {NoStop}%
\bibitem [{\citenamefont {Kong}\ \emph {et~al.}(2023)\citenamefont {Kong},
  \citenamefont {Minamikawa},\ and\ \citenamefont {Harada}}]{kong2023study}%
  \BibitemOpen
  \bibfield  {author} {\bibinfo {author} {\bibfnamefont {Y.~K.}\ \bibnamefont
  {Kong}}, \bibinfo {author} {\bibfnamefont {T.}~\bibnamefont {Minamikawa}}, \
  and\ \bibinfo {author} {\bibfnamefont {M.}~\bibnamefont {Harada}},\
  }\href@noop {} {\enquote {\bibinfo {title} {A study of neutron star matter
  based on a parity doublet model with $a_0(980)$ meson effect},}\ } (\bibinfo
  {year} {2023}),\ \Eprint {http://arxiv.org/abs/2306.08140} {arXiv:2306.08140
  [nucl-th]} \BibitemShut {NoStop}%
\bibitem [{\citenamefont {Fraga}\ \emph {et~al.}(2023)\citenamefont {Fraga},
  \citenamefont {da~Mata},\ and\ \citenamefont
  {Schaffner-Bielich}}]{Fraga:2023wtd}%
  \BibitemOpen
  \bibfield  {author} {\bibinfo {author} {\bibfnamefont {E.~S.}\ \bibnamefont
  {Fraga}}, \bibinfo {author} {\bibfnamefont {R.}~\bibnamefont {da~Mata}}, \
  and\ \bibinfo {author} {\bibfnamefont {J.}~\bibnamefont
  {Schaffner-Bielich}},\ }\href@noop {} {\  (\bibinfo {year} {2023})},\ \Eprint
  {http://arxiv.org/abs/2309.02368} {arXiv:2309.02368 [hep-ph]} \BibitemShut
  {NoStop}%
\bibitem [{\citenamefont {Steinheimer}\ \emph {et~al.}(2011)\citenamefont
  {Steinheimer}, \citenamefont {Schramm},\ and\ \citenamefont
  {St{\"o}cker}}]{Steinheimer_2011}%
  \BibitemOpen
  \bibfield  {author} {\bibinfo {author} {\bibfnamefont {J.}~\bibnamefont
  {Steinheimer}}, \bibinfo {author} {\bibfnamefont {S.}~\bibnamefont
  {Schramm}}, \ and\ \bibinfo {author} {\bibfnamefont {H.}~\bibnamefont
  {St{\"o}cker}},\ }\href {\doibase 10.1103/physrevc.84.045208} {\bibfield
  {journal} {\bibinfo  {journal} {Physical Review C}\ }\textbf {\bibinfo
  {volume} {84}} (\bibinfo {year} {2011}),\
  10.1103/physrevc.84.045208}\BibitemShut {NoStop}%
\bibitem [{\citenamefont {Fraga}\ \emph {et~al.}(2022)\citenamefont {Fraga},
  \citenamefont {da~Mata}, \citenamefont {Pitsinigkos},\ and\ \citenamefont
  {Schmitt}}]{fraga2022strange}%
  \BibitemOpen
  \bibfield  {author} {\bibinfo {author} {\bibfnamefont {E.~S.}\ \bibnamefont
  {Fraga}}, \bibinfo {author} {\bibfnamefont {R.}~\bibnamefont {da~Mata}},
  \bibinfo {author} {\bibfnamefont {S.}~\bibnamefont {Pitsinigkos}}, \ and\
  \bibinfo {author} {\bibfnamefont {A.}~\bibnamefont {Schmitt}},\ }\href@noop
  {} {\  (\bibinfo {year} {2022})},\ \Eprint {http://arxiv.org/abs/2206.09219}
  {arXiv:2206.09219 [nucl-th]} \BibitemShut {NoStop}%
\bibitem [{\citenamefont {Minamikawa}\ \emph
  {et~al.}(2023{\natexlab{b}})\citenamefont {Minamikawa}, \citenamefont {Gao},
  \citenamefont {kojo},\ and\ \citenamefont {Harada}}]{minamikawa2023parity}%
  \BibitemOpen
  \bibfield  {author} {\bibinfo {author} {\bibfnamefont {T.}~\bibnamefont
  {Minamikawa}}, \bibinfo {author} {\bibfnamefont {B.}~\bibnamefont {Gao}},
  \bibinfo {author} {\bibfnamefont {T.}~\bibnamefont {kojo}}, \ and\ \bibinfo
  {author} {\bibfnamefont {M.}~\bibnamefont {Harada}},\ }\href@noop {} {\
  (\bibinfo {year} {2023}{\natexlab{b}})},\ \Eprint
  {http://arxiv.org/abs/2306.15564} {arXiv:2306.15564 [hep-ph]} \BibitemShut
  {NoStop}%
\bibitem [{\citenamefont {Marczenko}\ \emph
  {et~al.}(2022{\natexlab{b}})\citenamefont {Marczenko}, \citenamefont
  {Redlich},\ and\ \citenamefont {Sasaki}}]{Marczenko_2022}%
  \BibitemOpen
  \bibfield  {author} {\bibinfo {author} {\bibfnamefont {M.}~\bibnamefont
  {Marczenko}}, \bibinfo {author} {\bibfnamefont {K.}~\bibnamefont {Redlich}},
  \ and\ \bibinfo {author} {\bibfnamefont {C.}~\bibnamefont {Sasaki}},\ }\href
  {\doibase 10.1103/physrevd.105.103009} {\bibfield  {journal} {\bibinfo
  {journal} {Physical Review D}\ }\textbf {\bibinfo {volume} {105}} (\bibinfo
  {year} {2022}{\natexlab{b}}),\ 10.1103/physrevd.105.103009}\BibitemShut
  {NoStop}%
\bibitem [{\citenamefont {Brandes}\ \emph {et~al.}(2021)\citenamefont
  {Brandes}, \citenamefont {Kaiser},\ and\ \citenamefont
  {Weise}}]{brandes_fluctuations_2021}%
  \BibitemOpen
  \bibfield  {author} {\bibinfo {author} {\bibfnamefont {L.}~\bibnamefont
  {Brandes}}, \bibinfo {author} {\bibfnamefont {N.}~\bibnamefont {Kaiser}}, \
  and\ \bibinfo {author} {\bibfnamefont {W.}~\bibnamefont {Weise}},\ }\href
  {\doibase 10.1140/epja/s10050-021-00528-2} {\bibfield  {journal} {\bibinfo
  {journal} {The European Physical Journal A}\ }\textbf {\bibinfo {volume}
  {57}},\ \bibinfo {pages} {243} (\bibinfo {year} {2021})},\ \bibinfo {note}
  {arXiv:2103.06096 [nucl-th]}\BibitemShut {NoStop}%
\bibitem [{\citenamefont {Berges}\ \emph {et~al.}(2003)\citenamefont {Berges},
  \citenamefont {Jungnickel},\ and\ \citenamefont
  {Wetterich}}]{berges_quark_2003}%
  \BibitemOpen
  \bibfield  {author} {\bibinfo {author} {\bibfnamefont {J.}~\bibnamefont
  {Berges}}, \bibinfo {author} {\bibfnamefont {D.-U.}\ \bibnamefont
  {Jungnickel}}, \ and\ \bibinfo {author} {\bibfnamefont {C.}~\bibnamefont
  {Wetterich}},\ }\href {\doibase 10.1142/S0217751X03014034} {\bibfield
  {journal} {\bibinfo  {journal} {Int. J. Mod. Phys. A}\ }\textbf {\bibinfo
  {volume} {18}},\ \bibinfo {pages} {3189} (\bibinfo {year} {2003})},\ \bibinfo
  {note} {arXiv:hep-ph/9811387}\BibitemShut {NoStop}%
\bibitem [{\citenamefont {Floerchinger}\ and\ \citenamefont
  {Wetterich}(2012)}]{Floerchinger_2012}%
  \BibitemOpen
  \bibfield  {author} {\bibinfo {author} {\bibfnamefont {S.}~\bibnamefont
  {Floerchinger}}\ and\ \bibinfo {author} {\bibfnamefont {C.}~\bibnamefont
  {Wetterich}},\ }\href {\doibase 10.1016/j.nuclphysa.2012.07.009} {\bibfield
  {journal} {\bibinfo  {journal} {Nuclear Physics A}\ }\textbf {\bibinfo
  {volume} {890-891}},\ \bibinfo {pages} {11} (\bibinfo {year}
  {2012})}\BibitemShut {NoStop}%
\bibitem [{\citenamefont {Tripolt}\ \emph {et~al.}(2021)\citenamefont
  {Tripolt}, \citenamefont {Jung}, \citenamefont {von Smekal},\ and\
  \citenamefont {Wambach}}]{Tripolt_2021}%
  \BibitemOpen
  \bibfield  {author} {\bibinfo {author} {\bibfnamefont {R.-A.}\ \bibnamefont
  {Tripolt}}, \bibinfo {author} {\bibfnamefont {C.}~\bibnamefont {Jung}},
  \bibinfo {author} {\bibfnamefont {L.}~\bibnamefont {von Smekal}}, \ and\
  \bibinfo {author} {\bibfnamefont {J.}~\bibnamefont {Wambach}},\ }\href
  {\doibase 10.1103/physrevd.104.054005} {\bibfield  {journal} {\bibinfo
  {journal} {Physical Review D}\ }\textbf {\bibinfo {volume} {104}} (\bibinfo
  {year} {2021}),\ 10.1103/physrevd.104.054005}\BibitemShut {NoStop}%
\bibitem [{\citenamefont {Gerber}\ and\ \citenamefont
  {Leutwyler}(1989)}]{gerber_hadrons_1989}%
  \BibitemOpen
  \bibfield  {author} {\bibinfo {author} {\bibfnamefont {P.}~\bibnamefont
  {Gerber}}\ and\ \bibinfo {author} {\bibfnamefont {H.}~\bibnamefont
  {Leutwyler}},\ }\href {\doibase 10.1016/0550-3213(89)90349-0} {\bibfield
  {journal} {\bibinfo  {journal} {Nuclear Physics B}\ }\textbf {\bibinfo
  {volume} {321}},\ \bibinfo {pages} {387} (\bibinfo {year}
  {1989})}\BibitemShut {NoStop}%
\bibitem [{\citenamefont {Brandt}\ \emph {et~al.}(2014)\citenamefont {Brandt},
  \citenamefont {Francis}, \citenamefont {Meyer},\ and\ \citenamefont
  {Robaina}}]{brandt_chiral_2014}%
  \BibitemOpen
  \bibfield  {author} {\bibinfo {author} {\bibfnamefont {B.~B.}\ \bibnamefont
  {Brandt}}, \bibinfo {author} {\bibfnamefont {A.}~\bibnamefont {Francis}},
  \bibinfo {author} {\bibfnamefont {H.~B.}\ \bibnamefont {Meyer}}, \ and\
  \bibinfo {author} {\bibfnamefont {D.}~\bibnamefont {Robaina}},\ }\href
  {\doibase 10.1103/PhysRevD.90.054509} {\bibfield  {journal} {\bibinfo
  {journal} {Physical Review D}\ }\textbf {\bibinfo {volume} {90}},\ \bibinfo
  {pages} {054509} (\bibinfo {year} {2014})},\ \bibinfo {note} {arXiv:1406.5602
  [hep-lat, physics:hep-ph, physics:nucl-th]}\BibitemShut {NoStop}%
\bibitem [{\citenamefont {Andronic}\ \emph {et~al.}(2018)\citenamefont
  {Andronic}, \citenamefont {Braun-Munzinger}, \citenamefont {Redlich},\ and\
  \citenamefont {Stachel}}]{Andronic_2018}%
  \BibitemOpen
  \bibfield  {author} {\bibinfo {author} {\bibfnamefont {A.}~\bibnamefont
  {Andronic}}, \bibinfo {author} {\bibfnamefont {P.}~\bibnamefont
  {Braun-Munzinger}}, \bibinfo {author} {\bibfnamefont {K.}~\bibnamefont
  {Redlich}}, \ and\ \bibinfo {author} {\bibfnamefont {J.}~\bibnamefont
  {Stachel}},\ }\href {\doibase 10.1038/s41586-018-0491-6} {\bibfield
  {journal} {\bibinfo  {journal} {Nature}\ }\textbf {\bibinfo {volume} {561}},\
  \bibinfo {pages} {321} (\bibinfo {year} {2018})}\BibitemShut {NoStop}%
\bibitem [{\citenamefont {Eser}\ and\ \citenamefont {Blaizot}()}]{nextpaper}%
  \BibitemOpen
  \bibfield  {author} {\bibinfo {author} {\bibfnamefont {J.}~\bibnamefont
  {Eser}}\ and\ \bibinfo {author} {\bibfnamefont {J.-P.}\ \bibnamefont
  {Blaizot}},\ }\href@noop {} {\bibinfo  {journal} {In preparation}\
  }\BibitemShut {NoStop}%
\bibitem [{\citenamefont {Workman}\ \emph {et~al.}(2022)\citenamefont {Workman}
  \emph {et~al.}}]{ParticleDataGroup:2022pth}%
  \BibitemOpen
\bibfield  {journal} {  }\bibfield  {author} {\bibinfo {author} {\bibfnamefont
  {R.~L.}\ \bibnamefont {Workman}} \emph {et~al.} (\bibinfo {collaboration}
  {Particle Data Group}),\ }\href {\doibase 10.1093/ptep/ptac097} {\bibfield
  {journal} {\bibinfo  {journal} {PTEP}\ }\textbf {\bibinfo {volume} {2022}},\
  \bibinfo {pages} {083C01} (\bibinfo {year} {2022})}\BibitemShut {NoStop}%
\bibitem [{\citenamefont {Drews}\ \emph {et~al.}(2013)\citenamefont {Drews},
  \citenamefont {Hell}, \citenamefont {Klein},\ and\ \citenamefont
  {Weise}}]{Drews:2013hha}%
  \BibitemOpen
  \bibfield  {author} {\bibinfo {author} {\bibfnamefont {M.}~\bibnamefont
  {Drews}}, \bibinfo {author} {\bibfnamefont {T.}~\bibnamefont {Hell}},
  \bibinfo {author} {\bibfnamefont {B.}~\bibnamefont {Klein}}, \ and\ \bibinfo
  {author} {\bibfnamefont {W.}~\bibnamefont {Weise}},\ }\href {\doibase
  10.1103/PhysRevD.88.096011} {\bibfield  {journal} {\bibinfo  {journal} {Phys.
  Rev. D}\ }\textbf {\bibinfo {volume} {88}},\ \bibinfo {pages} {096011}
  (\bibinfo {year} {2013})},\ \Eprint {http://arxiv.org/abs/1308.5596}
  {arXiv:1308.5596 [hep-ph]} \BibitemShut {NoStop}%
\bibitem [{\citenamefont {Motohiro}\ \emph {et~al.}(2015)\citenamefont
  {Motohiro}, \citenamefont {Kim},\ and\ \citenamefont
  {Harada}}]{Motohiro:2015taa}%
  \BibitemOpen
  \bibfield  {author} {\bibinfo {author} {\bibfnamefont {Y.}~\bibnamefont
  {Motohiro}}, \bibinfo {author} {\bibfnamefont {Y.}~\bibnamefont {Kim}}, \
  and\ \bibinfo {author} {\bibfnamefont {M.}~\bibnamefont {Harada}},\ }\href
  {\doibase 10.1103/PhysRevC.92.025201} {\bibfield  {journal} {\bibinfo
  {journal} {Phys. Rev. C}\ }\textbf {\bibinfo {volume} {92}},\ \bibinfo
  {pages} {025201} (\bibinfo {year} {2015})},\ \bibinfo {note} {[Erratum:
  Phys.Rev.C 95, 059903 (2017)]},\ \Eprint {http://arxiv.org/abs/1505.00988}
  {arXiv:1505.00988 [nucl-th]} \BibitemShut {NoStop}%
\bibitem [{\citenamefont {Gallas}\ \emph {et~al.}(2010)\citenamefont {Gallas},
  \citenamefont {Giacosa},\ and\ \citenamefont {Rischke}}]{Gallas:2009qp}%
  \BibitemOpen
  \bibfield  {author} {\bibinfo {author} {\bibfnamefont {S.}~\bibnamefont
  {Gallas}}, \bibinfo {author} {\bibfnamefont {F.}~\bibnamefont {Giacosa}}, \
  and\ \bibinfo {author} {\bibfnamefont {D.~H.}\ \bibnamefont {Rischke}},\
  }\href {\doibase 10.1103/PhysRevD.82.014004} {\bibfield  {journal} {\bibinfo
  {journal} {Phys. Rev. D}\ }\textbf {\bibinfo {volume} {82}},\ \bibinfo
  {pages} {014004} (\bibinfo {year} {2010})},\ \Eprint
  {http://arxiv.org/abs/0907.5084} {arXiv:0907.5084 [hep-ph]} \BibitemShut
  {NoStop}%
\bibitem [{\citenamefont {Gallas}\ and\ \citenamefont
  {Giacosa}(2014)}]{Gallas:2013ipa}%
  \BibitemOpen
  \bibfield  {author} {\bibinfo {author} {\bibfnamefont {S.}~\bibnamefont
  {Gallas}}\ and\ \bibinfo {author} {\bibfnamefont {F.}~\bibnamefont
  {Giacosa}},\ }\href {\doibase 10.1142/S0217751X14500985} {\bibfield
  {journal} {\bibinfo  {journal} {Int. J. Mod. Phys. A}\ }\textbf {\bibinfo
  {volume} {29}},\ \bibinfo {pages} {1450098} (\bibinfo {year} {2014})},\
  \Eprint {http://arxiv.org/abs/1308.4817} {arXiv:1308.4817 [hep-ph]}
  \BibitemShut {NoStop}%
\bibitem [{\citenamefont {Drischler}\ \emph {et~al.}(2016)\citenamefont
  {Drischler}, \citenamefont {Hebeler},\ and\ \citenamefont
  {Schwenk}}]{Drischler:2015eba}%
  \BibitemOpen
  \bibfield  {author} {\bibinfo {author} {\bibfnamefont {C.}~\bibnamefont
  {Drischler}}, \bibinfo {author} {\bibfnamefont {K.}~\bibnamefont {Hebeler}},
  \ and\ \bibinfo {author} {\bibfnamefont {A.}~\bibnamefont {Schwenk}},\ }\href
  {\doibase 10.1103/PhysRevC.93.054314} {\bibfield  {journal} {\bibinfo
  {journal} {Phys. Rev. C}\ }\textbf {\bibinfo {volume} {93}},\ \bibinfo
  {pages} {054314} (\bibinfo {year} {2016})},\ \Eprint
  {http://arxiv.org/abs/1510.06728} {arXiv:1510.06728 [nucl-th]} \BibitemShut
  {NoStop}%
\bibitem [{\citenamefont {Baym}\ and\ \citenamefont
  {Chin}(1976)}]{Baym:1975va}%
  \BibitemOpen
  \bibfield  {author} {\bibinfo {author} {\bibfnamefont {G.}~\bibnamefont
  {Baym}}\ and\ \bibinfo {author} {\bibfnamefont {S.~A.}\ \bibnamefont
  {Chin}},\ }\href {\doibase 10.1016/0375-9474(76)90513-3} {\bibfield
  {journal} {\bibinfo  {journal} {Nucl. Phys. A}\ }\textbf {\bibinfo {volume}
  {262}},\ \bibinfo {pages} {527} (\bibinfo {year} {1976})}\BibitemShut
  {NoStop}%
\bibitem [{\citenamefont {Friman}\ and\ \citenamefont
  {Weise}(2019)}]{Friman:2019ncm}%
  \BibitemOpen
  \bibfield  {author} {\bibinfo {author} {\bibfnamefont {B.}~\bibnamefont
  {Friman}}\ and\ \bibinfo {author} {\bibfnamefont {W.}~\bibnamefont {Weise}},\
  }\href {\doibase 10.1103/PhysRevC.100.065807} {\bibfield  {journal} {\bibinfo
   {journal} {Phys. Rev. C}\ }\textbf {\bibinfo {volume} {100}},\ \bibinfo
  {pages} {065807} (\bibinfo {year} {2019})},\ \Eprint
  {http://arxiv.org/abs/1908.09722} {arXiv:1908.09722 [nucl-th]} \BibitemShut
  {NoStop}%
\bibitem [{\citenamefont {Boguta}\ and\ \citenamefont
  {Bodmer}(1977)}]{boguta_relativistic_1977}%
  \BibitemOpen
  \bibfield  {author} {\bibinfo {author} {\bibfnamefont {J.}~\bibnamefont
  {Boguta}}\ and\ \bibinfo {author} {\bibfnamefont {A.}~\bibnamefont
  {Bodmer}},\ }\href {\doibase 10.1016/0375-9474(77)90626-1} {\bibfield
  {journal} {\bibinfo  {journal} {Nuclear Physics A}\ }\textbf {\bibinfo
  {volume} {292}},\ \bibinfo {pages} {413} (\bibinfo {year}
  {1977})}\BibitemShut {NoStop}%
\bibitem [{\citenamefont {Centelles}\ and\ \citenamefont
  {Vi{\~n}as}(1993)}]{centelles_semiclassical_1993}%
  \BibitemOpen
  \bibfield  {author} {\bibinfo {author} {\bibfnamefont {M.}~\bibnamefont
  {Centelles}}\ and\ \bibinfo {author} {\bibfnamefont {X.}~\bibnamefont
  {Vi{\~n}as}},\ }\href {\doibase 10.1016/0375-9474(93)90601-S} {\bibfield
  {journal} {\bibinfo  {journal} {Nuclear Physics A}\ }\textbf {\bibinfo
  {volume} {563}},\ \bibinfo {pages} {173} (\bibinfo {year}
  {1993})}\BibitemShut {NoStop}%
\bibitem [{\citenamefont {Hua}\ \emph {et~al.}(2000)\citenamefont {Hua},
  \citenamefont {Chossy},\ and\ \citenamefont {Stocker}}]{Hua:2000gd}%
  \BibitemOpen
  \bibfield  {author} {\bibinfo {author} {\bibfnamefont {G.}~\bibnamefont
  {Hua}}, \bibinfo {author} {\bibfnamefont {T.~v.}\ \bibnamefont {Chossy}}, \
  and\ \bibinfo {author} {\bibfnamefont {W.}~\bibnamefont {Stocker}},\ }\href
  {\doibase 10.1103/PhysRevC.61.014307} {\bibfield  {journal} {\bibinfo
  {journal} {Phys. Rev. C}\ }\textbf {\bibinfo {volume} {61}},\ \bibinfo
  {pages} {014307} (\bibinfo {year} {2000})}\BibitemShut {NoStop}%
\bibitem [{\citenamefont {Blaizot}\ and\ \citenamefont
  {Grammaticos}(1981)}]{blaizot1981breathing}%
  \BibitemOpen
  \bibfield  {author} {\bibinfo {author} {\bibfnamefont {J.}~\bibnamefont
  {Blaizot}}\ and\ \bibinfo {author} {\bibfnamefont {B.}~\bibnamefont
  {Grammaticos}},\ }\href@noop {} {\bibfield  {journal} {\bibinfo  {journal}
  {Nuclear Physics A}\ }\textbf {\bibinfo {volume} {355}},\ \bibinfo {pages}
  {115} (\bibinfo {year} {1981})}\BibitemShut {NoStop}%
\bibitem [{\citenamefont {Bohr}\ and\ \citenamefont
  {Mottelson}(1998)}]{bohr1998nuclear}%
  \BibitemOpen
  \bibfield  {author} {\bibinfo {author} {\bibfnamefont {A.~N.}\ \bibnamefont
  {Bohr}}\ and\ \bibinfo {author} {\bibfnamefont {B.~R.}\ \bibnamefont
  {Mottelson}},\ }\href@noop {} {\emph {\bibinfo {title} {Nuclear Structure (in
  2 volumes)}}}\ (\bibinfo  {publisher} {World Scientific Publishing Company},\
  \bibinfo {year} {1998})\BibitemShut {NoStop}%
\bibitem [{\citenamefont {Friar}\ and\ \citenamefont
  {Negele}(1975)}]{friar1975theoretical}%
  \BibitemOpen
  \bibfield  {author} {\bibinfo {author} {\bibfnamefont {J.}~\bibnamefont
  {Friar}}\ and\ \bibinfo {author} {\bibfnamefont {J.}~\bibnamefont {Negele}},\
  }\href@noop {} {\bibfield  {journal} {\bibinfo  {journal} {Advances in
  Nuclear Physics: Volume 8}\ ,\ \bibinfo {pages} {219}} (\bibinfo {year}
  {1975})}\BibitemShut {NoStop}%
\bibitem [{\citenamefont {Blaizot}\ \emph {et~al.}(1995)\citenamefont
  {Blaizot}, \citenamefont {Berger}, \citenamefont {Decharge},\ and\
  \citenamefont {Girod}}]{Blaizot:1995zz}%
  \BibitemOpen
  \bibfield  {author} {\bibinfo {author} {\bibfnamefont {J.~P.}\ \bibnamefont
  {Blaizot}}, \bibinfo {author} {\bibfnamefont {J.~F.}\ \bibnamefont {Berger}},
  \bibinfo {author} {\bibfnamefont {J.}~\bibnamefont {Decharge}}, \ and\
  \bibinfo {author} {\bibfnamefont {M.}~\bibnamefont {Girod}},\ }\href
  {\doibase 10.1016/0375-9474(95)00294-B} {\bibfield  {journal} {\bibinfo
  {journal} {Nucl. Phys. A}\ }\textbf {\bibinfo {volume} {591}},\ \bibinfo
  {pages} {435} (\bibinfo {year} {1995})}\BibitemShut {NoStop}%
\bibitem [{\citenamefont {Youngblood}\ \emph {et~al.}(1999)\citenamefont
  {Youngblood}, \citenamefont {Clark},\ and\ \citenamefont
  {Lui}}]{Youngblood:1999zza}%
  \BibitemOpen
  \bibfield  {author} {\bibinfo {author} {\bibfnamefont {D.~H.}\ \bibnamefont
  {Youngblood}}, \bibinfo {author} {\bibfnamefont {H.~L.}\ \bibnamefont
  {Clark}}, \ and\ \bibinfo {author} {\bibfnamefont {Y.~W.}\ \bibnamefont
  {Lui}},\ }\href {\doibase 10.1103/PhysRevLett.82.691} {\bibfield  {journal}
  {\bibinfo  {journal} {Phys. Rev. Lett.}\ }\textbf {\bibinfo {volume} {82}},\
  \bibinfo {pages} {691} (\bibinfo {year} {1999})}\BibitemShut {NoStop}%
\bibitem [{\citenamefont {Patel}\ \emph {et~al.}(2014)\citenamefont {Patel}
  \emph {et~al.}}]{Patel:2014cxa}%
  \BibitemOpen
  \bibfield  {author} {\bibinfo {author} {\bibfnamefont {D.}~\bibnamefont
  {Patel}} \emph {et~al.},\ }\href {\doibase 10.1016/j.physletb.2014.06.073}
  {\bibfield  {journal} {\bibinfo  {journal} {Phys. Lett. B}\ }\textbf
  {\bibinfo {volume} {735}},\ \bibinfo {pages} {387} (\bibinfo {year}
  {2014})},\ \Eprint {http://arxiv.org/abs/1406.6905} {arXiv:1406.6905
  [nucl-ex]} \BibitemShut {NoStop}%
\bibitem [{\citenamefont {Shlomo}\ \emph {et~al.}(2006)\citenamefont {Shlomo},
  \citenamefont {Kolomietz},\ and\ \citenamefont
  {Col{\`o}}}]{shlomo_deducing_2006}%
  \BibitemOpen
  \bibfield  {author} {\bibinfo {author} {\bibfnamefont {S.}~\bibnamefont
  {Shlomo}}, \bibinfo {author} {\bibfnamefont {V.~M.}\ \bibnamefont
  {Kolomietz}}, \ and\ \bibinfo {author} {\bibfnamefont {G.}~\bibnamefont
  {Col{\`o}}},\ }\href {\doibase 10.1140/epja/i2006-10100-3} {\bibfield
  {journal} {\bibinfo  {journal} {Eur. Phys. J. A}\ }\textbf {\bibinfo {volume}
  {30}},\ \bibinfo {pages} {23} (\bibinfo {year} {2006})}\BibitemShut {NoStop}%
\bibitem [{\citenamefont {Sasaki}\ and\ \citenamefont
  {Mishustin}(2010)}]{sasaki_thermodynamics_2010}%
  \BibitemOpen
  \bibfield  {author} {\bibinfo {author} {\bibfnamefont {C.}~\bibnamefont
  {Sasaki}}\ and\ \bibinfo {author} {\bibfnamefont {I.}~\bibnamefont
  {Mishustin}},\ }\href {\doibase 10.1103/PhysRevC.82.035204} {\bibfield
  {journal} {\bibinfo  {journal} {Phys. Rev. C}\ }\textbf {\bibinfo {volume}
  {82}},\ \bibinfo {pages} {035204} (\bibinfo {year} {2010})},\ \bibinfo {note}
  {arXiv:1005.4811 [hep-ph, physics:nucl-th]}\BibitemShut {NoStop}%
\bibitem [{\citenamefont {Mun}\ \emph {et~al.}(2023)\citenamefont {Mun},
  \citenamefont {Shin}, \citenamefont {Paeng}, \citenamefont {Harada},\ and\
  \citenamefont {Kim}}]{mun2023nuclear}%
  \BibitemOpen
  \bibfield  {author} {\bibinfo {author} {\bibfnamefont {M.-H.}\ \bibnamefont
  {Mun}}, \bibinfo {author} {\bibfnamefont {I.~J.}\ \bibnamefont {Shin}},
  \bibinfo {author} {\bibfnamefont {W.-G.}\ \bibnamefont {Paeng}}, \bibinfo
  {author} {\bibfnamefont {M.}~\bibnamefont {Harada}}, \ and\ \bibinfo {author}
  {\bibfnamefont {Y.}~\bibnamefont {Kim}},\ }\href@noop {} {\enquote {\bibinfo
  {title} {Nuclear structure in parity doublet model},}\ } (\bibinfo {year}
  {2023}),\ \Eprint {http://arxiv.org/abs/1805.03402} {arXiv:1805.03402
  [nucl-th]} \BibitemShut {NoStop}%
\bibitem [{\citenamefont {Koch}\ \emph {et~al.}(2023)\citenamefont {Koch},
  \citenamefont {Marczenko}, \citenamefont {Redlich},\ and\ \citenamefont
  {Sasaki}}]{koch2023fluctuations}%
  \BibitemOpen
  \bibfield  {author} {\bibinfo {author} {\bibfnamefont {V.}~\bibnamefont
  {Koch}}, \bibinfo {author} {\bibfnamefont {M.}~\bibnamefont {Marczenko}},
  \bibinfo {author} {\bibfnamefont {K.}~\bibnamefont {Redlich}}, \ and\
  \bibinfo {author} {\bibfnamefont {C.}~\bibnamefont {Sasaki}},\ }\href@noop {}
  {\enquote {\bibinfo {title} {Fluctuations and correlations of baryonic chiral
  partners},}\ } (\bibinfo {year} {2023}),\ \Eprint
  {http://arxiv.org/abs/2308.15794} {arXiv:2308.15794 [hep-ph]} \BibitemShut
  {NoStop}%
\bibitem [{\citenamefont {Yang}\ \emph {et~al.}(2018)\citenamefont {Yang},
  \citenamefont {Liang}, \citenamefont {Bi}, \citenamefont {Chen},
  \citenamefont {Draper}, \citenamefont {Liu},\ and\ \citenamefont
  {Liu}}]{Yang:2018nqn}%
  \BibitemOpen
  \bibfield  {author} {\bibinfo {author} {\bibfnamefont {Y.-B.}\ \bibnamefont
  {Yang}}, \bibinfo {author} {\bibfnamefont {J.}~\bibnamefont {Liang}},
  \bibinfo {author} {\bibfnamefont {Y.-J.}\ \bibnamefont {Bi}}, \bibinfo
  {author} {\bibfnamefont {Y.}~\bibnamefont {Chen}}, \bibinfo {author}
  {\bibfnamefont {T.}~\bibnamefont {Draper}}, \bibinfo {author} {\bibfnamefont
  {K.-F.}\ \bibnamefont {Liu}}, \ and\ \bibinfo {author} {\bibfnamefont
  {Z.}~\bibnamefont {Liu}},\ }\href {\doibase 10.1103/PhysRevLett.121.212001}
  {\bibfield  {journal} {\bibinfo  {journal} {Phys. Rev. Lett.}\ }\textbf
  {\bibinfo {volume} {121}},\ \bibinfo {pages} {212001} (\bibinfo {year}
  {2018})},\ \Eprint {http://arxiv.org/abs/1808.08677} {arXiv:1808.08677
  [hep-lat]} \BibitemShut {NoStop}%
\bibitem [{\citenamefont {Elliott}\ \emph {et~al.}(2013)\citenamefont
  {Elliott}, \citenamefont {Lake}, \citenamefont {Moretto},\ and\ \citenamefont
  {Phair}}]{Elliott:2013pna}%
  \BibitemOpen
  \bibfield  {author} {\bibinfo {author} {\bibfnamefont {J.~B.}\ \bibnamefont
  {Elliott}}, \bibinfo {author} {\bibfnamefont {P.~T.}\ \bibnamefont {Lake}},
  \bibinfo {author} {\bibfnamefont {L.~G.}\ \bibnamefont {Moretto}}, \ and\
  \bibinfo {author} {\bibfnamefont {L.}~\bibnamefont {Phair}},\ }\href
  {\doibase 10.1103/PhysRevC.87.054622} {\bibfield  {journal} {\bibinfo
  {journal} {Phys. Rev. C}\ }\textbf {\bibinfo {volume} {87}},\ \bibinfo
  {pages} {054622} (\bibinfo {year} {2013})}\BibitemShut {NoStop}%
\bibitem [{\citenamefont {Sainio}(2001)}]{sainio_pion-nucleon_2001}%
  \BibitemOpen
  \bibfield  {author} {\bibinfo {author} {\bibfnamefont {M.~E.}\ \bibnamefont
  {Sainio}},\ }\href {http://arxiv.org/abs/hep-ph/0110413} {\enquote {\bibinfo
  {title} {Pion-nucleon sigma-term - a review},}\ } (\bibinfo {year} {2001}),\
  \bibinfo {note} {arXiv:hep-ph/0110413}\BibitemShut {NoStop}%
\bibitem [{\citenamefont {Alarc\'on}(2021)}]{Alarcon:2021dlz}%
  \BibitemOpen
  \bibfield  {author} {\bibinfo {author} {\bibfnamefont {J.~M.}\ \bibnamefont
  {Alarc\'on}},\ }\href {\doibase 10.1140/epjs/s11734-021-00145-6} {\bibfield
  {journal} {\bibinfo  {journal} {Eur. Phys. J. ST}\ }\textbf {\bibinfo
  {volume} {230}},\ \bibinfo {pages} {1609} (\bibinfo {year} {2021})},\ \Eprint
  {http://arxiv.org/abs/2205.01108} {arXiv:2205.01108 [hep-ph]} \BibitemShut
  {NoStop}%
\bibitem [{\citenamefont {Cohen}\ \emph {et~al.}(1992)\citenamefont {Cohen},
  \citenamefont {Furnstahl},\ and\ \citenamefont {Griegel}}]{Cohen:1991nk}%
  \BibitemOpen
  \bibfield  {author} {\bibinfo {author} {\bibfnamefont {T.~D.}\ \bibnamefont
  {Cohen}}, \bibinfo {author} {\bibfnamefont {R.~J.}\ \bibnamefont
  {Furnstahl}}, \ and\ \bibinfo {author} {\bibfnamefont {D.~K.}\ \bibnamefont
  {Griegel}},\ }\href {\doibase 10.1103/PhysRevC.45.1881} {\bibfield  {journal}
  {\bibinfo  {journal} {Phys. Rev. C}\ }\textbf {\bibinfo {volume} {45}},\
  \bibinfo {pages} {1881} (\bibinfo {year} {1992})}\BibitemShut {NoStop}%
\bibitem [{\citenamefont {Hoferichter}\ \emph
  {et~al.}(2016{\natexlab{a}})\citenamefont {Hoferichter}, \citenamefont
  {Ruiz~de Elvira}, \citenamefont {Kubis},\ and\ \citenamefont
  {Mei\ss{}ner}}]{Hoferichter:2015hva}%
  \BibitemOpen
  \bibfield  {author} {\bibinfo {author} {\bibfnamefont {M.}~\bibnamefont
  {Hoferichter}}, \bibinfo {author} {\bibfnamefont {J.}~\bibnamefont {Ruiz~de
  Elvira}}, \bibinfo {author} {\bibfnamefont {B.}~\bibnamefont {Kubis}}, \ and\
  \bibinfo {author} {\bibfnamefont {U.-G.}\ \bibnamefont {Mei\ss{}ner}},\
  }\href {\doibase 10.1016/j.physrep.2016.02.002} {\bibfield  {journal}
  {\bibinfo  {journal} {Phys. Rept.}\ }\textbf {\bibinfo {volume} {625}},\
  \bibinfo {pages} {1} (\bibinfo {year} {2016}{\natexlab{a}})},\ \Eprint
  {http://arxiv.org/abs/1510.06039} {arXiv:1510.06039 [hep-ph]} \BibitemShut
  {NoStop}%
\bibitem [{\citenamefont {Friedman}\ and\ \citenamefont
  {Gal}(2019)}]{friedman_pion-nucleon_2019}%
  \BibitemOpen
  \bibfield  {author} {\bibinfo {author} {\bibfnamefont {E.}~\bibnamefont
  {Friedman}}\ and\ \bibinfo {author} {\bibfnamefont {A.}~\bibnamefont {Gal}},\
  }\href {\doibase 10.1016/j.physletb.2019.03.036} {\bibfield  {journal}
  {\bibinfo  {journal} {Physics Letters B}\ }\textbf {\bibinfo {volume}
  {792}},\ \bibinfo {pages} {340} (\bibinfo {year} {2019})}\BibitemShut
  {NoStop}%
\bibitem [{\citenamefont {Gasser}\ \emph {et~al.}(1991)\citenamefont {Gasser},
  \citenamefont {Leutwyler},\ and\ \citenamefont {Sainio}}]{Gasser:1990ce}%
  \BibitemOpen
  \bibfield  {author} {\bibinfo {author} {\bibfnamefont {J.}~\bibnamefont
  {Gasser}}, \bibinfo {author} {\bibfnamefont {H.}~\bibnamefont {Leutwyler}}, \
  and\ \bibinfo {author} {\bibfnamefont {M.~E.}\ \bibnamefont {Sainio}},\
  }\href {\doibase 10.1016/0370-2693(91)91393-A} {\bibfield  {journal}
  {\bibinfo  {journal} {Phys. Lett. B}\ }\textbf {\bibinfo {volume} {253}},\
  \bibinfo {pages} {252} (\bibinfo {year} {1991})}\BibitemShut {NoStop}%
\bibitem [{\citenamefont {Agadjanov}\ \emph {et~al.}(2023)\citenamefont
  {Agadjanov}, \citenamefont {Djukanovic}, \citenamefont {von Hippel},
  \citenamefont {Meyer}, \citenamefont {Ottnad},\ and\ \citenamefont
  {Wittig}}]{Agadjanov:2023jha}%
  \BibitemOpen
  \bibfield  {author} {\bibinfo {author} {\bibfnamefont {A.}~\bibnamefont
  {Agadjanov}}, \bibinfo {author} {\bibfnamefont {D.}~\bibnamefont
  {Djukanovic}}, \bibinfo {author} {\bibfnamefont {G.}~\bibnamefont {von
  Hippel}}, \bibinfo {author} {\bibfnamefont {H.~B.}\ \bibnamefont {Meyer}},
  \bibinfo {author} {\bibfnamefont {K.}~\bibnamefont {Ottnad}}, \ and\ \bibinfo
  {author} {\bibfnamefont {H.}~\bibnamefont {Wittig}},\ }\href@noop {} {\
  (\bibinfo {year} {2023})},\ \Eprint {http://arxiv.org/abs/2303.08741}
  {arXiv:2303.08741 [hep-lat]} \BibitemShut {NoStop}%
\bibitem [{\citenamefont {Hoferichter}\ \emph
  {et~al.}(2016{\natexlab{b}})\citenamefont {Hoferichter}, \citenamefont
  {Ruiz~de Elvira}, \citenamefont {Kubis},\ and\ \citenamefont
  {Mei\ss{}ner}}]{Hoferichter:2016ocj}%
  \BibitemOpen
  \bibfield  {author} {\bibinfo {author} {\bibfnamefont {M.}~\bibnamefont
  {Hoferichter}}, \bibinfo {author} {\bibfnamefont {J.}~\bibnamefont {Ruiz~de
  Elvira}}, \bibinfo {author} {\bibfnamefont {B.}~\bibnamefont {Kubis}}, \ and\
  \bibinfo {author} {\bibfnamefont {U.-G.}\ \bibnamefont {Mei\ss{}ner}},\
  }\href {\doibase 10.1016/j.physletb.2016.06.038} {\bibfield  {journal}
  {\bibinfo  {journal} {Phys. Lett. B}\ }\textbf {\bibinfo {volume} {760}},\
  \bibinfo {pages} {74} (\bibinfo {year} {2016}{\natexlab{b}})},\ \Eprint
  {http://arxiv.org/abs/1602.07688} {arXiv:1602.07688 [hep-lat]} \BibitemShut
  {NoStop}%
\bibitem [{\citenamefont {Hoferichter}\ \emph {et~al.}(2023)\citenamefont
  {Hoferichter}, \citenamefont {de~Elvira}, \citenamefont {Kubis},\ and\
  \citenamefont {Mei\ss{}ner}}]{Hoferichter:2023ptl}%
  \BibitemOpen
  \bibfield  {author} {\bibinfo {author} {\bibfnamefont {M.}~\bibnamefont
  {Hoferichter}}, \bibinfo {author} {\bibfnamefont {J.~R.}\ \bibnamefont
  {de~Elvira}}, \bibinfo {author} {\bibfnamefont {B.}~\bibnamefont {Kubis}}, \
  and\ \bibinfo {author} {\bibfnamefont {U.-G.}\ \bibnamefont {Mei\ss{}ner}},\
  }\href@noop {} {\  (\bibinfo {year} {2023})},\ \Eprint
  {http://arxiv.org/abs/2305.07045} {arXiv:2305.07045 [hep-ph]} \BibitemShut
  {NoStop}%
\bibitem [{\citenamefont {Birse}\ and\ \citenamefont
  {McGovern}(1992)}]{Birse:1992uxe}%
  \BibitemOpen
  \bibfield  {author} {\bibinfo {author} {\bibfnamefont {M.~C.}\ \bibnamefont
  {Birse}}\ and\ \bibinfo {author} {\bibfnamefont {J.~A.}\ \bibnamefont
  {McGovern}},\ }\href {\doibase 10.1016/0370-2693(92)91169-A} {\bibfield
  {journal} {\bibinfo  {journal} {Phys. Lett. B}\ }\textbf {\bibinfo {volume}
  {292}},\ \bibinfo {pages} {242} (\bibinfo {year} {1992})}\BibitemShut
  {NoStop}%
\bibitem [{\citenamefont {Birse}\ and\ \citenamefont
  {McGovern}(1993)}]{Birse:1993hx}%
  \BibitemOpen
  \bibfield  {author} {\bibinfo {author} {\bibfnamefont {M.~C.}\ \bibnamefont
  {Birse}}\ and\ \bibinfo {author} {\bibfnamefont {J.~A.}\ \bibnamefont
  {McGovern}},\ }\href {\doibase 10.1016/0370-2693(93)90925-8} {\bibfield
  {journal} {\bibinfo  {journal} {Phys. Lett. B}\ }\textbf {\bibinfo {volume}
  {309}},\ \bibinfo {pages} {231} (\bibinfo {year} {1993})}\BibitemShut
  {NoStop}%
\bibitem [{\citenamefont {Delorme}\ \emph {et~al.}(1996)\citenamefont
  {Delorme}, \citenamefont {Chanfray},\ and\ \citenamefont
  {Ericson}}]{Delorme:1996cc}%
  \BibitemOpen
  \bibfield  {author} {\bibinfo {author} {\bibfnamefont {J.}~\bibnamefont
  {Delorme}}, \bibinfo {author} {\bibfnamefont {G.}~\bibnamefont {Chanfray}}, \
  and\ \bibinfo {author} {\bibfnamefont {M.}~\bibnamefont {Ericson}},\ }\href
  {\doibase 10.1016/0375-9474(96)80001-8} {\bibfield  {journal} {\bibinfo
  {journal} {Nucl. Phys. A}\ }\textbf {\bibinfo {volume} {603}},\ \bibinfo
  {pages} {239} (\bibinfo {year} {1996})},\ \Eprint
  {http://arxiv.org/abs/nucl-th/9603005} {arXiv:nucl-th/9603005} \BibitemShut
  {NoStop}%
\bibitem [{\citenamefont {Chanfray}\ \emph {et~al.}(1998)\citenamefont
  {Chanfray}, \citenamefont {Delorme},\ and\ \citenamefont
  {Ericson}}]{Chanfray:1998hr}%
  \BibitemOpen
  \bibfield  {author} {\bibinfo {author} {\bibfnamefont {G.}~\bibnamefont
  {Chanfray}}, \bibinfo {author} {\bibfnamefont {J.}~\bibnamefont {Delorme}}, \
  and\ \bibinfo {author} {\bibfnamefont {M.}~\bibnamefont {Ericson}},\ }\href
  {\doibase 10.1016/S0375-9474(98)00237-1} {\bibfield  {journal} {\bibinfo
  {journal} {Nucl. Phys. A}\ }\textbf {\bibinfo {volume} {637}},\ \bibinfo
  {pages} {421} (\bibinfo {year} {1998})},\ \Eprint
  {http://arxiv.org/abs/nucl-th/9801020} {arXiv:nucl-th/9801020} \BibitemShut
  {NoStop}%
\bibitem [{\citenamefont {Dmitrasinovic}\ and\ \citenamefont
  {Myhrer}(2000)}]{Dmitrasinovic:1999mf}%
  \BibitemOpen
  \bibfield  {author} {\bibinfo {author} {\bibfnamefont {V.}~\bibnamefont
  {Dmitrasinovic}}\ and\ \bibinfo {author} {\bibfnamefont {F.}~\bibnamefont
  {Myhrer}},\ }\href {\doibase 10.1103/PhysRevC.61.025205} {\bibfield
  {journal} {\bibinfo  {journal} {Phys. Rev. C}\ }\textbf {\bibinfo {volume}
  {61}},\ \bibinfo {pages} {025205} (\bibinfo {year} {2000})},\ \Eprint
  {http://arxiv.org/abs/hep-ph/9911320} {arXiv:hep-ph/9911320} \BibitemShut
  {NoStop}%
\bibitem [{\citenamefont {Kaiser}\ \emph {et~al.}(2008)\citenamefont {Kaiser},
  \citenamefont {de~Homont},\ and\ \citenamefont {Weise}}]{Kaiser:2007nv}%
  \BibitemOpen
  \bibfield  {author} {\bibinfo {author} {\bibfnamefont {N.}~\bibnamefont
  {Kaiser}}, \bibinfo {author} {\bibfnamefont {P.}~\bibnamefont {de~Homont}}, \
  and\ \bibinfo {author} {\bibfnamefont {W.}~\bibnamefont {Weise}},\ }\href
  {\doibase 10.1103/PhysRevC.77.025204} {\bibfield  {journal} {\bibinfo
  {journal} {Phys. Rev. C}\ }\textbf {\bibinfo {volume} {77}},\ \bibinfo
  {pages} {025204} (\bibinfo {year} {2008})},\ \Eprint
  {http://arxiv.org/abs/0711.3154} {arXiv:0711.3154 [nucl-th]} \BibitemShut
  {NoStop}%
\bibitem [{\citenamefont {Haensch}\ \emph {et~al.}(2023)\citenamefont
  {Haensch}, \citenamefont {Rennecke},\ and\ \citenamefont {von
  Smekal}}]{haensch_medium_2023}%
  \BibitemOpen
  \bibfield  {author} {\bibinfo {author} {\bibfnamefont {M.}~\bibnamefont
  {Haensch}}, \bibinfo {author} {\bibfnamefont {F.}~\bibnamefont {Rennecke}}, \
  and\ \bibinfo {author} {\bibfnamefont {L.}~\bibnamefont {von Smekal}},\
  }\href {http://arxiv.org/abs/2308.16244} {\enquote {\bibinfo {title} {Medium
  {Induced} {Mixing} and {Critical} {Modes} in {QCD}},}\ } (\bibinfo {year}
  {2023}),\ \bibinfo {note} {arXiv:2308.16244 [hep-ph, physics:hep-th,
  physics:nucl-th]}\BibitemShut {NoStop}%
\bibitem [{\citenamefont {Kapusta}\ and\ \citenamefont
  {Gale}(2007)}]{kapusta2007finite}%
  \BibitemOpen
  \bibfield  {author} {\bibinfo {author} {\bibfnamefont {J.~I.}\ \bibnamefont
  {Kapusta}}\ and\ \bibinfo {author} {\bibfnamefont {C.}~\bibnamefont {Gale}},\
  }\href@noop {} {\emph {\bibinfo {title} {Finite-temperature field theory:
  Principles and applications}}}\ (\bibinfo  {publisher} {Cambridge university
  press},\ \bibinfo {year} {2007})\BibitemShut {NoStop}%
\bibitem [{\citenamefont {Larionov}\ and\ \citenamefont {von
  Smekal}(2022)}]{Larionov:2021ycq}%
  \BibitemOpen
  \bibfield  {author} {\bibinfo {author} {\bibfnamefont {A.~B.}\ \bibnamefont
  {Larionov}}\ and\ \bibinfo {author} {\bibfnamefont {L.}~\bibnamefont {von
  Smekal}},\ }\href {\doibase 10.1103/PhysRevC.105.034914} {\bibfield
  {journal} {\bibinfo  {journal} {Phys. Rev. C}\ }\textbf {\bibinfo {volume}
  {105}},\ \bibinfo {pages} {034914} (\bibinfo {year} {2022})},\ \Eprint
  {http://arxiv.org/abs/2109.03556} {arXiv:2109.03556 [nucl-th]} \BibitemShut
  {NoStop}%
\bibitem [{\citenamefont {Marczenko}\ \emph {et~al.}(2023)\citenamefont
  {Marczenko}, \citenamefont {Redlich},\ and\ \citenamefont
  {Sasaki}}]{Marczenko_2023}%
  \BibitemOpen
  \bibfield  {author} {\bibinfo {author} {\bibfnamefont {M.}~\bibnamefont
  {Marczenko}}, \bibinfo {author} {\bibfnamefont {K.}~\bibnamefont {Redlich}},
  \ and\ \bibinfo {author} {\bibfnamefont {C.}~\bibnamefont {Sasaki}},\ }\href
  {\doibase 10.1103/physrevd.107.054046} {\bibfield  {journal} {\bibinfo
  {journal} {Physical Review D}\ }\textbf {\bibinfo {volume} {107}} (\bibinfo
  {year} {2023}),\ 10.1103/physrevd.107.054046}\BibitemShut {NoStop}%
\bibitem [{\citenamefont {Skokov}\ \emph {et~al.}(2010)\citenamefont {Skokov},
  \citenamefont {Friman}, \citenamefont {Nakano}, \citenamefont {Redlich},\
  and\ \citenamefont {Schaefer}}]{Skokov:2010sf}%
  \BibitemOpen
  \bibfield  {author} {\bibinfo {author} {\bibfnamefont {V.}~\bibnamefont
  {Skokov}}, \bibinfo {author} {\bibfnamefont {B.}~\bibnamefont {Friman}},
  \bibinfo {author} {\bibfnamefont {E.}~\bibnamefont {Nakano}}, \bibinfo
  {author} {\bibfnamefont {K.}~\bibnamefont {Redlich}}, \ and\ \bibinfo
  {author} {\bibfnamefont {B.~J.}\ \bibnamefont {Schaefer}},\ }\href {\doibase
  10.1103/PhysRevD.82.034029} {\bibfield  {journal} {\bibinfo  {journal} {Phys.
  Rev. D}\ }\textbf {\bibinfo {volume} {82}},\ \bibinfo {pages} {034029}
  (\bibinfo {year} {2010})},\ \Eprint {http://arxiv.org/abs/1005.3166}
  {arXiv:1005.3166 [hep-ph]} \BibitemShut {NoStop}%
\end{thebibliography}%

\end{document}